\documentclass{article}

\usepackage[english]{babel}

\usepackage[letterpaper,top=2cm,bottom=2cm,left=2.7cm,right=2.7cm,marginparwidth=1.75cm]{geometry}

\usepackage{amsmath}
\usepackage{amssymb}
\usepackage{amsthm}
\usepackage{amsfonts}
\usepackage{mathtools}
\usepackage[colorlinks=true,allcolors=blue,hypertexnames=false]{hyperref}
\usepackage{booktabs}
\usepackage{float}
\usepackage{tikz}
\usepackage{microtype}
\usepackage{placeins}
\usepackage{flafter}
\usepackage{comment}

\usepackage[T1]{fontenc}

\newtheorem{theorem}{Theorem}[section]
\newtheorem{proposition}[theorem]{Proposition}
\newtheorem{lemma}[theorem]{Lemma}
\newtheorem{corollary}[theorem]{Corollary}
\theoremstyle{definition}
\newtheorem{definition}[theorem]{Definition}
\theoremstyle{remark}
\newtheorem{remark}[theorem]{Remark}

\newcommand{\vac}{|\Omega\rangle}
\newcommand{\C}{\mathbb C}
\newcommand{\R}{\mathbb R}
\newcommand{\PP}{\mathbb P}
\newcommand{\Cl}{\operatorname{Cl}}
\newcommand{\End}{\operatorname{End}}
\newcommand{\Spin}{\operatorname{Spin}}
\newcommand{\Span}{\operatorname{Span}}
\newcommand{\ket}[1]{\lvert #1\rangle}
\newcommand{\norm}[1]{\lVert #1\rVert}
\newcommand{\symdiff}{\mathbin{\triangle}}

\newcommand{\affnum}[1]{\textsuperscript{\normalfont\fontsize{8}{9}\selectfont #1}}%
\definecolor{spinorsupport}{RGB}{32,79,130}
\definecolor{spinormomentum}{RGB}{164,67,24}
\tikzset{
  spinor vertex/.style={circle,draw=black!65,fill=white,
    line width=.65pt,inner sep=2.2pt},
  spinor selected/.style={circle,draw=spinorsupport,fill=spinorsupport!16,
    line width=1.15pt,inner sep=2.8pt},
  spinor momentum/.style={circle,draw=spinormomentum,fill=spinormomentum,
    inner sep=2.5pt},
  spinor nonedge/.style={draw=spinorsupport,line width=1.05pt,
    dash pattern=on 4pt off 2.5pt},
  spinor heading/.style={font=\normalsize\bfseries,anchor=west},
  spinor formula/.style={font=\normalsize,align=center},
  spinor state/.style={font=\normalsize,text=spinorsupport,
    fill=spinorsupport!5,rounded corners=2pt,
    inner xsep=7pt,inner ysep=6pt}
}

\newcommand{\DfourWeightFigure}{%
\begin{figure}[H]
\centering
\begin{tikzpicture}[x=1cm,y=1cm,font=\normalsize]
  \node[spinor heading] at (0,2.75) {(a) Fock weights};
  \node[spinor heading] at (8.1,2.75) {(b) The support $\Sigma_8$};
  \begin{scope}[shift={(3.7,0)}]
    \foreach \v/\a in {v1/90,v2/45,v3/0,v4/-45,v5/-90,v6/-135,v7/180,v8/135}
      \coordinate (\v) at (\a:1.82);
    \foreach \a/\b in {
      v1/v2,v1/v3,v1/v4,v1/v6,v1/v7,v1/v8,
      v2/v3,v2/v4,v2/v5,v2/v7,v2/v8,
      v3/v4,v3/v5,v3/v6,v3/v8,
      v4/v5,v4/v6,v4/v7,
      v5/v6,v5/v7,v5/v8,v6/v7,v6/v8,v7/v8}
      \draw[black!38,line width=.45pt] (\a)--(\b);
    \draw[spinor nonedge] (v1)--(v5)
      node[midway,fill=white,inner sep=3pt,text=spinorsupport] {$d_H=4$};
    \foreach \v in {v1,v5}\node[spinor selected] at (\v) {};
    \foreach \v in {v2,v3,v4,v6,v7,v8}\node[spinor vertex] at (\v) {};
    \node[above=5pt,text=spinorsupport] at (v1) {$\ket\varnothing$};
    \node[below=5pt,text=spinorsupport] at (v5) {$\ket{1234}$};
    \node[above right=3pt] at (v2) {$\ket{12}$};
    \node[right=5pt] at (v3) {$\ket{13}$};
    \node[below right=3pt] at (v4) {$\ket{14}$};
    \node[below left=3pt] at (v6) {$\ket{34}$};
    \node[left=5pt] at (v7) {$\ket{24}$};
    \node[above left=3pt] at (v8) {$\ket{23}$};
  \end{scope}
  \node[spinor formula] at (11.35,1.95)
    {$\eta_\varnothing=\tfrac12(1,1,1,1)$,\qquad
     $\eta_{1234}=-\eta_\varnothing$};
  \node[spinor formula] at (11.35,1.2)
    {$p_\varnothing=p_{1234}=\tfrac12$};
  \draw[spinor nonedge] (8.9,.35)--(13.8,.35);
  \node[spinor selected] at (8.9,.35) {};
  \node[spinor selected] at (13.8,.35) {};
  \node[spinor momentum] at (11.35,.35) {};
  \node[below=6pt] at (8.9,.35) {$\eta_\varnothing$};
  \node[below=6pt] at (13.8,.35) {$\eta_{1234}$};
  \node[below=6pt,text=spinormomentum] at (11.35,.35) {$\beta=0$};
  \node[spinor formula] at (11.35,-.95)
    {$\beta=\tfrac12\eta_\varnothing+\tfrac12\eta_{1234}$};
  \node[spinor state] at (11.35,-1.8)
    {$\displaystyle\psi_8=\frac{\ket\varnothing+\ket{1234}}{\sqrt2}$};
  \draw[black!60,line width=.65pt] (.25,-2.8)--(1.1,-2.8);
  \node[anchor=west] at (1.3,-2.8) {root edge: $d_H=2$};
  \draw[spinor nonedge] (8.1,-2.8)--(8.95,-2.8);
  \node[anchor=west] at (9.15,-2.8) {selected nonedge: $d_H=4$};
\end{tikzpicture}
\caption{The $N=8$ construction. (a) A vertex labelled $\ket I$
represents the weight $\eta_I$ of \eqref{eq:general-fock-weight} and
$d_H(I,J)=|I\symdiff J|$. Solid lines are all root edges.
After fixing the vacuum, $\ket{1234}$ is its only root-distinct partner.
The other three complementary pairs are also nonedges, but are not
highlighted. (b) Equal probabilities place the momentum at the origin,
as in \eqref{eq:D4-antipodal-midpoint}. The displayed state is normalized.}
\label{fig:D4-root-graph}
\end{figure}%
}

\newcommand{\DsixWeightFigure}{%
\begin{figure}[htbp]
\centering
\begin{tikzpicture}[x=1cm,y=1cm,font=\normalsize]
  \node[spinor heading] at (0,4.7) {(a) The support $\Sigma_{12}$};
  \node[spinor heading] at (8.4,4.7) {(b) Coefficients and momentum};
  \coordinate (A) at (3.7,3.55);
  \coordinate (B) at ({3.7-3.3/sqrt(3)},.25);
  \coordinate (C) at ({3.7+3.3/sqrt(3)},.25);
  \coordinate (beta) at (3.7,1.35);
  \fill[spinorsupport!4] (A)--(B)--(C)--cycle;
  \draw[black!35,line width=.55pt] (A)--(3.7,.25);
  \draw[black!35,line width=.55pt] (B)--({3.7+1.65/sqrt(3)},1.9);
  \draw[black!35,line width=.55pt] (C)--({3.7-1.65/sqrt(3)},1.9);
  \draw[spinor nonedge] (A)--(B)
    node[midway,sloped,above=4pt,fill=white,inner sep=2pt] {$d_H=4$};
  \draw[spinor nonedge] (A)--(C)
    node[midway,sloped,above=4pt,fill=white,inner sep=2pt] {$d_H=4$};
  \draw[spinor nonedge] (B)--(C)
    node[midway,below=5pt] {$d_H=4$};
  \foreach \v in {A,B,C}\node[spinor selected] at (\v) {};
  \node[above=6pt,align=center] at (A)
    {$\eta_\varnothing$\quad$(\ket\varnothing)$};
  \node[below left=6pt,align=center] at (B)
    {$\eta_{1234}$\\[2pt]$(\ket{1234})$};
  \node[below right=6pt,align=center] at (C)
    {$\eta_{3456}$\\[2pt]$(\ket{3456})$};
  \node[spinor momentum] at (beta) {};
  \node[above right=3pt,fill=spinorsupport!4,inner sep=2pt,
    text=spinormomentum] at (beta) {$\beta$};
  \node[spinor formula] at (3.7,-1)
    {$\displaystyle\beta=
       \frac{\eta_\varnothing+\eta_{1234}+\eta_{3456}}{3}$};
  \node[spinor formula] at (11.7,3.7)
    {$\displaystyle G_{\Sigma_{12}}=\frac12
       \begin{pmatrix}3&-1&-1\\-1&3&-1\\-1&-1&3\end{pmatrix}$};
  \node[spinor formula] at (11.7,2.5)
    {$p_\varnothing=p_{1234}=p_{3456}=\tfrac13$};
  \node[spinor formula,text=spinormomentum] at (11.7,1.55)
    {$\beta=\tfrac16(1,1,-1,-1,1,1)$};
  \draw[->,black!70,line width=.7pt] (11.7,1.12)--(11.7,.55);
  \node[anchor=west,align=left,font=\normalsize] at (12,.84)
    {change signs in\\coordinates $3,4$};
  \node[spinor formula] at (11.7,.18)
    {$\beta^+=\tfrac16(1,1,1,1,1,1)$};
  \node[spinor formula] at (11.7,-.85)
    {$\norm\beta^2=\norm{\beta^+}^2=\tfrac16$};
  \node[spinor state] at (7.6,-2.03)
    {$\displaystyle\psi_{12}=
       \frac{\ket\varnothing+\ket{1234}+\ket{3456}}{\sqrt3}$};
\end{tikzpicture}
\caption{The $N=12$ three-term critical representative.
(a) The weights belonging to \eqref{eq:D6-triple-support} span a triangle;
the dashed sides indicate Hamming distance four, so none is a root edge.
The medians meet at the barycentre $\beta$, the nonzero momentum value
in the displayed Fock frame. (b) In the order
$\varnothing,1234,3456$, the Gram system gives equal probabilities.
Changing the signs of coordinates $3,4$ is a $D_6$ Weyl transformation
and gives \eqref{eq:D6-triple-beta}. The diagram is schematic; the
coordinates shown on the right are the actual momentum values.}
\label{fig:D6-root-graph}
\end{figure}%
}
\newcommand{\MomentumMethodFigure}{
\begin{figure}[t] 
\centering 
\begin{tikzpicture}[font=\small, 
  flow/.style={->,>=stealth,line width=.6pt,black!55}, 
  crit/.style={draw=spinormomentum,line width=1.1pt}] 
  \node[anchor=west,font=\normalsize\bfseries] at (-0.5,4.45) {(a) In $\PP(S)$}; 
  \path[fill=spinorsupport!7] 
    (0,0) .. controls (2,-0.5) and (4,0.9) .. (6.2,0.3) 
    .. controls (6.8,1.5) and (6.6,2.8) .. (5.6,3.5) 
    .. controls (4,4.2) and (1.8,3.6) .. (0.7,3.2) 
    .. controls (0.1,2.4) and (-0.3,0.9) .. (0,0); 
  \draw[spinorsupport!50,densely dotted,line width=.6pt] 
    (6.2,0.3) .. controls (6.8,1.5) and (6.6,2.8) .. (5.6,3.5) 
    .. controls (4,4.2) and (1.8,3.6) .. (0.7,3.2) 
    .. controls (0.1,2.4) and (-0.3,0.9) .. (0,0); 
  \draw[spinorsupport,line width=1.6pt] 
    (0,0) .. controls (2,-0.5) and (4,0.9) .. (6.2,0.3); 
  \draw[flow] (0.8,2.95) .. controls (1.5,3.0) and (2.2,2.45) .. (2.78,2.03); 
  \draw[flow] (1.3,1.0) .. controls (1.9,1.1) and (2.4,1.45) .. (2.86,1.82); 
  \draw[flow] (2.6,3.45) .. controls (3.0,3.4) and (3.3,2.95) .. (3.42,2.46); 
  \draw[flow] (5.9,3.0) .. controls (5.4,3.0) and (4.8,2.65) .. (4.33,2.36); 
  \draw[flow] (5.9,1.2) .. controls (5.4,1.3) and (4.8,1.8) .. (4.4,2.12); 
  \draw[flow] (4.6,0.95) .. controls (4.3,1.1) and (3.95,1.4) .. (3.78,1.74); 
  \draw[flow] plot[smooth] coordinates {(1.20,0.07) (1.68,0.11) (2.17,0.19) (2.62,0.28)}; 
  \draw[flow] plot[smooth] coordinates {(5.55,0.60) (5.03,0.63) (4.52,0.61) (4.05,0.54)}; 
  \draw[crit,fill=spinormomentum!12,rotate around={12:(3.6,2.1)}] 
    (3.6,2.1) ellipse (0.75 and 0.3); 
  \node[font=\footnotesize,text=spinormomentum!80!black] at (3.6,2.1) {$K\cdot[\psi_c]$}; 
  \fill[spinormomentum] (3.333,0.25) circle (2.4pt); 
  \node[below,font=\footnotesize,text=spinormomentum!80!black] at (3.333,0.12) {$K\cdot[\varphi_c]$}; 
  \fill[black!80] (0,0) circle (2.4pt); 
  \node[below left,font=\footnotesize] at (0.05,-0.02) {pure spinors}; 
  \node[text=spinorsupport] at (4.7,4.22) {$G\cdot[\psi]$}; 
  \node[right,text=spinorsupport] at (6.2,0.18) {$G\cdot[\varphi]$}; 
  \draw[->,>=stealth,line width=.8pt,black!60] (7.0,2.1) -- node[above] {$\mu$} (7.95,2.1); 
  \begin{scope}[shift={(8.35,0)},scale=0.9] 
    \node[anchor=west,font=\normalsize\bfseries] at (-1.9,4.94) {(b) In $\mathfrak t_+$}; 
    \fill[black!3] (0,0) -- (10:5.6) arc (10:75:5.6) -- cycle; 
    \draw[black!55,line width=.6pt] (10:5.6) -- (0,0) -- (75:5.6); 
    \filldraw[fill=spinorsupport!10,draw=spinorsupport,line width=.8pt] 
      (2.2,0.8) -- (3.7,1.0) -- (4.3,2.4) -- (3.2,3.6) -- (0.9,2.6) -- cycle; 
    \filldraw[fill=spinorsupport!30,draw=spinorsupport,line width=.8pt] 
      (4.3,2.4) -- (2.7,2.7) -- (3.6,1.3) -- cycle; 
    \draw[spinormomentum!70,densely dashed] (10:2.252) arc (10:75:2.252); 
    \draw[spinormomentum!45,densely dashed] (22:3.731) arc (22:44:3.731); 
    \draw[spinormomentum,line width=.7pt,dashed] (0,0) -- (1.826,1.318); 
    \draw[spinormomentum,line width=.5pt] (1.648,1.189) -- (1.777,1.011) -- (1.955,1.140); 
    \fill[spinormomentum] (1.826,1.318) circle (2.4pt); 
    \fill[spinormomentum!75] (3.139,2.018) circle (2.2pt); 
    \fill[black!80] (4.3,2.4) circle (2.2pt); 
    \node[left,text=spinormomentum!80!black] at (1.80,1.36) {$\beta_\psi$}; 
    \node[below left,font=\footnotesize,text=spinormomentum!80!black] at (3.16,2.02) {$\beta_\varphi$}; 
    \node[right,font=\footnotesize] at (4.3,2.4) {$\eta_\varnothing$}; 
    \node[text=spinorsupport] at (1.75,2.45) {$\Delta_\psi$}; 
    \node[font=\footnotesize,text=spinorsupport] at (3.62,2.12) {$\Delta_\varphi$}; 
    \node[below left] at (0,0) {$0$}; 
  \end{scope} 
\end{tikzpicture} 
\caption{ Schematic picture of the method. (a) The compact group $K$ 
preserves $\norm\mu^2$, and the negative gradient flow of $\norm\mu^2$ moves 
each point within its $G$-orbit. When the orbit contains a critical point, as 
in all our cases, every flow line ends on the single $K$-orbit of critical 
points, where $\norm\mu^2$ is smallest 
\cite{Kirwan_cohomology_84,GeorgoulasRobbinSalamon2021}. A smaller orbit 
$G\cdot[\varphi]$ in the closure has its own critical $K$-orbit, and the pure 
spinors form the smallest orbit, which lies in every closure. (b) Moved into 
the positive Weyl chamber, the momentum values of an orbit closure fill a 
convex polytope \cite{Brion1987}. The critical value $\beta_\psi$ is the point 
of this polytope closest to the origin. An orbit in the closure has a 
smaller polytope, so its closest point cannot be nearer to the origin, and in 
our cases it is strictly farther; this is 
Proposition~\ref{prop:momentum-degeneration}. Every polytope contains the 
highest weight $\eta_\varnothing$, the momentum value of the pure spinors.} 
\label{fig:method} 
\end{figure}
} 

\newcommand{\AllDimensionsFigure}{%
\begin{figure}[t]
\centering
\begin{tikzpicture}[font=\small,x=0.95cm,y=6cm,
  orbit/.style={circle,fill=spinorsupport,inner sep=0pt,minimum size=4.4pt},
  open/.style={fill=spinormomentum}]
  \fill[black!6] (12.62,0) rectangle (13.38,1);
  \node[rotate=90,font=\scriptsize,black!60] at (13,0.5) {infinitely many orbits};
  \foreach \e/\t in {0/0,0.25/\frac14,0.5/\frac12,0.75/\frac34,1/1} {
    \draw[black!12] (1.45,\e) -- (14.4,\e);
    \node[anchor=east,font=\footnotesize] at (1.4,\e) {$\t$};
  }
  \draw[black!55,line width=.5pt] (1.45,-0.02) -- (1.45,1.02);
  \node[rotate=90] at (0.45,0.5) {$\frac4n\norm\beta^2$};
  \node[anchor=south west,font=\scriptsize,black!60] at (1.5,0) {$\mu=0$};
  \foreach \n in {2,...,14} \node[font=\footnotesize] at (\n,-0.065) {$\n$};
  \node at (8,-0.14) {dimension $N$};
  \node[anchor=east,font=\scriptsize,black!60] at (1.4,1.12) {orbits};
  \foreach \n/\c in {2/1,3/1,4/1,5/1,6/1,7/2,8/2,9/3,10/2,11/5,12/4,13/\infty,14/9}
    \node[font=\scriptsize,black!60] at (\n,1.12) {$\c$};
  \foreach \n/\e in {7/1,8/1,9/1,9/0.25,10/1,11/1,11/0.4,11/0.2,11/0.1,
                     12/1,12/0.3333,12/0.1111,
                     14/1,14/0.4286,14/0.2381,13.86/0.1429,14.14/0.1429,
                     14/0.0857,13.8/0.0476,14.22/0.0204}
    \node[orbit] at (\n,\e) {};
  \foreach \n/\e in {2/1,3/1,4/1,5/1,6/1,7/0,8/0,9/0,10/0.2,11/0,12/0,14/0}
    \path[open] (\n,\e) +(0,3.8pt) -- +(3.8pt,0) -- +(0,-3.8pt) -- +(-3.8pt,0) -- cycle;
  \foreach \x/\e/\ly/\lab in {14.06/1/1/I,14.06/0.4286/0.4286/II,14.06/0.2381/0.245/III,
                              14.2/0.1429/0.165/{IV, V},14.06/0.0857/0.105/VI,
                              13.86/0.0476/0.055/VII,14.28/0.0204/0.005/VIII,14.1/0/-0.045/IX}
    \draw[black!40,line width=.4pt] (\x,\e) -- (14.55,\ly)
      node[anchor=west,font=\scriptsize,black!80] {\lab};
  \node[orbit] at (2.1,0.62) {};
  \node[anchor=west,font=\footnotesize] at (2.3,0.62) {an orbit};
  \path[open] (2.1,0.55) +(0,3.8pt) -- +(3.8pt,0) -- +(0,-3.8pt) -- +(-3.8pt,0) -- cycle;
  \node[anchor=west,font=\footnotesize] at (2.3,0.55) {the open orbit};
\end{tikzpicture}
\caption{The critical values $\norm\beta^2$, normalized to $\frac4n\norm\beta^2$,
in all dimensions with finitely many orbits
(Tables~\ref{tab:finite-even-low}, \ref{tab:s14-orbits}
and~\ref{tab:finite-odd}; $N=1$, where $\PP(S)$ is a point, is left out).
The normalized value is $1$ for the pure spinors and $0$ when $\mu=0$. Each
marker is one nonzero projective orbit, and the numbers at the top count them;
in dimension fourteen markers at nearly equal heights are spread sideways.
Diamonds mark the open orbit. It reaches $\mu=0$ in every dimension except
$2\le N\le6$ and $N=10$; by the Kempf--Ness theorem \cite{KempfNess1979} these
are exactly the cases in which the spin module has no nonconstant polynomial
invariant. In dimension fourteen the orbits IV and V have the same
$\norm\beta^2=\frac14$ but different chamber values $(\frac12,0^6)$ and
$((\frac14)^4,0^3)$. Dimension thirteen has infinitely many orbits
(Section~\ref{sec:finite-boundary}).}
\label{fig:all-dimensions}
\end{figure}%
}
\newcommand{\SpinFourteenDynkinFigure}{%
\begin{figure}[t]
\centering
\begin{tikzpicture}[
  cv/.style={circle,draw=spinorsupport,fill=spinorsupport!10,line width=.8pt,
    inner sep=1.3pt,minimum size=14pt,font=\scriptsize},
  bond/.style={draw=spinorsupport,line width=1.1pt},
  lab/.style={font=\small}]
  \node[cv] (a1) at (0,0) {1};
  \node[lab] at (0,-.62) {I: $A_1$};
  \node[cv] at (2.0,0) {1}; \node[cv] at (2.6,0) {1};
  \node[lab] at (2.3,-.62) {II: $2A_1$};
  \node[cv] at (4.6,0) {1}; \node[cv] at (5.2,0) {1}; \node[cv] at (5.8,0) {1};
  \node[lab] at (5.2,-.62) {III: $3A_1$};
  \node[cv] (c1) at (7.9,0) {2}; \node[cv] (c2) at (8.5,0) {2}; \draw[bond] (c1)--(c2);
  \node[lab] at (8.2,-.62) {IV: $A_2$};
  \node[cv] at (10.4,0) {1}; \node[cv] at (11.0,0) {1}; \node[cv] at (11.6,0) {1}; \node[cv] at (12.2,0) {1};
  \node[lab] at (11.3,-.62) {V: $4A_1$};
  \node[cv] (e1) at (0.6,-1.75) {2}; \node[cv] (e2) at (1.2,-1.75) {2}; \draw[bond] (e1)--(e2);
  \node[cv] at (1.9,-1.75) {1};
  \node[lab] at (1.25,-2.37) {VI: $A_2{+}A_1$};
  \node[cv] (g1) at (3.9,-1.75) {2}; \node[cv] (g2) at (4.5,-1.75) {2}; \draw[bond] (g1)--(g2);
  \node[cv] at (5.2,-1.75) {1}; \node[cv] at (5.8,-1.75) {1};
  \node[lab] at (4.85,-2.37) {VII: $A_2{+}2A_1$};
  \node[cv] (i1) at (7.6,-1.75) {2}; \node[cv] (i2) at (8.2,-1.75) {2}; \draw[bond] (i1)--(i2);
  \node[cv] at (8.9,-1.75) {1}; \node[cv] at (9.5,-1.75) {1}; \node[cv] at (10.1,-1.75) {1};
  \node[lab] at (8.85,-2.37) {VIII: $A_2{+}3A_1$};
  \node[cv] (h1) at (11.3,-1.75) {2}; \node[cv] (h2) at (11.9,-1.75) {2}; \draw[bond] (h1)--(h2);
  \node[cv] (h3) at (12.6,-1.75) {2}; \node[cv] (h4) at (13.2,-1.75) {2}; \draw[bond] (h3)--(h4);
  \node[lab] at (12.25,-2.37) {IX: $2A_2$};
\end{tikzpicture}
\caption{The reduced supports in dimension fourteen as Dynkin diagrams of
their carrier algebras. Vertices are pure constituents; two constituents are
joined when their occupation strings differ in six positions, i.e.\ when
their annihilators meet in a line, and are not joined when they differ in
four positions (annihilators meeting in three dimensions). The numbers are
the unnormalized critical probabilities $c_a$ of
Lemma~\ref{lem:dynkin-coefficients}; the critical state is
$\sum_a\sqrt{c_a}\ket{I_a}$ and $\norm\beta^2=2/C-1/4$ with $C=\sum_ac_a$.}
\label{fig:s14-dynkin}
\end{figure}%
}

\newcommand{\SpinFourteenClosureFigure}{%
\begin{figure}[t]
\centering
\begin{tikzpicture}[
  orb/.style={draw=spinorsupport,fill=spinorsupport!6,rounded corners=2pt,
    align=center,font=\small,inner sep=3pt,minimum width=1.25cm},
  arr/.style={->,>=stealth,line width=.7pt,black!70}]
  \node[orb] (I)    at (0,0)     {I\\[1pt]\scriptsize $21$, $\frac74$};
  \node[orb] (II)   at (1.8,0)   {II\\[1pt]\scriptsize $34$, $\frac34$};
  \node[orb] (III)  at (3.6,0)   {III\\[1pt]\scriptsize $42$, $\frac5{12}$};
  \node[orb] (IV)   at (5.4,.85) {IV\\[1pt]\scriptsize $43$, $\frac14$};
  \node[orb] (V)    at (5.4,-.85){V\\[1pt]\scriptsize $49$, $\frac14$};
  \node[orb] (VI)   at (7.2,0)   {VI\\[1pt]\scriptsize $53$, $\frac3{20}$};
  \node[orb] (VII)  at (9.0,0)   {VII\\[1pt]\scriptsize $58$, $\frac1{12}$};
  \node[orb] (VIII) at (10.8,0)  {VIII\\[1pt]\scriptsize $62$, $\frac1{28}$};
  \node[orb] (IX)   at (12.6,0)  {IX\\[1pt]\scriptsize $63$, $0$};
  \draw[arr] (I)--(II); \draw[arr] (II)--(III);
  \draw[arr] (III)--(IV); \draw[arr] (III)--(V);
  \draw[arr] (IV)--(VI); \draw[arr] (V)--(VI);
  \draw[arr] (VI)--(VII); \draw[arr] (VII)--(VIII); \draw[arr] (VIII)--(IX);
\end{tikzpicture}
\caption{Closure order of the nine nonzero projective orbits in dimension
fourteen (Proposition~\ref{prop:s14-closure}). An arrow
$\mathcal O'\to\mathcal O$ means $\mathcal O'\subset\overline{\mathcal O}$;
each box shows the orbit dimension and the momentum minimum
$m(\mathcal O)=\norm\beta^2$. The orbits IV and V are incomparable.}
\label{fig:s14-closure}
\end{figure}%
}

\makeatletter
\renewcommand\subsubsection{\@startsection{subsubsection}{3}{\z@}%
  {-2.5ex\@plus-1ex\@minus-.2ex}%
  {0.8ex\@plus.2ex}%
  {\normalfont\normalsize\bfseries}}
\makeatother

\title{%
    {\LARGE\bfseries\scshape
    The finite projective classification of complex\\[0.2em]
    spinors via momentum maps}%
}

\hypersetup{pdfauthor={Bartłomiej Sikorski, Adam Sawicki}}
\author{%
    {\large\scshape
    Bart\l{}omiej Sikorski\affnum{1} and Adam Sawicki\affnum{2,3}}
    \\[1.2em]
    {\small \affnum{1}Faculty of Physics, University of Warsaw}\\
    {\small Pasteura 5, 02-093 Warsaw, Poland}
    \\[0.7em]
    {\small \affnum{2}Center for Theoretical Physics,
    Polish Academy of Sciences}\\
    {\small Al. Lotnik\'ow 32/46, 02-668 Warsaw, Poland}\\[0.7em]
    {\small \affnum{3}Guangdong Technion--Israel Institute of Technology}\\
    {\small 241 Daxue Road, Jinping District, Shantou, Guangdong Province, China}   
}

\date{}

\begin{document}
\maketitle

\begin{abstract}
We describe the projective orbits of complex spinors from the point of
view of the momentum map. We treat the irreducible spin modules in odd
dimensions 1,3,5,7,9,11 and the half-spin modules in even dimensions
2,4,6,8,10,12,14, which are the only cases with finitely many orbits. 
In the Clifford--Fock model, we construct representatives as sums
of pure basis spinors, and the coefficients are chosen so that the representative is a critical point of the squared norm of the momentum map. The resulting momentum values in a fixed Weyl chamber distinguish the orbits.
Completeness follows from Vinberg's theory of graded Lie algebras and stabilizer actions, independently of the classical spinor
classifications.
We recover the classical counts of 1,1,1,2,3,5 orbits in the odd cases and 1,1,1,2,2,4,9 in the even cases, and compute the projective dimension of every orbit.
In dimension fourteen we also determine the closure order of the orbits, using the fact that the minimum of the squared momentum norm increases
strictly under degeneration.
\end{abstract}

\section{Introduction}
\label{sec:introduction}

Spinor orbits arise in algebraic geometry, representation theory and
mathematical physics. Pure spinors, whose annihilators are maximal totally null subspaces of the vector representation, were introduced by Cartan and studied
algebraically by Chevalley \cite{Cartan1966,Chevalley1954}. They form the
smallest orbit of the complex spin group on spinors, and the subject of this
paper is the classification of all the orbits. The fermionic Fock model makes this connection clear,
since its monomial basis consists of pure spinors and translates each annihilator to the occupation number
\cite{spinorial_chessboard,Trautman_Budinich,TrautmanTrautman}.

We classify nonzero spinors up to complex spin transformations and an
overall nonzero scalar. A point $[\psi]$ represents the complex line
$\C\psi$, and its orbit is the set of lines obtained by spin transformations.
Equivalently, we study the action of
$G=\Spin(N,\C)$ on projective space, with compact form
$K=\Spin(N)$. In even dimension $N=2n$ we use one half-spin module $S_N^+=\bigwedge^{\mathrm{even}}\C^n$; in odd
dimension $N=2n+1$ we use the irreducible module
$S_N=\bigwedge^\bullet\C^n$. There are finitely many projective orbits exactly for $N=1,\ldots,12$ and $N=14$; we call these values of $N$ the finite range. The cases through twelve were classified
by Igusa \cite{Igusa1970uptodim12}, and the fourteen-dimensional case by
Popov \cite{Popov_dim14RUS,Popov_dim14ENG}. Dimension thirteen has continuous projective
families, as shown by Kac and Vinberg \cite{Gatti_dim13}.

The construction has two steps. First we choose the pure spinors that
make up a representative: Fock basis vectors, selected by their
occupation patterns and by how their annihilators intersect.
We work in a common Fock basis and choose weights
such that no difference of two distinct weights is a root. This condition,
called \emph{root-distinctness}, makes every non-Cartan component of the
momentum map vanish by Wildberger's weight-space argument
\cite{Wildberger1992TheMM} (see \cite[Lemma~4]{Tsanov2019} for the statement
for arbitrary root-distinct subsets). For spinors, it forces any two chosen occupation patterns to differ in at least four positions in type $D_n$ and at least three in type $B_n$.

We then determine the coefficients using the Kirwan--Ness critical-point
method \cite{Kirwan_cohomology_84,Ness1984}, in the form developed for
entanglement classification
\cite{SOK_criticalset,SOK_convexity,MaciazekSawicki2015}. The critical-point
equation for $\norm\mu^2$ requires all chosen weights to have the same
inner product with the momentum value. Since that value is their weighted
average, with the squared coefficient moduli as weights, this condition
and normalization give a linear system for the squared moduli. Affine
independence of the chosen weights makes the solution unique; positivity
of the resulting squared moduli gives a critical representative.

In low dimensions the possible supports are found by counting sizes and
intersections of occupation patterns. In dimension fourteen we also use
an identity which replaces four Fock states with a common annihilating
null direction by a sum of two pure spinors
(Lemma~\ref{lem:s14-four-to-two}). Counting how the annihilators of the
remaining constituents intersect leaves nine patterns, and each of them
determines the support up to the Weyl group. All constituents are taken
from one fixed Fock basis.

We construct critical representatives with pairwise distinct momentum
values in the closed positive Weyl chamber. Ness's theorem implies that
their complex orbits are distinct. To prove completeness, we use
Vinberg's general carrier-algebra theorem \cite{Vinberg1979} and a
finite calculation with root systems to obtain root-distinct normal forms
in the even dimensions. In odd dimensions, stabilizer actions
on a vector quadric give the required orbit bounds.

The momentum minimum also constrains orbit closures: it increases strictly
under a proper degeneration between orbits containing critical points.
Together with torus degenerations, this determines the full closure
order in dimension fourteen. In even dimension, the momentum map is
proportional to the Majorana covariance matrix. Minimizing its squared norm
maximizes the linear entropy of the quasiparticle occupations within each
orbit of invertible Gaussian transformations.

In the following statements, $S_N$ denotes the chosen spin or
half-spin module, and $\mu$ is the momentum map of the $K$-action on
$\PP(S_N)$.

\begin{theorem}
\label{thm:finite-orbit-main}
Let $N\in\{1,\ldots,12\}\cup\{14\}$.
\begin{enumerate}
\item The numbers of $G$-orbits on $\PP(S_N)$ are
\begin{equation*}
\begin{array}{c|ccccccc}
N\text{ even}&2&4&6&8&10&12&14\\
\#\,G\backslash\PP(S_N)&1&1&1&2&2&4&9
\end{array}
\qquad
\begin{array}{c|cccccc}
N\text{ odd}&1&3&5&7&9&11\\
\#\,G\backslash\PP(S_N)&1&1&1&2&3&5
\end{array}
\end{equation*}
\item The vectors listed in Tables~\ref{tab:finite-even-low},
      \ref{tab:s14-orbits} and \ref{tab:finite-odd} are critical points of
      $\norm\mu^2$, lie in pairwise distinct $G$-orbits and meet every
      $G$-orbit. Their momentum values and projective orbit dimensions are as listed there.
\item Every $G$-orbit contains exactly one $K$-orbit of critical points of
      $\norm\mu^2$. On it $\norm\mu^2$ attains its minimum over the
      closure of the $G$-orbit.
\end{enumerate}
\end{theorem}

\begin{theorem}
\label{thm:finite-range}
For $N=13$ and for every $N\ge15$ the group $G$ has infinitely many orbits
on $\PP(S_N)$. Hence the representations in
Theorem~\ref{thm:finite-orbit-main} are exactly the spin and half-spin
representations with finitely many projective orbits.
\end{theorem}

\begin{corollary}
\label{cor:critical-values}
For $N$ as in Theorem~\ref{thm:finite-orbit-main}, the critical set of
$\norm\mu^2$ on $\PP(S_N)$ is the union of the $K$-orbits of the listed
representatives. In particular, the critical values of $\norm\mu^2$ are
exactly the numbers $\norm\beta^2$ in the tables.
\end{corollary}

The uniqueness assertion uses Ness's theorem for critical points of the
squared momentum norm \cite[Theorem~7.1]{Ness1984}, while existence
in every orbit follows from the construction of the representatives and
the completeness proof. The theorem concerns projective orbits. Affine orbits of nonzero vectors can still occur in families, even in the finite projective range, for example, each nonzero level set of the
invariant quadratic form is a single orbit. All tables use positive chirality in even dimension. Clifford multiplication
by a real unit vector exchanges the two chiralities and intertwines their
actions after the corresponding automorphism of $\Spin(2n,\C)$, so it
induces a bijection of their projective orbit sets. The corresponding chamber values differ in the sign of the last coordinate.

The paper is organized as follows. 
Section~\ref{sec:fock-model} sets the
Clifford-Fock conventions and develops the geometry of pure spinors.
Section~\ref{sec:spin-representations} describes the spin action, its roots
and its weights.  Section~\ref{sec:momentum-map} develops the momentum-map and
closest-point machinery used throughout.  Section~\ref{sec:interpretation} explains its spectral and fermionic-Gaussian SLOCC interpretations. Section~\ref{sec:normal-forms} proves the existence of root-distinct
normal forms and computes their critical coefficients.
Sections~\ref{sec:even-orbits}, \ref{sec:spin14} and \ref{sec:odd-orbits}
treat the half-spinors up to dimension twelve, dimension fourteen, and the
odd dimensions. Section~\ref{sec:finite-boundary} proves
Theorem~\ref{thm:finite-range}, and Section~\ref{sec:momentum-orbit-geometry}
relates momentum minima to orbit degenerations.
Appendix~\ref{app:finite-calculations} contains the carrier-algebra
argument, the finite grading calculation and the occupation data, and Appendix~\ref{app:exact-verification}
describes the exact computations of orbit dimensions and the enumerations
used in the remarks.

\section{Fock space description of spinors}
\label{sec:fock-model}

We recall the standard realization of spinors in fermionic Fock space and fix the notation and conventions used throughout the paper.
Our exposition was partly inspired by Derezi{\'n}ski's lecture notes \cite{Derezinski2023Clifford}.
\subsection{Fock space and the canonical anticommutation relations}
\label{sec:fock-car}
Let $W\simeq\C^n$ be the one-particle space with its standard Hermitian
inner product, conjugate-linear in the first argument. Its exterior power
$\bigwedge^kW$ consists of antisymmetric $k$-particle states: exchanging
two factors in a wedge product changes its sign, so a repeated factor gives
zero. The fermionic Fock space collects all particle numbers:
\begin{equation}
 \mathcal H=\bigwedge W
 =\bigoplus_{k=0}^{n}\bigwedge^k W .
\end{equation}
Here $\bigwedge^0W=\C$. Fix an orthonormal basis $w_1,\ldots,w_n$
of $W$ and give $\mathcal H$ the Hermitian inner product for which the
ordered wedge monomials are orthonormal. The star in $w_j^*$ denotes
the dual covector, a complex-linear map $W\to\C$ defined by
$w_j^*(w_i)=\delta_{ji}$, where $\delta_{ji}$ is $1$ if $j=i$ and $0$
otherwise. Creation adds a wedge
factor, whereas contraction removes one:
\begin{equation*}
 a_j^\dagger\omega=w_j\wedge\omega,
 \qquad
 a_j\omega=\iota_{w_j^*}\omega.
\end{equation*}
For a covector $\ell$, contraction is defined by
\[
 \iota_\ell(v_1\wedge\cdots\wedge v_k)
 =\sum_{r=1}^k(-1)^{r-1}\ell(v_r)
 v_1\wedge\cdots\wedge\widehat{v_r}\wedge\cdots\wedge v_k,
\]
where the hat means omission, and contraction of a scalar is zero.
With this inner product, $a_j^\dagger$ is the Hermitian adjoint of $a_j$;
thus the dual-basis star and the operator-adjoint dagger have different meanings.
The wedge signs imply the canonical anticommutation relations
\begin{equation}
 \{a_i,a_j\}=0,\qquad
 \{a_i^\dagger,a_j^\dagger\}=0,\qquad
 \{a_i,a_j^\dagger\}=\delta_{ij}\mathbf 1 .
 \label{eq:car}
\end{equation}
Here $\{A,B\}=AB+BA$ and $\mathbf1$ is the identity operator. Starting from the
vacuum $\vac=1\in\bigwedge^0W$, successive creations give the
orthonormal Fock basis
\begin{equation}
 \ket I
 =a_{i_1}^\dagger\cdots a_{i_k}^\dagger\vac,
 \qquad
 I=\{i_1<\cdots<i_k\}\subseteq\{1,\ldots,n\}.
 \label{basis_of_fock_space}
\end{equation}
Let $\epsilon_j(I)$ count the occupied modes before position $j$:
\begin{equation*}
 \epsilon_j(I)=\#\{i\in I:i<j\}.
\end{equation*}
Inserting or removing $w_j$ requires this many interchanges, each
contributing a minus sign. Consequently,
\begin{align}
 a_j^\dagger\ket I
 &=
 \begin{cases}
  0,&j\in I,\\
  (-1)^{\epsilon_j(I)}\ket{I\cup\{j\}},&j\notin I,
 \end{cases}
 \label{eq:creation-on-fock-basis}\\
 a_j\ket I
 &=
 \begin{cases}
  (-1)^{\epsilon_j(I)}\ket{I\setminus\{j\}},&j\in I,\\
  0,&j\notin I.
 \end{cases}
 \label{eq:annihilation-on-fock-basis}
\end{align}
These formulas also verify \eqref{eq:car}. Applying annihilation and then
creation restores the state exactly when mode $j$ is occupied; otherwise
it gives zero. Hence the number operator $N_j=a_j^\dagger a_j$ acts by
\begin{equation}
 N_j\ket I=\chi_I(j)\ket I.
 \label{eq:number-operator-action}
\end{equation}
The indicator $\chi_I(j)$ is the scalar $1$ if $j\in I$ and $0$
otherwise; unlike $\mathbf1$, it is not an identity operator. Thus each
mode is empty or occupied once, and the $2^n$ subsets label a basis of
$\mathcal H$. The degree $|I|$ is the particle number of $\ket I$.
A general spinor can mix degrees; the even and odd sums of the
$k$-particle sectors will instead become the two chiral spin modules.

\subsection{The Clifford representation on Fock space}
\label{sec:clifford-fock-representation}

Let $V_{\mathbb R}\simeq\mathbb R^N$, where $N=2n$ or $2n+1$, carry
its Euclidean inner product.  We write
$V=V_{\mathbb R}\otimes_{\mathbb R}\C$ and extend the form complex
bilinearly to a symmetric form $g$ on $V$, so in orthonormal coordinates
$g(z,w)=\sum_r z_rw_r$, without complex conjugation. It is nondegenerate:
if $g(u,v)=0$ for every $v$, then $u=0$. It is not a positive Hermitian form:
when $N\ge2$,
orthonormal real vectors $e_1,e_2$ give $u=e_1+ie_2\ne0$ with
$g(u,u)=1+i^2=0$, although its Hermitian norm squared is $2$. Such a vector is called \emph{null} or
\emph{isotropic}.

The Clifford algebra $\Cl_N^\C=\Cl(V,g)$ is built from formal words
$v_1\cdots v_k$ in $V$ and their finite complex linear combinations.
Multiplication joins words; the empty word is the unit $1_{\Cl}$.
We impose linearity in each vector and
identify products using
\begin{equation}
 vw+wv=2g(v,w)1_{\Cl},\qquad v,w\in V,
 \label{eq:clifford-relation}
\end{equation}
also inside longer words. Here $vw$ is an algebra product, whereas $g(v,w)$
is a scalar. Setting $w=v$ gives $v^2=g(v,v)1_{\Cl}$. For an orthonormal
real basis $e_1,\ldots,e_N$, this says $e_r^2=1_{\Cl}$ and
$e_re_s=-e_se_r$ when $r\ne s$.

A representation on $\mathcal H$ is a complex-linear map
$\gamma:\Cl(V,g)\to\End(\mathcal H)$ preserving products and sending
$1_{\Cl}$ to the identity operator $\mathbf1_{\mathcal H}$. Here
$\End(\mathcal H)$ is the algebra of linear operators on $\mathcal H$;
a spinor is a vector of $\mathcal H$. Writing $\gamma_r=\gamma(e_r)$, we construct
the representation by assigning the first $2n$ basis vectors the Hermitian
Majorana operators
\begin{equation}
 \gamma_{2j-1}=a_j+a_j^\dagger,\qquad
 \gamma_{2j}=i(a_j-a_j^\dagger),
 \qquad 1\le j\le n.
\label{gamma_by_a}
\end{equation}
These obey
\begin{equation}
 \gamma_r\gamma_s+\gamma_s\gamma_r=2\delta_{rs}\mathbf 1 .
\end{equation}
For example,
\begin{equation*}
 \gamma_{2j-1}^2=(a_j+a_j^\dagger)^2
 =\{a_j,a_j^\dagger\}=\mathbf1,
 \qquad
 \gamma_{2j}^2=-(a_j-a_j^\dagger)^2=\mathbf1;
\end{equation*}
all mixed anticommutators vanish by \eqref{eq:car}. Since these are the
defining relations, the assignment extends to a representation of
$\Cl_{2n}^\C$ on $\mathcal H$. Conversely,
\begin{equation}
 a_j=\frac12(\gamma_{2j-1}-i\gamma_{2j}),
 \qquad
 a_j^\dagger=\frac12(\gamma_{2j-1}+i\gamma_{2j}).
 \label{eq:gamma-to-fock}
\end{equation}
The operators $a_j,a_j^\dagger$ generate all matrix units on the Fock basis.
For $J=\{j_1<\cdots<j_\ell\}$, the rank-one operator
$\ket I\!\langle J\rvert$ sends $\ket J$ to $\ket I$ and all other basis
vectors to zero. With the vacuum projector
\begin{equation*}
 P_0=\prod_{j=1}^n(\mathbf1-N_j)
\end{equation*}
it is given by
\begin{equation*}
 \ket I\!\langle J\rvert
 =a_{i_1}^\dagger\cdots a_{i_k}^\dagger
  P_0\,a_{j_\ell}\cdots a_{j_1}.
\end{equation*}
Hence $\gamma$ maps onto $\End(\mathcal H)$, whose dimension is $4^n$.
The defining relations reorder every abstract word into a linear combination
of $e_{i_1}\cdots e_{i_k}$ with $i_1<\cdots<i_k$. There are $2^{2n}=4^n$
such words, so surjectivity forces them to be independent and makes $\gamma$
an isomorphism:
$\Cl_{2n}^\C\simeq\End(\mathcal H)$.  In particular, the Fock model is irreducible (it has no nonzero proper
invariant subspace) and is, up to equivalence, the unique irreducible
complex Clifford module in even dimension.

In odd dimension the normalized volume element
\begin{equation*}
 \omega_{2n+1}=i^n e_1e_2\cdots e_{2n+1}
\end{equation*}
commutes with every generator and satisfies
$\omega_{2n+1}^2=1_{\Cl}$. Its two complementary central projections
$\frac12(1_{\Cl}\pm\omega_{2n+1})$ split the algebra into two ideals.
In each ideal, fixing the volume sign expresses the last generator through
the first $2n$, giving a copy of $\Cl_{2n}^\C$. Thus
$\Cl_{2n+1}^\C\simeq\End(\mathcal H)\oplus\End(\mathcal H)$, with two
irreducible modules distinguished by the sign chosen for the last
generator. Put $F=\sum_jN_j$, the total particle-number operator. Since
$F\ket I=|I|\ket I$, the parity operator $P=(-1)^F$ acts by $(-1)^{|I|}$.
One choice for the last generator is
\begin{equation}
 \gamma_{2n+1}
 =(-1)^F
 =\prod_{j=1}^n(\mathbf1-2a_j^\dagger a_j),
 \label{eq:parity-gamma}
\end{equation}
and the other choice is its negative.  Since every $a_j$ and
$a_j^\dagger$ changes fermion number by one, $(-1)^F$ anticommutes with
all the generators in \eqref{gamma_by_a}; it also squares to the identity.
The two choices give $\gamma(\omega_{2n+1})=\pm\mathbf1_{\mathcal H}$.
Conjugation by $P$ intertwines their restrictions to the even Clifford
algebra, spanned by even-length words. Both therefore give equivalent
spin representations of $\Spin(2n+1)$, although a bilinear $\gamma_rP$
with $r\le2n$ changes fermion parity.

\subsection{Chirality and the two half-spin modules}
\label{sec:chirality-half-spin}

For $N=2n$, the chirality operator
\begin{equation}
 \Gamma=i^n\gamma_1\gamma_2\cdots\gamma_{2n}
 =(-1)^{\sum_j a_j^\dagger a_j}
 \label{eq:chirality}
\end{equation}
is the fermion-parity operator. Indeed,
\begin{equation}
 i\gamma_{2j-1}\gamma_{2j}=\mathbf1-2N_j.
\end{equation}
Multiplying these commuting identities gives
$\Gamma=\prod_j(\mathbf1-2N_j)=(-1)^F$. Moreover,
$\Gamma\gamma_r=-\gamma_r\Gamma$ for every Clifford generator, because
$\gamma_r$ must pass through $2n-1$ other generators.  It follows that
$\Gamma$ commutes with every even Clifford element.  Its eigenvalues are
$\pm1$, so the Dirac module splits into two invariant half-spin modules:
\begin{equation}
 \mathcal H=S^+\oplus S^-,
 \qquad
 S^+=\bigwedge\nolimits^{\mathrm{even}}W,\qquad
 S^-=\bigwedge\nolimits^{\mathrm{odd}}W .
 \label{eq:half-spin}
\end{equation}
Both have dimension $2^{n-1}$.  With our convention, the vacuum has positive chirality 
\eqref{eq:chirality}, so $S^+$ consists of the states with an even number of fermions and chirality is simply fermion parity.

\subsection{Null bases and pure spinors}
\label{sec:null-bases-pure-spinors}
A vector is null when its Clifford square is zero, but its Clifford
action need not be the zero operator. We therefore ask which vectors
kill a given nonzero spinor $\psi$ and define its annihilator by
\begin{equation}
 M(\psi)=\{v\in V:v\cdot\psi=0\}.
 \label{eq:annihilator}
\end{equation}
Here $v\cdot\psi$ means $\gamma(v)\psi$; it is the Clifford action,
not the bilinear form $g$. For $v,w\in M(\psi)$, the Clifford relation gives
\begin{equation}
     2g(v,w)\psi=(vw+wv)\psi=0.
\end{equation}
Thus $M(\psi)$ is \emph{totally null}: $g$ vanishes on every pair of
its vectors, so $M(\psi)\subseteq M(\psi)^\perp$. Here $\perp$ means
orthogonality with respect to $g$. Nondegeneracy gives
$\dim M(\psi)^\perp=N-\dim M(\psi)$, hence
$\dim M(\psi)\le\lfloor N/2\rfloor=n$.
A spinor is \emph{pure} (or \emph{simple}) when this bound is attained.
In even dimension it then has definite chirality, so one also speaks of
a pure Weyl spinor. Up to nonzero scalar multiples, pure spinors
correspond to maximal totally null subspaces. The space of these
subspaces is the \emph{orthogonal Grassmannian}; in even dimension its
two connected components correspond to the two chiralities
\cite{Trautman_Budinich,TrautmanTrautman}.

The creation and annihilation operators are themselves Clifford actions of
null vectors.  Namely, with
\begin{equation*}
 p_j=\frac1{\sqrt2}(e_{2j-1}-ie_{2j}),
 \qquad
 q_j=\frac1{\sqrt2}(e_{2j-1}+ie_{2j}),
\end{equation*}
one has
\begin{equation*}
 \gamma(p_j)=\sqrt2\,a_j,\qquad
 \gamma(q_j)=\sqrt2\,a_j^\dagger,
\end{equation*}
and
\begin{equation}
   g(p_i,p_j)=g(q_i,q_j)=0, \qquad g(p_i,q_j)=\delta_{ij}.
\end{equation}
These pairings define a \emph{Witt basis}: two complementary null
$n$-planes paired nondegenerately by $g$. In odd dimension the displayed
vectors are supplemented by the orthogonal unit vector $e_{2n+1}$.

We will also write $a_j,a_j^\dagger$ for the corresponding vector
directions $p_j,q_j$ when describing subspaces of $V$. Every Fock basis
vector is pure, with annihilator
\begin{equation}
 M(\ket I)
 =\Span\{a_j^\dagger:j\in I\}
  \oplus\Span\{a_j:j\notin I\}.
 \label{eq:fock-annihilator}
\end{equation}
Indeed, each displayed generator kills $\ket I$, and the $n$ generators
span a totally null $n$-plane, which is maximal.  Thus $I$ determines the maximal null plane $M(\ket I)$.  For two subsets
$I$ and $J$, the intersection $M(\ket I)\cap M(\ket J)$ contains exactly
one of $a_j,a_j^\dagger$ when $I$ and $J$ agree at position $j$, and
neither when they differ. Writing
$I\symdiff J=(I\setminus J)\cup(J\setminus I)$ for the symmetric
difference, we obtain
\begin{equation}
 \dim\bigl(M(\ket I)\cap M(\ket J)\bigr)
 =n-|I\symdiff J|.
 \label{eq:general-annihilator-intersection}
\end{equation}
We use \eqref{eq:general-annihilator-intersection} to classify
annihilator configurations in Section~\ref{sec:spin14}.

In even dimension there is also a useful local formula for pure spinors
of positive chirality. If
$Z=(Z_{ij})$ is skew-symmetric, meaning $Z_{ij}=-Z_{ji}$, then
\begin{equation}
 \psi_Z=
 \exp\left(\frac12\sum_{i,j=1}^n
 Z_{ij}a_i^\dagger a_j^\dagger\right)\vac
 \label{eq:pure-gaussian-chart}
\end{equation}
is a finite sum, because each application of the operator in its exponent raises degree
by two. Its vacuum coefficient is $1$, so it is nonzero. It is
annihilated by the $n$ operators
\begin{equation}
 a_k-\sum_{j=1}^nZ_{kj}a_j^\dagger,
 \qquad 1\leq k\leq n.
 \label{eq:pure-gaussian-annihilators}
\end{equation}
To verify this, put
$B=\frac12\sum Z_{ij}a_i^\dagger a_j^\dagger$ and write
$[A,B]=AB-BA$ for the commutator. The canonical
anticommutation relations give
$[a_k,B]=\sum_jZ_{kj}a_j^\dagger$ and all higher iterated commutators
vanish.  Hence
$a_ke^B=e^B(a_k+\sum_jZ_{kj}a_j^\dagger)$, which proves
\eqref{eq:pure-gaussian-annihilators}.  These operators span a maximal null
space, so $\psi_Z$ is pure. Conversely, any maximal null plane
transverse to $\Span\{q_1,\ldots,q_n\}$ (with zero intersection) is a graph with basis
$p_k-\sum_jZ_{kj}q_j$; the null-pairing conditions say exactly that $Z$
is skew-symmetric. Thus \eqref{eq:pure-gaussian-chart} constructs the pure
spinor corresponding to every plane in this open chart.

Pure spinors form one closed orbit in each projectivized irreducible spin
module; in $S^+$ and in odd dimension it contains the vacuum.
For $n\ge4$, the positive half-spinor $\ket\varnothing+\ket{1234}$ is not
pure: its annihilator has dimension less than $n$.

The \emph{nullity} $\dim M(\psi)$ is invariant under Spin. In dimension
fourteen it does not separate all orbits, so we also use the momentum
value and the intersections of the annihilators of pure constituents.

For $N=2n$, a maximal annihilator has a basis of operators $b_i$ linear in
$a_j,a_j^\dagger$. Nullness gives $\{b_i,b_j\}=0$; orthonormalizing their
coefficient vectors gives $\{b_i,b_j^\dagger\}=\delta_{ij}\mathbf1$.
These are Bogoliubov quasiparticle annihilators, with $\psi$ as their
vacuum \cite{Araki1970}. Thus normalized pure positive half-spinors are
exactly the even pure fermionic Gaussian states \cite{HacklBianchi2021}.
Formula~\eqref{eq:pure-gaussian-chart} describes their chart with nonzero
vacuum amplitude \cite{SL2014}.

Hermitian quadratic Majorana Hamiltonians generate the compact
$\Spin(2n)$ action, preserving fermion parity while allowing particle
number to change \cite{Bravyi2005,SL2014}. The complex group permits
invertible nonunitary Gaussian transformations \cite{BalianBrezin1969};
its projective orbits are the equivalence classes studied here. Since the
momentum records quadratic expectations, it is proportional to the
Majorana covariance matrix (Section~\ref{sec:interpretation}).
Table~\ref{table:dictionary} summarizes these correspondences.

\begin{table}[ht]
\centering
\renewcommand{\arraystretch}{1.15}
\begin{tabular}{ll}
\toprule
Spinor/Clifford language & Fermionic Fock language\\
\midrule
null basis of $V$ & creation and annihilation operators\\
orthonormal basis of $V$ & Majorana operators\\
distinguished pure spinor & vacuum\\
pure positive half-spinors & even pure fermionic Gaussian states\\
$\dim M(\psi)$ & number of independent linear annihilation conditions\\
$\Spin(2n)$ action & unitary evolution under quadratic Hamiltonians\\
positive/negative half-spinors & even/odd fermion parity\\
$\Spin(2n)$ momentum map & Majorana covariance matrix\\
\bottomrule
\end{tabular}
\caption{Dictionary between Clifford-algebra and fermionic Fock space
descriptions.}
\label{table:dictionary}
\end{table}
\FloatBarrier

\section{Spin representations, roots and weights}
\label{sec:spin-representations}

\subsection{From the Clifford algebra to
\texorpdfstring{$\Spin(N)$}{Spin(N)}}
\label{sec:spin-cover}

The same word construction over $\R$ gives the real Clifford algebra
$\Cl(V_{\mathbb R})$. Its even products of unit vectors realize rotations:
\begin{equation}
 \Spin(N)=
 \{u_1u_2\cdots u_{2k}:
   u_a\in V_{\mathbb R},\ g(u_a,u_a)=1\}.
 \label{eq:spin-products}
\end{equation}
Thus $\Spin(N)$ consists of products of an even number of real unit
vectors.  The inverse of
$x=u_1\cdots u_{2k}$ is $x^{-1}=u_{2k}\cdots u_1$.
For one unit vector $u$, the Clifford relation gives
\begin{equation}
 -uvu^{-1}=v-2g(u,v)u.
\end{equation}
The right-hand side is the orthogonal reflection in the hyperplane
perpendicular to $u$.  For an even product the minus signs cancel, so
\begin{equation}
 R_x(v)=xvx^{-1}
\end{equation}
is a composition of an even number of reflections and hence belongs to
$SO(N)$, the group of real orthogonal transformations of determinant one.
Conversely, the Cartan--Dieudonn\'e theorem expresses every
element of $SO(N)$ as such an even product (see
\cite[Chapter~I]{Chevalley1954}). The elements acting trivially
on $V_{\mathbb R}$ are $1$ and $-1$. Every rotation therefore has exactly
two spin lifts, differing by sign; this is expressed by the exact sequence
\begin{equation}
 1\longrightarrow\{\pm1\}\longrightarrow
 \Spin(N)\xrightarrow{\,R\,}SO(N)\longrightarrow1.
 \label{eq:spin-double-cover}
\end{equation}

Restricting $\gamma$ to this subgroup gives the spin representations
$\Pi_N=\gamma|_{\Spin(N)}$, where $U(\mathcal H)$ denotes the group of
operators preserving its Hermitian inner product:
\begin{align}
 \Pi_{2n}&:\Spin(2n)\longrightarrow U(\mathcal H),
 \\
 \Pi_{2n+1}&:\Spin(2n+1)\longrightarrow U(\mathcal H).
\end{align}
Indeed, $\gamma(u)$ is Hermitian and squares to the identity for every real
unit vector $u$, so every product in \eqref{eq:spin-products} acts
unitarily.  Since the central element $-1$ lies in the kernel of $R$ and acts
as $-\mathbf1_{\mathcal H}$, the spin representation does not descend
to $SO(N)$.

Allowing $u_a\in V$ with $g(u_a,u_a)=1$ gives $G=\Spin(N,\C)$, whose orbits
we classify; $K=\Spin(N)$ is its compact form. The group $K$
preserves the Hermitian inner product on $\mathcal H$; this is what permits
its action on projective space to be described by a momentum map
(Section~\ref{sec:momentum-map}).

\subsection{Derivation of the Lie-algebra representation}
\label{sec:spin-lie-algebra-derivation}

The Lie algebra $\mathfrak k\simeq\mathfrak{so}(N)$ of $K$ consists of real skew-symmetric
matrices $A=(A_{rs})$ generating the curves $e^{tA}$ in $SO(N)$. The
corresponding infinitesimal spin action, denoted by $\rho_*(A)$, satisfies
$\Pi(e^{t\widetilde A})=e^{t\rho_*(A)}$, where $\widetilde A$ is the lift of
$A$ through the differential of $R$. It is given by
\begin{equation}
 \rho_*(A)=\frac14\sum_{r,s=1}^N A_{rs}\gamma_r\gamma_s
 =\frac12\sum_{r<s}A_{rs}\gamma_r\gamma_s .
 \label{Hamiltonians}
\end{equation}
For $r\ne s$,
$(\gamma_r\gamma_s)^\dagger
=\gamma_s\gamma_r=-\gamma_r\gamma_s$; hence
$\rho_*(A)$ is anti-Hermitian, as a differential of a unitary
representation.

To verify the representation property, use the Clifford relations:
\begin{equation}
 [\gamma_r\gamma_s,\gamma_t]
 =2(\delta_{st}\gamma_r-\delta_{rt}\gamma_s).
\end{equation}
Substitution into \eqref{Hamiltonians} yields, for every basis vector
$e_t$,
\begin{equation}
 [\rho_*(A),\gamma_t]
 =\sum_{r=1}^NA_{rt}\gamma_r
 =\gamma(Ae_t).
 \label{eq:spin-differential-vector-action}
\end{equation}
By complex linearity, this holds for every $v\in V$.  Applying
\eqref{eq:spin-differential-vector-action} to each factor in
$\rho_*(B)$ gives
\begin{align}
 [\rho_*(A),\rho_*(B)]
 &=\frac14\sum_{s,t}B_{st}
 \bigl([\rho_*(A),\gamma_s]\gamma_t
       +\gamma_s[\rho_*(A),\gamma_t]\bigr)
 =\rho_*([A,B]).
\end{align}
Thus $\rho_*:\mathfrak{so}(N)\to\mathfrak u(\mathcal H)$
preserves commutators $[X,Y]=XY-YX$, as required of a Lie-algebra
representation. Here $\mathfrak u(\mathcal H)$ is the space of
anti-Hermitian operators; multiplying one by $-i$ gives the Hermitian
observable whose expectation value we shall use.

The group-level conjugation formula follows by solving a differential
equation.  Set
\begin{equation*}
 F(t)=e^{t\rho_*(A)}\gamma(v)e^{-t\rho_*(A)}.
\end{equation*}
Equation \eqref{eq:spin-differential-vector-action} gives
$F'(t)=[\rho_*(A),F(t)]$ and $F(0)=\gamma(v)$. The function
$\gamma(e^{tA}v)$ solves the same linear initial-value problem, whence
\begin{equation}
 e^{t\rho_*(A)}\gamma(v)e^{-t\rho_*(A)}
 =\gamma(e^{tA}v).
\end{equation}
Thus Clifford bilinears exponentiate to the spin action and cover the
corresponding orthogonal transformations. For $N=2n$ these are quadratic
fermionic transformations; the odd-dimensional extension is discussed in
Section~\ref{sec:interpretation}.

For a concrete example, let
$L_{rs}=E_{rs}-E_{sr}$, $r<s$, where the matrix unit $E_{rs}$ maps $e_s$ to
$e_r$ and all other basis vectors to zero. Then
\begin{equation*}
 \rho_*(L_{rs})=\frac12\gamma_r\gamma_s,
 \qquad
 e^{t\rho_*(L_{rs})}
 =\cos\frac t2+\gamma_r\gamma_s\sin\frac t2,
\end{equation*}
because $(\gamma_r\gamma_s)^2=-\mathbf1$.  Under $R$ this element maps
to the rotation $e^{tL_{rs}}$ of the $(r,s)$-plane.  At $t=2\pi$ the
orthogonal rotation is the identity, while its spin lift equals $-1$.

For $N=2n$, every $\rho_*(A)$ contains an even number of the first
$2n$ Clifford generators, so it commutes with $\Gamma$.  The even Clifford algebra acts as
$\End(S^+)\oplus\End(S^-)$: the matrix units constructed above have
even Clifford degree when their two basis labels have the same
parity. Since even Clifford monomials are products of bilinears, the
restrictions of $\Pi_{2n}$ to $S^+$ and $S^-$ are irreducible.  In odd dimension,
$(\Cl_{2n+1}^\C)^0\simeq\Cl_{2n}^\C\simeq\End(\mathcal H)$, so the
representation of $\Spin(2n+1)$ on all of $\mathcal H$ is irreducible.

\subsection{The Cartan algebra and the spinor weights}
\label{sec:cartan-spinor-weights}

Let
$J_j=E_{2j,2j-1}-E_{2j-1,2j}$, $1\le j\le n$, be the standard
rotation generators. Rotations in these disjoint planes commute, and
span a maximal abelian subalgebra $\mathfrak t$, called a Cartan algebra.
Its dimension $n$ is the rank of $K$; exponentiating it gives a maximal
torus $T$, the subgroup of independent plane rotations. The corresponding
Hermitian observables are
\begin{equation}
 \widehat H_j=-i\rho_*(J_j)
 =\frac{i}{2}\gamma_{2j-1}\gamma_{2j}
 =\frac12(1-2a_j^\dagger a_j).
 \label{eq:cartan-observables}
\end{equation}
The normalization by $1/2$ is the standard one in which the spinor
weights have coordinates $\pm\frac12$.  By
\eqref{eq:number-operator-action},
\begin{equation}
 \widehat H_j\ket I
 =
 \begin{cases}
  \frac12\ket I,&j\notin I,\\
  -\frac12\ket I,&j\in I.
 \end{cases}
\end{equation}
The tuple of these simultaneous eigenvalues is the \emph{weight} of
$\ket I$. We write it in the coordinate basis $e_1,\ldots,e_n$ of the
weight space $\mathfrak t^*$, the real dual of $\mathfrak t$; these $e_j$
are distinct from the physical
basis vectors $e_r\in V$. Thus
\begin{equation}
 \eta_I
 =\frac12\sum_{j\notin I}e_j-\frac12\sum_{j\in I}e_j.
 \label{eq:general-fock-weight}
\end{equation}
For $\Spin(2n)$, the positive and negative half-spin modules contain,
respectively, the weights with $|I|$ even and odd.  For $\Spin(2n+1)$,
all subsets occur.  In particular, all weight spaces in the representations
considered here are one-dimensional.

Root operators in the Lie algebra shift weights by vectors called roots. We
choose the positive roots to be $e_i\pm e_j$ for $i<j$, together
with $e_i$ in odd dimension. A highest-weight vector is one annihilated
by every positive-root operator, so its weight cannot be raised further.
The root operators below show that the vacuum has this property in $S^+$
and in the odd spin representation, with weight
\begin{equation*}
 \left(\frac12,\ldots,\frac12\right).
\end{equation*}
For $S^-$, one may take $\ket n=a_n^\dagger\vac$, whose highest weight
is
\begin{equation*}
 \left(\frac12,\ldots,\frac12,-\frac12\right).
\end{equation*}
Thus the parity description \eqref{eq:half-spin} agrees with the usual
highest-weight description.

\subsection{Roots and their explicit Fock space operators}
\label{sec:explicit-root-operators}

We extend $\rho_*$ complex-linearly to
$\mathfrak g=\mathfrak k\otimes_{\R}\C\simeq\mathfrak{so}(N,\C)$.
A root $\alpha$ records the weight shift caused by a Lie-algebra
operator: $[\widehat H_j,X_\alpha]=\alpha_jX_\alpha$ implies that
$X_\alpha$ sends weight $\eta$ to $\eta+\alpha$ whenever its action is
nonzero. The corresponding one-dimensional subspace of the complexified
Lie algebra is the \emph{root space} $\mathfrak g_\alpha$; a nonzero element
of it is a root vector $E_\alpha$. For $\mathfrak{so}(2n,\C)$ and $\mathfrak{so}(2n+1,\C)$,
respectively called types $D_n$ and $B_n$, the roots are
\begin{align}
 D_n:\quad&\Delta=\{\pm e_i\pm e_j:1\le i<j\le n\},
 \label{eq:roots-Dn}\\
 B_n:\quad&\Delta=\{\pm e_i\pm e_j:1\le i<j\le n\}
             \cup\{\pm e_i:1\le i\le n\}.
 \label{eq:roots-Bn}
\end{align}
The $2n(n-1)$ roots of $D_n$ all have length $\sqrt2$; $B_n$ adds the
$2n$ short roots of length one.
Their Fock operators follow from
\eqref{eq:cartan-observables} and the canonical anticommutation relations:
\begin{equation}
 [\widehat H_k,a_i^\dagger]=-\delta_{ki}a_i^\dagger,
 \qquad
 [\widehat H_k,a_i]=\delta_{ki}a_i.
\end{equation}
Thus creation in mode $i$ changes a weight by $-e_i$, while
annihilation changes it by $+e_i$.  Since the complexified spin Lie algebra
is represented by Clifford bilinears, \eqref{eq:gamma-to-fock} gives, for
$i\ne j$, the four exact identities
\begin{align}
 (\gamma_{2i-1}+i\gamma_{2i})
 (\gamma_{2j-1}+i\gamma_{2j})
 &=4a_i^\dagger a_j^\dagger,
 &&\text{root }-e_i-e_j,\notag\\
 (\gamma_{2i-1}-i\gamma_{2i})
 (\gamma_{2j-1}-i\gamma_{2j})
 &=4a_i a_j,
 &&\text{root }e_i+e_j,\notag\\
 (\gamma_{2i-1}+i\gamma_{2i})
 (\gamma_{2j-1}-i\gamma_{2j})
 &=4a_i^\dagger a_j,
 &&\text{root }-e_i+e_j,\notag\\
 (\gamma_{2i-1}-i\gamma_{2i})
 (\gamma_{2j-1}+i\gamma_{2j})
 &=4a_i a_j^\dagger,
 &&\text{root }e_i-e_j.
 \label{eq:Dn-root-operators}
\end{align}
These operators span the $D_n$ root spaces.
The first operator creates two particles, the second annihilates two, and
the last two transfer a particle between modes.  Whenever their action is
nonzero, exactly two occupation numbers change.

For $B_n$, the same bilinears give all long-root spaces.  The remaining
Clifford bilinears contain the last generator
$P=\gamma_{2n+1}=(-1)^F$.  Their root combinations are
\begin{align}
 (\gamma_{2i-1}-i\gamma_{2i})P
 &=2a_iP,
 &&\text{root }e_i,\label{eq:Bn-positive-short-root}\\
 (\gamma_{2i-1}+i\gamma_{2i})P
 &=2a_i^\dagger P,
 &&\text{root }-e_i.
 \label{eq:Bn-negative-short-root}
\end{align}
The parity factor changes only the sign of a nonzero matrix element; the
occupation string itself changes in the single mode $i$.  These are the
short roots $\pm e_i$.

As a check, the Cartan generators and the root operators span a space of the right dimension.
In even dimension there are $n$ Cartan generators and four root operators for
each pair $i<j$, giving
\begin{equation*}
 n+4\binom n2=n(2n-1)=\dim\mathfrak{so}(2n).
\end{equation*}
In odd dimension the $2n$ short-root operators raise the total to
\begin{equation*}
 n+4\binom n2+2n=n(2n+1)=\dim\mathfrak{so}(2n+1).
\end{equation*}
Thus these operators give the full Cartan--root decomposition before
restriction to a half-spin module. In the lowest even ranks, restriction
may have a kernel; orbit dimensions must use the actual represented
operators, as in Appendix~\ref{app:exact-verification}.

The \emph{Hamming distance} between occupation strings is the number of
modes in which they differ. By \eqref{eq:general-fock-weight}, two weights
in a fixed half-spin module differ by a $D_n$ root exactly when this
distance is two; in the odd module a $B_n$ root permits distance one or
two. This translates root adjacency into the binary-code conditions of
Sections~\ref{sec:even-orbits} and~\ref{sec:odd-orbits}.

\subsection{Weyl symmetry, root-distinct supports and chambers}
\label{sec:weyl-root-distinct}

For positive rank, after choosing the Cartan algebra the residual symmetries
are its normalizer modulo $T$: the Weyl group $W=N_K(T)/T$. Here $N_K(T)$ consists
of elements carrying $T$ to itself by conjugation. On weight coordinates,
this finite group is
\begin{equation*}
 W(D_n)=S_n\ltimes
 \{(\varepsilon_1,\ldots,\varepsilon_n)\in\{\pm1\}^n:
   \varepsilon_1\cdots\varepsilon_n=1\}
\end{equation*}
in even type and
\begin{equation*}
 W(B_n)=S_n\ltimes\{\pm1\}^n
\end{equation*}
in odd type.  Here $S_n$ is the permutation group relabelling the $n$ modes. A sign change in
coordinate $j$ interchanges $+\frac12$ and $-\frac12$ in
\eqref{eq:general-fock-weight}, hence toggles the $j$-th occupation.
Consequently $W(D_n)$ acts on the subsets $I$ by permutations and
symmetric difference with an even subset, while $W(B_n)$ allows symmetric
difference with an arbitrary subset.

The \emph{support} consists of Fock labels with nonzero coefficients.
These coordinate symmetries also relate states: their spin lifts permute
the Fock basis vectors up to signs. Moving a support by the Weyl group can
therefore be performed by a compact spin transformation.

\begin{lemma}[Weyl group in the Fock basis]
\label{lem:weyl-fock}
For $1\le p\ne q\le n$ put
\begin{equation*}
 \sigma_{pq}=\exp\Bigl(\tfrac\pi2\bigl(a_p^\dagger a_q-a_q^\dagger a_p\bigr)\Bigr),
 \qquad
 R_{pq}=\exp\Bigl(\tfrac\pi2\bigl(a_p^\dagger a_q^\dagger-a_qa_p\bigr)\Bigr),
\end{equation*}
and, in odd dimension, $R_p=\gamma_{2p-1}\gamma_{2n+1}=\exp\bigl(\tfrac\pi2\gamma_{2p-1}\gamma_{2n+1}\bigr)$.
These elements lie in $K$ and normalize the maximal torus, and
\begin{equation*}
 \sigma_{pq}\ket I=\pm\ket{\tau_{pq}(I)},\qquad
 R_{pq}\ket I=\pm\ket{\tau_{pq}(I\symdiff\{p,q\})},\qquad
 R_p\ket I=\pm\ket{I\symdiff\{p\}},
\end{equation*}
where $\tau_{pq}$ exchanges $p$ and $q$. They realize the reflections in
$e_p-e_q$, $e_p+e_q$ and $e_p$, and hence generate $W(D_n)$, respectively
$W(B_n)$. Consequently, if two supports are Weyl-equivalent, every spinor
supported on the first is $K$-equivalent to a spinor supported on the second
with the same coefficient moduli.
\end{lemma}

\begin{proof}
The generators are anti-Hermitian real combinations of Clifford bilinears,
so the exponentials lie in $K$. The generator
$X=a_p^\dagger a_q-a_q^\dagger a_p$ vanishes on Fock vectors in which $p$ and
$q$ are both empty or both occupied, and satisfies $X^2=-1$ on the span of
the others; hence $\exp(\frac\pi2X)=X$ there, which exchanges the
occupations of $p$ and $q$. The generator
$Y=a_p^\dagger a_q^\dagger-a_qa_p$ vanishes when exactly one of $p,q$ is
occupied and satisfies $Y^2=-1$ on the span of the others, so
$\exp(\frac\pi2Y)$ toggles both occupations when they are equal and fixes
the vector otherwise; this is $I\mapsto\tau_{pq}(I\symdiff\{p,q\})$. Finally
$(\gamma_{2p-1}\gamma_{2n+1})^2=-1$, and $\gamma_{2p-1}\gamma_{2n+1}$ changes
only the occupation of mode $p$. Each element maps weight spaces to weight
spaces, hence normalizes the torus, and acts on weights by the indicated
reflection.
\end{proof}

Physically, $\sigma_{pq}$ swaps the modes $p$ and $q$, and $R_{pq}$ is a
particle--hole transformation of the pair $p,q$: it turns an empty pair into
an occupied one and vice versa.

Following the root-distinctness method of Wildberger
\cite{Wildberger1992TheMM}, in its general subset form
\cite{Tsanov2019}, a set of weights
$\Sigma$ is called \emph{root-distinct} if
\begin{equation}
 \eta-\nu\notin\Delta
 \qquad\text{for all distinct }\eta,\nu\in\Sigma.
 \label{eq:root-distinct}
\end{equation}
Otherwise the pair is called \emph{root-neighbouring}. The point of the
definition is that a root operator maps the $\eta$-weight space to the
$(\eta+\alpha)$-weight space, so on a root-distinct support it has no matrix
elements between two weights of the support. This will make all non-Cartan
components of the momentum map vanish
(Lemma~\ref{lemma_root_distinct_support}).

We identify $\mathfrak t^*$ with $\R^n$ with orthonormal coordinate vectors
$e_j$. The invariant inner product also embeds it in $\mathfrak k^*$,
where $\mathfrak k$ is the Lie algebra of $K$. The coadjoint action is the dual of conjugation on the
Lie algebra. Every coadjoint $K$-orbit meets $\mathfrak t^*$ in a Weyl
orbit, so ordering its coordinates selects one representative in the
closed positive Weyl chamber. Explicitly,
\begin{align}
 \mathfrak t_+(D_n)
 &=\{x_1\ge x_2\ge\cdots\ge x_{n-1}\ge|x_n|\},\label{eq:Dn-chamber}\\
 \mathfrak t_+(B_n)
 &=\{x_1\ge x_2\ge\cdots\ge x_n\ge0\}.
 \label{eq:Bn-chamber}
\end{align}
For $n=1$ we still write $D_1$, although $\Spin(2)$ is just a circle: there
are no roots, the Weyl group is trivial and the chamber is the whole real
line. For $n=0$ in odd dimension the module is $\C$, the Lie algebra is zero,
and we write $0$ for the (empty) momentum value.

\section{Momentum maps and critical points}
\label{sec:momentum-map}

Let $S$ be one of the spin modules above. Its projective space
$\mathcal M=\PP(S)$ consists of rays $[\psi]=\C\psi$ with $\psi\ne0$,
so multiplying a spinor by a nonzero scalar does not change its point.
The Hermitian structure induces the Fubini--Study symplectic form on
$\mathcal M$. Relative to this form, the expectation of a Hermitian
generator produces the corresponding infinitesimal unitary motion;
this is the Hamiltonian property of the $K$-action. The momentum map
collects these expectation values into a real linear functional on
$\mathfrak k$, hence a point of its dual space $\mathfrak k^*$.
We normalize $\mu:\mathcal M\to\mathfrak k^*$ by
\begin{equation}
 \langle\mu([\psi]),A\rangle
 =-i\,\frac{\langle\psi,\rho_*(A)\psi\rangle}
              {\langle\psi,\psi\rangle},
 \qquad A\in\mathfrak k .
 \label{eq:momentum-map-definition}
\end{equation}
Here $\mathfrak k=\mathfrak{spin}(N)\simeq\mathfrak{so}(N)$, and the
inner product
\begin{equation*}
 (A,B)_{\mathfrak k}=-\frac12\operatorname{tr}(AB)
\end{equation*}
identifies $\mathfrak k^*$ with $\mathfrak k$ and defines the norm $\norm\mu$.
On the left of \eqref{eq:momentum-map-definition}, the brackets mean evaluation
of a functional; on the right they denote the Hermitian inner product.
The matrices
$L_{rs}=E_{rs}-E_{sr}$, $r<s$, are orthonormal, as are the Cartan
generators $J_j=-L_{2j-1,2j}$. The Hermitian inner product on $S$ is
conjugate-linear in its first argument.
Thus each component is the expectation of the Hermitian operator
$-i\rho_*(A)$. The quotient is unchanged by rescaling $\psi$, so it is
well defined on rays. Unitary conjugation gives $K$-equivariance:
\begin{equation}
 \mu(k[\psi])=\operatorname{Ad}^*_k\mu([\psi]).
 \label{eq:momentum-equivariance}
\end{equation}
Here $\operatorname{Ad}^*$ is the dual of the adjoint action on the Lie
algebra. It preserves the chosen inner product, so $\norm\mu^2$ is
$K$-invariant, and every momentum value has a representative in the
positive Weyl chamber.  In even dimension, formula \eqref{eq:momentum-map-definition} is a
fixed multiple of the Majorana covariance matrix; in the matrix basis
$E_{rs}-E_{sr}$, its entries are
\begin{equation}
 \mu_{rs}([\psi])
 =-\frac{i}{2}\,
   \frac{\langle\psi,\gamma_r\gamma_s\psi\rangle}
        {\langle\psi,\psi\rangle}.
 \label{eq:majorana-momentum}
\end{equation}

\subsection{Weight supports and root components}

Expand $\psi$ in the orthonormal Fock basis, indexed by its weights.
Its \emph{weight support} $\Sigma(\psi)$ records exactly the nonzero
coefficients:
\begin{equation}
 \psi=\sum_{\eta\in\Sigma(\psi)}c_\eta\ket\eta,\qquad
 \Sigma(\psi)=\{\eta:c_\eta\ne0\}.
 \label{eq:weight-support}
\end{equation}
Projecting orthogonally to $\mathfrak t^*$ retains the expectations of
the commuting Cartan generators. Since these are diagonal in the weight
basis, their values are
\begin{equation}
 \mu_{\mathfrak t}([\psi])
 =\sum_{\eta\in\Sigma(\psi)}p_\eta\eta,\qquad
 p_\eta=\frac{|c_\eta|^2}{\sum_\nu|c_\nu|^2}.
 \label{eq:cartan-momentum}
\end{equation}
The $p_\eta$ are probabilities obtained from the squared coefficient
moduli; they are positive and sum to one. Thus the Cartan projection belongs to the
\emph{convex hull}, the set of combinations with nonnegative coefficients
summing to one; root components may still contribute to the full momentum.

\begin{lemma}[Root components]
\label{lemma_cartan_roots}
Let
\begin{equation*}
 \mathfrak g=\mathfrak t^\C
 \oplus\bigoplus_{\alpha\in\Delta}\mathfrak g_\alpha
\end{equation*}
be the root decomposition of $\mathfrak g=\mathfrak k^\C$.  Then
$\mu([\psi])\in\mathfrak t^*$ if and only if
\begin{equation}
 \langle\psi,\rho_*(E_\alpha)\psi\rangle=0
 \qquad\text{for every }E_\alpha\in\mathfrak g_\alpha .
 \label{eq:root-components}
\end{equation}
\end{lemma}

\begin{proof}
Put $F_\alpha=\rho_*(E_\alpha)$. The Hermitian operators
$F_\alpha+F_\alpha^\dagger$ and $i(F_\alpha-F_\alpha^\dagger)$ give the
observable directions associated with $\pm\alpha$. Their expectations
vanish together exactly when $\langle\psi,F_\alpha\psi\rangle=0$.
Together, these pairs test every direction in $\mathfrak k$
orthogonal to $\mathfrak t$, proving the claim.
\end{proof}

\medskip

\begin{lemma}[Root-distinct supports]
\label{lemma_root_distinct_support}
If $\Sigma(\psi)$ is root-distinct, then
$\mu([\psi])\in\mathfrak t^*$, and the full momentum value is the convex
combination in \eqref{eq:cartan-momentum}.
\end{lemma}

\begin{proof}
A root operator sends the $\eta$-weight space to the
$(\eta+\alpha)$-weight space.  Root-distinctness therefore makes
$\rho_*(E_\alpha)\psi$ orthogonal to $\psi$ for every root $\alpha$.
Lemma~\ref{lemma_cartan_roots} eliminates all non-Cartan components, while
\eqref{eq:cartan-momentum} gives the remaining component.
\end{proof}

\medskip

Root-distinctness is sufficient but not necessary, since different terms
in \eqref{eq:root-components} may cancel. In the finite cases treated here,
every orbit contains a spinor with root-distinct support; this is proved separately in
Sections~\ref{sec:root-distinct-exhaustiveness}
and~\ref{sec:odd-exhaustiveness}.

The distinction between the Cartan projection and the full momentum map is
already visible for $S^+=\bigwedge^{\mathrm{even}}\C^2$.  For
\begin{equation*}
 \psi=c_0\ket\varnothing+c_{12}\ket{12},
\end{equation*}
the two weights are $(\frac12,\frac12)$ and
$(-\frac12,-\frac12)$, so
\begin{equation*}
 \mu_{\mathfrak t}([\psi])
 =\frac{|c_0|^2-|c_{12}|^2}
        {2(|c_0|^2+|c_{12}|^2)}(e_1+e_2).
\end{equation*}
Their difference is the $D_2$ root $e_1+e_2$.  Correspondingly, the root
operator $a_1a_2$ has expectation proportional to
$\overline{c_0}c_{12}$.  Unless one coefficient vanishes, the non-Cartan
momentum component is therefore nonzero.  For $|c_0|=|c_{12}|\ne0$, the Cartan projection vanishes while the full
momentum map is nonzero.

The torus convexity theorem is due to Atiyah
\cite{Atiyah1982ConvexityAC} and to Guillemin--Sternberg
\cite{GuilleminSternberg1982}. Kirwan's non-Abelian extension
\cite{Kirwan1984ConvexityIII} implies that
$\mu(\mathcal M)\cap\mathfrak t_+$ is a convex polytope, since $\mathcal M$ is compact and connected (see also Sjamaar's treatment \cite{Sjamaar1998Convexity}). For a fixed support,
varying the coefficient moduli in \eqref{eq:weight-support} moves its Cartan
momentum value in the convex hull of that support.  Infinitesimally,
\begin{equation}
 \operatorname{pr}_{\mathfrak t^*}
 \bigl(\operatorname{Im}d\mu_{[\psi]}\bigr)
 =\Span_{\R}\{\eta-\nu:\eta,\nu\in\Sigma(\psi)\}.
\end{equation}
Here $d\mu_{[\psi]}$ is the differential, $\operatorname{Im}$ its image,
and $\operatorname{pr}_{\mathfrak t^*}$ orthogonal projection.
Indeed, the first-order variations of the probabilities satisfy
$\sum_\eta\dot p_\eta=0$, and their images are
$\sum_\eta\dot p_\eta\eta$. Since all $p_\eta$, $\eta\in\Sigma(\psi)$, are positive, every such variation occurs, and the images span the differences of the support weights.

\subsection{Critical points of \texorpdfstring{$\norm\mu^2$}{the squared momentum norm}}
\label{sec:critical-machinery}

For $\xi\in\mathfrak k$, let
$\widehat\xi=-i\rho_*(\xi)$ denote its Hermitian action on $S$.
A point is critical for $\norm\mu^2$ when its derivative vanishes in
every tangent direction. The following equation expresses this condition
as an eigenvector problem
\cite{Ness1984,Kirwan_cohomology_84, SOK_criticalset,SOK_convexity}.

\begin{proposition}[Critical-point equation]
\label{prop:critical-general}
A point $[\psi]\in\PP(S)$ is critical for $\norm\mu^2$ if and only if
\begin{equation}
 \widehat{\mu([\psi])}\psi
 =\norm{\mu([\psi])}^2\psi .
 \label{eq:critical-general}
\end{equation}
\end{proposition}

\begin{proof}
Normalize $\psi$. After removing changes of norm and phase, tangent
variations at $[\psi]$ are represented by vectors in $\psi^\perp$.  For $\dot\psi\perp\psi$, differentiation gives
\begin{equation*}
 d\norm\mu^2_{[\psi]}(\dot\psi)
 =4\operatorname{Re}
   \langle\dot\psi,\widehat{\mu([\psi])}\psi\rangle.
\end{equation*}
The differential vanishes on $\psi^\perp$ if and only if
$\widehat{\mu([\psi])}\psi\in\C\psi$. Taking its inner
product with $\psi$ and using
\eqref{eq:momentum-map-definition} shows that the proportionality constant is
$\langle\mu,\mu\rangle=\norm\mu^2$.
\end{proof}

\medskip

\begin{proposition}[Closest-point rule]
\label{prop:closest-point}
Suppose $\mu([\psi])=\beta\in\mathfrak t^*$. Then $[\psi]$ is critical
if and only if
\begin{equation}
 (\eta,\beta)=\norm\beta^2
 \qquad\text{for every }\eta\in\Sigma(\psi).
\label{eq:affine-hyperplane}
\end{equation}
\par\noindent
Equivalently, the support lies in the affine set
\begin{equation*}
 H_\beta=\{\eta:(\eta,\beta)=\norm\beta^2\}.
\end{equation*}
For $\beta\ne0$ this is a hyperplane; for $\beta=0$ it is
the whole Cartan dual.
Whenever the critical-point condition holds, $\beta$ is the unique
point of smallest norm in the affine
hull of the support, namely the set of real combinations whose
coefficients sum to one. It belongs to the convex hull, in fact to its
relative interior (the interior within the affine hull), because all
support probabilities are positive.
\end{proposition}

\begin{proof}
Because $\beta$ is in the Cartan algebra, $\widehat\beta$ acts on
$\ket\eta$ by the scalar $(\eta,\beta)$.  Comparing the nonzero weight
components in \eqref{eq:critical-general} proves
\eqref{eq:affine-hyperplane}.  Subtracting the equations for two support
weights shows that $\beta$ is orthogonal to every direction
$\eta-\nu$ in their affine hull.  Hence $\beta$ is the orthogonal
projection of the origin onto that affine hull.  Finally,
\eqref{eq:cartan-momentum} writes $\beta$ as a convex combination of the
support weights.
\end{proof}

\medskip

This closest-point rule is equivalent to Kirwan's method of minimal combinations of
weights \cite{Kirwan_cohomology_84}; see
\cite{MaciazekSawicki2015} for its use in the
critical-point algorithm. Take $\Sigma=\{\eta_1,\ldots,\eta_k\}$
to be affinely independent: the differences $\eta_i-\eta_1$, $i>1$,
are linearly independent. Its convex hull is then a simplex, and each
point of its affine hull has unique barycentric coordinates
$\beta=\sum_i p_i\eta_i$ with $\sum_i p_i=1$.
Let $G_\Sigma=((\eta_i,\eta_j))$ be the Gram matrix and $\mathbf1$
the column of $k$ ones. Minimizing $\frac12p^TG_\Sigma p$ subject to
$\mathbf1^Tp=1$ gives the Lagrange-multiplier equations
$(\eta_i,\beta)=\lambda$, or
\begin{equation}
 \begin{pmatrix}
  G_\Sigma&-\mathbf1\\
  \mathbf1^T&0
 \end{pmatrix}
 \begin{pmatrix}p\\ \lambda\end{pmatrix}
 =
 \begin{pmatrix}0\\1\end{pmatrix}.
 \label{eq:bordered-gram-general}
\end{equation}
The bordered matrix is nonsingular if and only if $\Sigma$
is affinely independent. Indeed, a real kernel vector $(q,c)$ satisfies
$G_\Sigma q=c\mathbf1$ and $\mathbf1^Tq=0$, so
$0=q^TG_\Sigma q=\norm{\sum_iq_i\eta_i}^2$. Affine independence
forces $q=0$, and then $c=0$, while an affine dependence gives a nonzero
kernel vector in the converse direction.  If the solution has $p_i>0$ and the root equations
\eqref{eq:root-components} vanish, then
\begin{equation}
 \psi_\Sigma=\sum_{i=1}^k\sqrt{p_i}\ket{\eta_i}
 \label{eq:state-from-probabilities}
\end{equation}
is critical and $\lambda=\norm\beta^2$. Choosing the support is thus a combinatorial problem; once it is chosen,
the squared moduli of the coefficients are found by solving a linear
system.
\begin{proposition}[Ness uniqueness]
\label{prop:ness-uniqueness}
If two critical points of $\norm\mu^2$ lie in the same complex
$G$-orbit, they lie in the same compact $K$-orbit. Their chamber
momentum values therefore agree.
\end{proposition}

This is \cite[Theorem~7.1]{Ness1984}; it holds for nonzero critical
values as well as for $\mu=0$. In general a complex orbit need not
contain a critical point. In the finite cases every orbit does, as
follows from the list of representatives together with the orbit bounds
of Sections~\ref{sec:even-orbits} and~\ref{sec:odd-orbits}. Equal chamber
momentum values need not imply $G$-equivalence: in dimension thirteen there are infinitely
many orbits with $\mu=0$ (Section~\ref{sec:finite-boundary}).

\MomentumMethodFigure
The probabilities determine the coefficient moduli. On an affinely
independent support, the Cartan torus also adjusts their ratios and
phases, as the following argument shows.

\begin{lemma}[Coefficient rescaling]
\label{lem:torus-rescaling}
On an affinely independent weight support, all choices of nonzero complex
coefficients are projectively equivalent under the complex Cartan torus.
All choices of phases with fixed moduli are equivalent under the compact
Cartan torus, up to a common phase.
\end{lemma}
\begin{proof}
For $h\in\C^n$, the torus element $\exp(\sum_jh_j\widehat H_j)$
multiplies $c_a/c_1$ by $e^{(h,\eta_a-\eta_1)}$, with the weight pairing
extended bilinearly. Affine independence
makes $h\mapsto((h,\eta_a-\eta_1))_{a=2}^k$ surjective. Choose its
values as logarithms of the required changes in the ratios. For phases
alone these values, and hence a solution $h$, can be chosen imaginary,
giving an element of the compact torus.
\end{proof}

\subsection{Constructive form of the method}
\label{sec:constructive-procedure}

For the finite cases treated below, these results turn the search for
critical orbits into the following procedure.
\begin{enumerate}
\item Encode the weights by Fock occupation strings and translate root
      adjacency into the corresponding Hamming-distance condition.
\item Classify candidate supports up to the Weyl group.  In dimension fourteen we do this through intersections of the annihilators of the constituents, using the reduction condition of Definition~\ref{def:reduced-coordinate}.
\item For each candidate support, solve
      \eqref{eq:bordered-gram-general}.  Discard a support if the solution has
      nonpositive entries: zero entries reduce the support, while a
      negative entry means that the affine minimum is outside the simplex.
      In the latter case its convex minimum must be sought on a proper face.
\item Check the non-Cartan equations \eqref{eq:root-components}.  They vanish
      automatically for a root-distinct support, while a non-root-distinct
      support requires an explicit cancellation check.
\item Move the resulting momentum value to the positive Weyl chamber and use
      Ness's uniqueness theorem to separate critical complex orbits.
\end{enumerate}
Completeness requires that the selected supports meet
every orbit. This is proved in Sections~\ref{sec:root-distinct-exhaustiveness} and~\ref{sec:odd-exhaustiveness}. Section~\ref{sec:s14-incidence} classifies the reduced supports in dimension fourteen.

\begin{remark}[Comparison with qubits]
\label{rem:qubit-comparison}
For $L$ qubits every minimal combination of weights in the positive chamber,
with one exception, is the momentum value of a critical point
\cite[Theorem~10]{MaciazekSawicki2015}. The main reason is that a state built
from a minimal combination $\beta$ with $\beta_j>0$ for every $j$ has diagonal
one-qubit reduced density matrices \cite[Lemma~9]{MaciazekSawicki2015}, so the
Cartan part of its momentum is the whole momentum. For spinors this is far from true. The
positive chamber contains $6$, $12$ and $26$ minimal combinations of weights
for $D_4$, $D_5$ and $D_6$, and $5$, $10$ and $26$ for $B_3$, $B_4$ and $B_5$
(Appendix~\ref{app:minimal-combinations}), but only $2,2,4$ and $2,3,5$ of them
are critical values. The difference
comes from pairs of weights that differ by a root: they produce non-Cartan
components of the momentum map, and in general these do not vanish. This is
why we work with root-distinct supports and prove completeness by a separate
argument.
\end{remark}

\section{Geometric and fermionic interpretations}
\label{sec:interpretation}

\subsection{Compact and complex spin transformations}

For $N=2n$, the compact group $K$ acts by reversible fermionic Gaussian
transformations, generated by Hermitian quadratic expressions in the
$2n$ Majorana operators. Its complexification allows invertible
nonunitary Gaussian transformations. Accordingly, its projective orbits
are the equivalence classes under invertible Gaussian filtering, sometimes
called fermionic Gaussian SLOCC \cite{SL2014}.

For $N=2n+1$, the additional generator is
$\gamma_{2n+1}=P=(-1)^F$. The short-root generators
$a_jP$ and $a_j^\dagger P$ change fermion parity. The full
$\Spin(2n+1)$ action therefore extends the parity-preserving Gaussian
group on $n$ modes.

The function $\norm\mu^2$ is $K$-invariant but in general not $G$-invariant.
If a complex orbit contains a critical point, then by Ness's theorem
the critical points in it form a single $K$-orbit. Their common chamber
value is therefore an invariant of the orbit, and in the finite cases it
distinguishes the orbits.

\subsection{Covariance matrices and the spectral polytope}

In even dimension define the physical covariance matrix
\cite{Bravyi2005} by
\begin{equation*}
 C_{rs}=\frac{i}{2}\,
 \frac{\langle\psi,[\gamma_r,\gamma_s]\psi\rangle}
      {\langle\psi,\psi\rangle}.
\end{equation*}
It is real skew-symmetric, with $C_{rs}=-2\mu_{rs}$ for $r<s$
in the convention of \eqref{eq:majorana-momentum}. The Lie algebra element
representing the momentum is $M=\sum_{r<s}\mu_{rs}L_{rs}$. At a Cartan
value $\beta$, its $j$-th block is
$\bigl(\begin{smallmatrix}0&-\beta_j\\\beta_j&0\end{smallmatrix}\bigr)$,
and the corresponding covariance block has upper entry $2\beta_j$.
Thus $\norm\mu^2=\sum_{r<s}\mu_{rs}^2=\sum_j\beta_j^2$.
Orthogonal conjugation reduces the matrix to these blocks.
The numbers
$\frac12\pm\beta_j$ are the occupations of the modes after a suitable
Bogoliubov transformation, which, in general, differ from the eigenvalues of
the number-conserving one-particle density matrix $(\langle a_i^\dagger a_j\rangle)$.

For odd dimension the same skew-matrix construction uses the extended
Clifford system. Its last row and column contain expectations of
$\gamma_rP$, so the full momentum is not just the ordinary covariance
matrix on the $2n$ physical Majoranas.

For $N=3$, the one-mode state
$\psi=(\vac+\ket1)/\sqrt2=\exp(\frac\pi4\gamma_1P)\vac$
is a pure spinor, but has no definite parity and hence is not a standard
pure fermionic Gaussian state. Its physical covariance vanishes:
$C_{12}=1-2\langle a_1^\dagger a_1\rangle=0$.
Yet \eqref{eq:majorana-momentum} gives $\mu_{12}=\mu_{13}=0$,
$\mu_{23}=-\frac12$, and $\norm\mu^2=\frac14$.
The nonzero component comes from the parity-changing observable $-i\gamma_2P$.

The momentum map image of the torus is the convex hull of all Fock weights. The
non-Abelian momentum polytope is
$\mu(\PP(S))\cap\mathfrak t_+$. On a root-distinct support the momentum value is the
barycentre of the weights, weighted by the probabilities $p_\eta$, and the
critical representative is the one whose barycentre is closest to the origin.

\subsection{Nullity and intersections of annihilators}
Besides the momentum value, a spinor has a classical geometric
invariant, its nullity $\dim M(\psi)$. In Section~\ref{sec:spin14} we use
a refinement of it. If a spinor is written as a sum of pure spinors, the
dimensions of the pairwise intersections of their annihilators do not
change when all of them are moved by the same spin transformation. These
numbers describe the relative position of the constituents, and in
dimension fourteen we classify the possible patterns. The orbit of the
resulting critical spinor is then identified by its chamber value.

\subsection{Physical meaning of the critical points}
\label{sec:physical-meaning}

For qubits, $\norm\mu^2$ is an affine function of the linear entropy of the
one-qubit reduced states, and this is why the critical points found in
\cite{MaciazekSawicki2015} are the locally maximally entangled states. For
spinors the momentum norm can be interpreted in the same way. Let $k\in K$ bring
$\mu([\psi])$ to the Cartan algebra, $\mu([k\psi])=\beta$, and put
$\lambda_j=\frac12-\beta_j$. Then
\begin{equation}
 \norm{\mu([\psi])}^2=\sum_{j=1}^n\Bigl(\lambda_j-\frac12\Bigr)^2 ,
 \label{eq:momentum-occupations}
\end{equation}
and for $N=2n$ the numbers $\lambda_j$ are the occupations of the
quasiparticle modes described above. Since $\widehat H_j$ has eigenvalues
$\pm\frac12$, every $\lambda_j$ lies between $0$ and $1$, and
$\norm\mu^2\le\frac n4$. Equality holds exactly for the pure spinors, which
for $N=2n$ are the fermionic Gaussian states. Indeed, they form the $K$-orbit of the
vacuum, where $\norm\mu^2=\frac n4$; conversely, if every $\lambda_j$ is $0$ or
$1$, then $k\psi$ is an eigenvector of every $\widehat H_j$, hence a weight
vector and a multiple of a Fock basis vector. At the other extreme, $\mu=0$
means $\lambda_j=\frac12$ for every $j$. For $N=2n\ge2$, up to normalization,
$1-\frac4n\norm\mu^2$ is the $\mathfrak{so}(2n)$ generalized entanglement of
Barnum, Knill, Ortiz, Somma and Viola \cite{BarnumEtAl2004,SommaEtAl2004}, and
it is closely related to the one-body entanglement of Gigena and Rossignoli
\cite{GigenaRossignoli2015}; see also \cite{OszmaniecKus2013}. By Ness's
theorem \cite{Ness1984} the critical representatives minimize
$\norm\mu^2$ on their $G$-orbits. In even dimension they therefore maximize
this generalized entanglement within each Gaussian-SLOCC class, analogously
to the locally maximally entangled critical states of \cite{MaciazekSawicki2015}.

\section{Normal forms, critical coefficients and orbit dimensions}
\label{sec:normal-forms}

In this section we prepare the case analysis of
Sections~\ref{sec:even-orbits}--\ref{sec:odd-orbits}. In
Section~\ref{sec:root-distinct-exhaustiveness} we show that every orbit
in dimensions ten to fourteen has a representative on a root-distinct
support; the proof, which uses Vinberg's carrier algebras, is given in
Appendix~\ref{app:carriers}. Lemma~\ref{lem:dynkin-coefficients} gives
the critical coefficients on such supports, and
Lemma~\ref{lem:s14-four-to-two} shows when four Fock states can be
replaced by two. In Section~\ref{sec:orbit-dimensions} we compute the
orbit dimensions
\subsection{Root-distinct normal forms}
\label{sec:root-distinct-exhaustiveness}

We now show that every half-spinor orbit in dimensions ten, twelve and
fourteen contains a representative with a root-distinct support. Note that
this is a statement about orbits and it does not say that every critical point
has a root-distinct support.

The proof uses Vinberg's theory of carrier algebras, and here we only explain
the idea and leave the details to Appendix~\ref{app:carriers}. The half-spin
modules in dimensions ten, twelve and fourteen are the degree-one parts of
$\mathbb Z$-gradings of the exceptional Lie algebras $\mathfrak e_6$,
$\mathfrak e_7$ and $\mathfrak e_8$ whose degree-zero parts are
$\mathfrak{so}_{2n}\oplus\C$, and the roots of degree one restrict to the Fock
weights. Vinberg showed that every element of degree one can be moved by the
degree-zero group into general position in a small graded subalgebra, its
\emph{carrier algebra}. In our gradings the carriers are sums of copies of
$\mathfrak{sl}_2$ and $\mathfrak{sl}_3$, and the degree-one part of a carrier is
spanned by the root vectors of its simple roots. These root vectors are Fock
basis vectors, so the carrier tells us which few Fock states a normal form
needs.

\begin{proposition}[Existence of coordinate normal forms]
\label{prop:root-distinct-normal-forms}
For $N=10,12,14$, every projective half-spinor orbit has a representative
with an affinely independent, root-distinct Fock support. For $N=14$
the support can moreover be chosen so that no four constituents have a
common annihilating null direction.
\end{proposition}

The support consists of the degree-one simple roots $\rho_1,\ldots,\rho_k$ of a carrier algebra.
These roots are root-distinct even in the ambient exceptional root system. Indeed, if $\rho_a-\rho_b$ were an ambient root, then
\begin{equation}
 [E_{\rho_a},E_{-\rho_b}]
 \in\mathfrak l_{\rho_a-\rho_b}
\end{equation}
would be nonzero. Since both root vectors on
the left belong to the carrier, the difference
would also be a root of the carrier, which is
impossible for two distinct simple roots.

As $\rho_a-\rho_b$ has degree zero, this proves
root-distinctness of the corresponding spinor
weights. Linear independence of the lifted roots
also implies affine independence of these weights,
as shown in Appendix~\ref{app:carriers}.

For $N=14$, the completeness of the carrier
provides an additional restriction: four
constituents cannot have a common annihilating
null direction. The proof is given in
Appendix~\ref{app:carriers}.


Once a root-distinct support is known, finding the critical coefficients is
a small linear problem. In dimensions ten to fourteen it can even be solved
once and for all and the answer can be read off from a graph attached to the
support.

\begin{lemma}[Critical coefficients of root-distinct supports]
\label{lem:dynkin-coefficients}
Let $n\in\{5,6,7\}$ and let $\Sigma=\{I_1,\ldots,I_k\}$ be a root-distinct
support of even words in $n$ modes, so that
$d_{ab}:=\lvert I_a\symdiff I_b\rvert\in\{4,6\}$ for $a\ne b$. Let $\Gamma$
be the graph on $\{1,\ldots,k\}$ with an edge $ab$ whenever $d_{ab}=6$.
\begin{enumerate}
\item $\Gamma$ is a disjoint union of isolated vertices and single edges.
\item Put $c_a=1$ if $a$ is isolated and $c_a=2$ if $a$ lies on an edge, and
      $C=\sum_ac_a$. Then
      \begin{equation}
       \psi_\Sigma=\sum_{a=1}^k\sqrt{c_a}\,\ket{I_a}
       \label{eq:dynkin-critical-state}
      \end{equation}
      is a critical point of $\norm\mu^2$ with probabilities $p_a=c_a/C$ and
      \begin{equation}
       \norm{\beta}^2=\frac2C-\frac{8-n}4 .
       \label{eq:dynkin-norm}
      \end{equation}
\item The weights $\eta_{I_1},\ldots,\eta_{I_k}$ are affinely independent.
\end{enumerate}
\end{lemma}

\begin{proof}
The idea is that, after a shift by a multiple of the all-ones matrix, the Gram
matrix of the support becomes the Cartan matrix of a sum of copies of $A_1$
and $A_2$. Indeed, by \eqref{eq:general-fock-weight}
$(\eta_{I_a},\eta_{I_b})=\frac{n-2d_{ab}}4$, so
\begin{equation}
 G_\Sigma+\tfrac{8-n}4\,J=2I-A_\Gamma ,
 \label{eq:shifted-gram}
\end{equation}
where $J$ is the all-ones matrix and $A_\Gamma$ the adjacency matrix of
$\Gamma$.
(1) If $d_{ab}=d_{bc}=6$ with $a\ne c$, then
$I_a\symdiff I_c=X\symdiff Y$ for two six-element subsets $X,Y$ of an
$n$-element set. Hence $\lvert X\cap Y\rvert\ge12-n$ and
$d_{ac}\le2n-12\le2$, contradicting root-distinctness.
(2) By (1), $(2I-A_\Gamma)c=2\cdot\mathbf 1$. By \eqref{eq:shifted-gram},
$G_\Sigma c=\bigl(2-\frac{8-n}4C\bigr)\mathbf 1$, so $p=c/C$ solves
\eqref{eq:bordered-gram-general} with $\lambda=\frac2C-\frac{8-n}4$.
Lemma~\ref{lemma_root_distinct_support} and
Proposition~\ref{prop:closest-point} show that $\psi_\Sigma$ is critical with
$\norm\beta^2=\lambda$.
(3) The matrix $2I-A_\Gamma$ has eigenvalues $2$ on isolated vertices and
$1,3$ on edges, so it is positive definite. By \eqref{eq:shifted-gram} it is
the Gram matrix of the vectors
$\rho_a=(\eta_{I_a};\frac12,\ldots,\frac12)\in\R^8$, which are therefore
linearly independent. An affine relation $\sum_at_a\eta_{I_a}=0$,
$\sum_at_a=0$, would give $\sum_at_a\rho_a=0$.
\end{proof}

For example, in dimension twelve ($n=6$, so $\frac{8-n}4=\frac12$) the support
$\{\varnothing,1234,3456\}$ has no pair at distance six. Its graph has no
edges, the three coefficients are equal, and
$\norm\beta^2=\frac23-\frac12=\frac16$. For $\{\varnothing,123456\}$ the graph
is a single edge, $C=4$ and $\norm\beta^2=0$. Both values appear in
Table~\ref{tab:finite-even-low}.

When the support comes from a carrier algebra, the graph $\Gamma$ is the Dynkin
diagram of the carrier (Remark~\ref{rem:carrier-dynkin}). This is why the
supports in Table~\ref{tab:s14-orbits} and Figure~\ref{fig:s14-dynkin} are
labelled by carriers.


A sum of two root-neighbouring Fock monomials is pure. Indeed, in a half-spin module two root-neighbouring weights differ in exactly
two occupation numbers. A Weyl transformation brings their support to $\{\varnothing,12\}$, and, for $c_0\ne0$,
\begin{equation}
 c_0\ket\varnothing+c_{12}\ket{12}
 =c_0\exp\left(
 \frac{c_{12}}{c_0}a_1^\dagger a_2^\dagger
 \right)\vac,
\end{equation}
which is a pure spinor. This is the coordinate case of \cite[Proposition~5]{Trautman_Budinich}, which refers to \cite[Proposition~III.1.12]{Chevalley1954}.

In an odd-dimensional spin module, weights may also differ by a short root, corresponding to a change in a single occupation number. Such a pair can be brought to $\{\varnothing,1\}$. Writing $P=(-1)^F$, we have
\begin{equation}
 c_0\ket\varnothing+c_1\ket1
 =c_0\exp\left(
 \frac{c_1}{c_0}a_1^\dagger P
 \right)\vac.
\end{equation}
Since $a_1^\dagger P$ is a short-root
generator of $\mathfrak{spin}(2n+1,\C)$, this state is also pure. If $c_0=0$, purity follows directly from the purity of every Fock basis vector.

Thus, in either parity, a root-neighbouring
pair produces only a pure spinor.
The next lemma gives a less obvious reduction:
four Fock states in a special configuration
can be transformed into a sum of two pure
spinors. The corresponding orbit equivalence
goes back to Igusa
\cite{Igusa1970uptodim12}, and its role
in the combinatorics of pure-spinor
decompositions was observed by
Bhoja and Krasnov \cite{BhojaKrasnov2025}.
We give an explicit spin transformation
realizing this equivalence.

\begin{lemma}[Four-to-two reduction]
\label{lem:s14-four-to-two}
\begin{enumerate}
\item Let $n=7$. If four root-distinct Fock basis vectors have a common
      annihilating null direction, then a Weyl transformation sends their
      words to $\varnothing,1234,1256,3456$.
\item Let $n\in\{6,7\}$. Every linear combination of
      $\ket\varnothing,\ket{1234},\ket{1256},\ket{3456}$ with four nonzero
      coefficients is a sum of two pure spinors and is projectively
      equivalent to $\ket\varnothing+\ket{123456}$ under
      $\Spin(2n,\C)$.
\end{enumerate}
\end{lemma}

\begin{proof}
(1) Send the first word to $\varnothing$ by a Weyl transformation. The common
intersection is a nonzero coordinate subspace of
$M(\vac)=\Span\{a_1,\ldots,a_7\}$, so it contains some $a_j$. After a
permutation of the modes, no word contains the mode $7$. The remaining three
words lie in six modes and have size four or six. A six-set has distance two
from every four-set, so all three have size four. Their pairwise
intersections have size two, so their two-element complements are disjoint
and partition the six modes, which gives the stated words.

(2) The trick is to treat each of the pairs $(1,2),(3,4),(5,6)$, which in
our four words are either empty or doubly occupied, as a single qubit. The
support then becomes $\{000,110,101,011\}$. The difference vectors $(1,1,0),(1,0,1),(0,1,1)$ have
determinant $-2$, so the weights are affinely independent.
By Lemma~\ref{lem:torus-rescaling}, after a complex Cartan transformation
and an overall nonzero rescaling, we may use the representative
\begin{equation}
 \ket{000}+\ket{110}+\ket{101}+\ket{011}.
 \label{eq:s14-even-parity}
\end{equation}
For a mode pair $(p,q)$ put
$R_{pq}(\theta)=\exp\bigl(\theta(a_p^\dagger a_q^\dagger-a_qa_p)\bigr)$. On
its empty and doubly occupied states, $R_{pq}(\pi/4)$ acts by
$2^{-1/2}\bigl(\begin{smallmatrix}1&-1\\1&1\end{smallmatrix}\bigr)$. Writing
\eqref{eq:s14-even-parity} as
$\frac12\bigl((\ket0+\ket1)^{\otimes3}+(\ket0-\ket1)^{\otimes3}\bigr)$ and
applying this matrix to each factor shows that the commuting product
$R_{12}(\pi/4)R_{34}(\pi/4)R_{56}(\pi/4)$ sends \eqref{eq:s14-even-parity} to
$\sqrt2(\ket{000}+\ket{111})=\sqrt2(\vac+\ket{123456})$, a sum of two pure
spinors.
\end{proof}

In particular, a decomposition into Fock basis constituents that is minimal
among all pure-spinor decompositions contains no quadruple as in (1). For
$n=6$ the hypothesis of (1) is never satisfied, since four words in five
modes cannot be pairwise root-distinct; the support in (2) is the four-word
support met in dimension twelve (Section~\ref{sec:even-twelve}).

\subsection{Orbit dimensions}
\label{sec:orbit-dimensions}

The tangent space of the vector orbit $G\psi$ at $\psi$ is
$\mathfrak g\psi=\{X\psi:X\in\mathfrak g\}$, and that of the projective orbit
$G[\psi]$ is $(\mathfrak g\psi+\C\psi)/\C\psi$. So if $X_1,\ldots,X_d$ span the
image of $\mathfrak g$ in $\End(S)$ and
\begin{equation}
 A_\psi=(X_1\psi\ \cdots\ X_d\psi),\qquad
 B_\psi=(\psi\ X_1\psi\ \cdots\ X_d\psi),
 \label{eq:orbit-matrices}
\end{equation}
then
\begin{equation}
 \dim_\C G\psi=\operatorname{rank}_\C A_\psi,\qquad
 \dim_\C G[\psi]=\operatorname{rank}_\C B_\psi-1.
 \label{eq:exact-orbit-rank}
\end{equation}
For the vacuum in even dimension, $X\vac$ is a multiple of $\vac$ plus a
combination of the $\binom n2$ states $\ket{ij}$, so $\dim G[\vac]=\binom n2$;
for $N=14$ these are the $21$ pair-creation directions of the pure-spinor chart
\eqref{eq:pure-gaussian-chart}. For the other representatives we take for
$X_1,\ldots,X_d$ the bilinears $\gamma_r\gamma_s$, $r<s$.
Lemma~\ref{lem:torus-rescaling} allows us to set all coefficients on an
affinely independent support equal to one, and the ranks are then computed
exactly over $\mathbb Q(i)$ (Appendix~\ref{app:orbit-dimensions}).

\section{Half-spinors in dimension at most twelve}
\label{sec:even-orbits}

Let $N=2n$, let
\[
 S_N^+=\bigwedge\nolimits^{\mathrm{even}}\C^n,
 \qquad G_N=\Spin(2n,\C),
\]
and use positive chirality. By Section~\ref{sec:explicit-root-operators}, a
root-distinct Fock support is an even binary code of minimum distance at
least four, and by Section~\ref{sec:normal-forms} every orbit in dimensions
ten and twelve has a representative with such a support. We now classify
these supports and find their critical coefficients. In these dimensions
everything can be done by hand, and the calculations illustrate the method before we turn to dimension fourteen.

\subsection{Pure spinors in dimensions two, four and six}
\label{sec:even-pure-cases}

The pure-spinor variety is the image of one component of the projective
maximal isotropic Grassmannian under the pure-spinor embedding, so it
is closed. It has dimension $n(n-1)/2$: in the chart
\eqref{eq:pure-gaussian-chart} its coordinates are the entries of a skew
$n\times n$ matrix. For $n=1,2,3$ this equals
$\dim\PP(S_N^+)=2^{n-1}-1$. A closed subvariety of the same dimension
as the irreducible projective space must be the whole space. In the Fock picture this is reflected in the fact that no two even words in
at most three modes are root-distinct.

The spin group is transitive on the projectivized half-spin module in these
dimensions.  A representative is the vacuum, and its momentum value is
\[
 \beta_{\mathrm{pure}}=
 \left(\left(\frac12\right)^n\right),
 \qquad
 \norm{\beta_{\mathrm{pure}}}^2=\frac n4 .
\]
This gives one orbit in each of dimensions $N=2,4,6$, of projective
dimensions $0,1,3$, respectively.

\subsection{Dimension eight}
\label{sec:even-eight}

For $n=4$, normalize one support weight to
$\eta_\varnothing=(\frac12,\frac12,\frac12,\frac12)$.
The only even occupation string at Hamming distance at least four from it is
$1234$.  Therefore the only root-distinct support with more than one word is the antipodal
pair
\[
 \Sigma_8=\{\eta_\varnothing,\eta_{1234}\}.
\]
The closest point of its segment is its midpoint,
\begin{equation}
 \frac12\eta_\varnothing+\frac12\eta_{1234}=0.
 \label{eq:D4-antipodal-midpoint}
\end{equation}
The corresponding critical representative is
$\ket\varnothing+\ket{1234}$. To see that there are no other orbits we use triality. The symmetric
Chevalley pairing $q$ on $S_8^+$ is nondegenerate and invariant
\cite{Chevalley1954}, so $\Spin(8,\C)$ maps into $SO(S_8^+,q)$. The
corresponding map of Lie algebras $\mathfrak{so}_8\to\mathfrak{so}(S_8^+)$ is
injective, because $\mathfrak{so}_8$ is simple and the representation is
nontrivial, and both algebras have dimension $28$; hence the image of
$\Spin(8,\C)$ is all of $SO(S_8^+,q)$. By Witt's theorem this group is
transitive on the isotropic lines and on each nonzero level set of $q$ (a
reflection in the orthogonal complement corrects the determinant if
necessary). After projectivization all nonzero levels become one orbit. So
there are exactly two orbits: the pure spinors, which form the quadric $q=0$,
and all the others, represented by $\ket\varnothing+\ket{1234}$.
Figure~\ref{fig:D4-root-graph}(a) displays the
root-adjacency graph in the same Fock notation: the six two-occupation
states are root-neighbours of the vacuum, and $\ket{1234}$ is its only
root-distinct partner. Panel (b) shows the midpoint construction and the
normalized critical state.

\DfourWeightFigure

\subsection{Dimension ten}
\label{sec:even-ten}

For $n=5$, every word root-distinct from the vacuum has even size at least
four, hence has size four.  The stabilizer of the vacuum permutes the five
modes, so every such word is equivalent to $1234$.  Two distinct
four-subsets of a five-element set have Hamming distance two and are
root-neighbouring.  Consequently no root-distinct triple exists.

The unique nonsingleton root-distinct support is therefore
$\{\varnothing,1234\}$.  Its midpoint is
\begin{equation}
 \left(\frac12(\eta_\varnothing+\eta_{1234})\right)^+
 =\left(\frac12,0,0,0,0\right)
\end{equation}
after moving to the positive Weyl chamber.  The two critical representatives
are $\ket\varnothing$ and
$\ket\varnothing+\ket{1234}$.

\subsection{Dimension twelve}
\label{sec:even-twelve}

For $n=6$, a word root-distinct from the vacuum has size four or six.
This gives two inequivalent pairs:
\[
 \{\varnothing,1234\},
 \qquad
 \{\varnothing,123456\}.
\]
Their chamber closest points are, respectively,
\[
 \left(\frac12,\frac12,0,0,0,0\right)
 \quad\text{and}\quad 0.
\]

If a support contains the six-letter word $123456$, no four-subset can join
it: their distance would be two. Otherwise every nonempty word has size
four. Distinct four-subsets are compatible if and only if their complements
are disjoint two-subsets of the six modes. There can be at most three such
complements. Consequently the possible support sizes are two, three and
four, and there is a single triple type,

\begin{equation}
 \Sigma_{12}=\{\varnothing,1234,3456\}.
 \label{eq:D6-triple-support}
\end{equation}
Every pair has Hamming distance four.  The three weights therefore have the
same Gram products, so the bordered system
\eqref{eq:bordered-gram-general} gives equal probabilities. In the displayed
Fock frame their momentum is
\begin{equation}
 \beta=\frac13(\eta_\varnothing+\eta_{1234}+\eta_{3456})
       =\frac16(1,1,-1,-1,1,1).
\end{equation}
The $D_6$ Weyl transformation that changes the signs of coordinates
$3,4$ sends this point to the positive chamber. Hence
\begin{equation}
 \beta^+=
 \left(\left(\frac16\right)^6\right),
 \qquad \norm\beta^2=\frac16.
 \label{eq:D6-triple-beta}
\end{equation}
Therefore, the new critical representative is
$\ket\varnothing+\ket{1234}+\ket{3456}$.
Figure~\ref{fig:D6-root-graph} shows the support triangle, its Gram matrix,
and the Weyl transformation from $\beta$ to $\beta^+$.

\DsixWeightFigure

The remaining root-distinct support, $\{\varnothing,1234,3456,1256\}$,
gives no new orbit: by Lemma~\ref{lem:s14-four-to-two}(2)
every combination with four nonzero coefficients is projectively
equivalent to $\ket\varnothing+\ket{123456}$. Thus the same orbit can be represented on different root-distinct
supports, even with different numbers of constituents, so the number of
constituents is not an orbit invariant.

\subsection{Half-spinor orbit representatives through dimension twelve}
\label{sec:even-low-summary}

Table~\ref{tab:finite-even-low} collects the representatives. Here and in all
subsequent tables, $(x^r,y^s)$ denotes repeated coordinates, so $0^r$
means $r$ zero entries. A displayed representative is unnormalized;
$\beta^+$ is the chamber translate of its momentum, and need not be its
Cartan momentum value in the displayed Fock frame. The last column gives projective
orbit dimensions, computed by \eqref{eq:exact-orbit-rank} as explained in
Section~\ref{sec:orbit-dimensions}.

\begin{table}[t]
\centering
\renewcommand{\arraystretch}{1.15}
\small
\begin{tabular}{c l c c c}
\toprule
$N$ & critical representative $\psi$ & $\beta^+$ &
$\norm\beta^2$ & $\dim G_N[\psi]$\\
\midrule
2  & $\ket\varnothing$ & $(\frac12)$ & $\frac14$ & 0\\
4  & $\ket\varnothing$ & $((\frac12)^2)$ & $\frac12$ & 1\\
6  & $\ket\varnothing$ & $((\frac12)^3)$ & $\frac34$ & 3\\
\midrule
8  & $\ket\varnothing$ & $((\frac12)^4)$ & $1$ & 6\\
8  & $\ket\varnothing+\ket{1234}$ & $(0^4)$ & $0$ & 7\\
\midrule
10 & $\ket\varnothing$ & $((\frac12)^5)$ & $\frac54$ & 10\\
10 & $\ket\varnothing+\ket{1234}$ & $(\frac12,0^4)$ & $\frac14$ & 15\\
\midrule
12 & $\ket\varnothing$ & $((\frac12)^6)$ & $\frac32$ & 15\\
12 & $\ket\varnothing+\ket{1234}$ & $((\frac12)^2,0^4)$ & $\frac12$ & 24\\
12 & $\ket\varnothing+\ket{1234}+\ket{3456}$ & $((\frac16)^6)$ & $\frac16$ & 30\\
12 & $\ket\varnothing+\ket{123456}$ & $(0^6)$ & $0$ & 31\\
\bottomrule
\end{tabular}
\caption{Critical representatives, chamber values and projective orbit dimensions
for the half-spin modules for even $N\le 12$. The list agrees with Igusa's
classification \cite{Igusa1970uptodim12}.}
\label{tab:finite-even-low}
\end{table}

We can now show that the table is complete. By
Proposition~\ref{prop:root-distinct-normal-forms}, every orbit in dimensions
ten and twelve has a representative on one of the supports just enumerated.
The additional four-word support in dimension twelve does not occur as a
normal form and lies in the orbit of $\ket\varnothing+\ket{123456}$. In dimensions two to eight the transitivity arguments above leave no room for
other orbits. So Table~\ref{tab:finite-even-low} meets every orbit. Its
chamber values are distinct, so by Proposition~\ref{prop:ness-uniqueness} no
two entries in the same dimension represent the same orbit. This gives the
counts $1,1,1,2,2,4$ without using Igusa's classification, and comparison with
\cite{Igusa1970uptodim12} recovers his list.

\FloatBarrier

\section{The fourteen-dimensional half-spin representation}
\label{sec:spin14}

Dimension fourteen is the richest of the finite cases. There are nine
nonzero orbits, and their representatives need up to five Fock states, so the
classification can no longer be done by inspection. Let
\[
 S^+=\bigwedge\nolimits^{\mathrm{even}}\C^7,\qquad
 K=\Spin(14),\qquad G=\Spin(14,\C).
\]
The following theorem lists the nine nonzero projective orbits together with
their critical representatives.
The supports are those found by Kac and Vinberg \cite{Gatti_dim13}; we
choose the coefficients so that each representative is critical and
compute its momentum value.

\begin{theorem}
\label{thm:s14}
The nine vectors in Table~\ref{tab:s14-orbits} are critical points of
$\norm\mu^2$, lie in pairwise distinct $G$-orbits, and represent all nonzero
projective orbits in $\PP(S^+)$. Their orbit dimensions are also
given in Table~\ref{tab:s14-orbits}, and their closure order in
Proposition~\ref{prop:s14-closure}.
\end{theorem}

\begin{table}[t]
\centering
\renewcommand{\arraystretch}{1.18}
\small
\begin{tabular}{c c l c c c}
\toprule
type & carrier & critical representative $\psi$ & $\beta^+$ &
$\norm\beta^2$ &
$\dim G[\psi]$\\
\midrule
I & $A_1$ & $\ket\varnothing$
  & $((\frac12)^7)$ & $\frac74$ & 21\\
II & $2A_1$ & $\ket\varnothing+\ket{1234}$
  & $((\frac12)^3,0^4)$ & $\frac34$ & 34\\
III & $3A_1$ & $\ket\varnothing+\ket{1234}+\ket{3456}$
  & $(\frac12,(\frac16)^6)$ & $\frac5{12}$ & 42\\
IV & $A_2$ & $\ket\varnothing+\ket{123456}$
  & $(\frac12,0^6)$ & $\frac14$ & 43\\
V & $4A_1$ & $\ket\varnothing+\ket{1234}+\ket{3456}+\ket{1367}$
  & $((\frac14)^4,0^3)$ & $\frac14$ & 49\\
VI & $A_2{+}A_1$ & $\sqrt2\,(\ket\varnothing+\ket{123456})+\ket{1237}$
  & $(\frac3{10},(\frac1{10})^5,-\frac1{10})$ & $\frac3{20}$ & 53\\
VII & $A_2{+}2A_1$ & $\sqrt2\,(\ket\varnothing+\ket{123456})+\ket{1237}+\ket{1567}$
  & $((\frac16)^3,0^4)$ & $\frac1{12}$ & 58\\
VIII & $A_2{+}3A_1$ & $\sqrt2\,(\ket\varnothing+\ket{123456})+\ket{1237}+\ket{1567}+\ket{2467}$
  & $((\frac1{14})^6,-\frac1{14})$ & $\frac1{28}$ & 62\\
IX & $2A_2$ & $\ket\varnothing+\ket{123456}+\ket{1237}+\ket{4567}$
  & $(0^7)$ & $0$ & 63\\
\bottomrule
\end{tabular}
\caption{Critical representatives of the nine nonzero projective orbits in
dimension fourteen, numbered as in \cite[Table~1]{Gatti_dim13} (by increasing
orbit dimension). The carrier is the carrier algebra of
Theorem~\ref{thm:vinberg-carrier}; $\beta^+$ is the positive-chamber
momentum value and $(x^r,y^s)$ denotes repeated entries. Since $W(D_7)$ changes only an even number of
signs, the last coordinate of $\beta^+$ stays negative for VI and VIII. The
last column gives the projective orbit dimensions
(Section~\ref{sec:popov-comparison-integrated}).}
\label{tab:s14-orbits}
\end{table}

The proof has the three steps described in the introduction. First,
Proposition~\ref{prop:root-distinct-normal-forms} reduces the problem to
\emph{reduced} coordinate supports (Definition~\ref{def:reduced-coordinate}):
sets of Fock states whose annihilators never share a direction four at a
time. Second, an elementary count of how the annihilators intersect, leaves only nine possible
incidence patterns \cite{BhojaKrasnov2025}, and each of them is realized by exactly one support up to
the Weyl group. Third, Lemma~\ref{lem:dynkin-coefficients} gives the critical
coefficients.

\subsection{Coordinate supports}
\label{sec:s14-weights}

For an even subset $I\subseteq\{1,\ldots,7\}$ put $M_I=M(\ket I)$. By
\eqref{eq:general-annihilator-intersection},
\begin{equation}
 \dim(M_I\cap M_J)=7-\lvert I\symdiff J\rvert,
 \label{eq:s14-intersection}
\end{equation}
and by Section~\ref{sec:explicit-root-operators} a pair is root-distinct if
and only if $\lvert I\symdiff J\rvert\in\{4,6\}$. It is convenient to picture
a support as a complete graph whose vertices are its words, and to label the
edge $IJ$ by $e_{IJ}=\dim(M_I\cap M_J)\in\{3,1\}$. The edges with label $1$ are
exactly the edges of the graph $\Gamma$ of Lemma~\ref{lem:dynkin-coefficients}.

\begin{definition}[Reduced coordinate support]
\label{def:reduced-coordinate}
A set of distinct even words in seven modes is a \emph{reduced} support if
it is root-distinct and no four of the planes $M_I$ have a nonzero common
intersection.
\end{definition}

Equivalently, no mode is empty in four of the words and no mode is occupied
in four of them, since a nonzero intersection of coordinate planes contains a
coordinate direction $a_j$ or $a_j^\dagger$. By
Proposition~\ref{prop:root-distinct-normal-forms} every nonzero orbit has a
representative with a reduced, affinely independent support.

\begin{remark}
\label{rem:property-A}
Reducedness is our version of property (A) of Kac and Vinberg
\cite[Section~1.2]{Gatti_dim13}, which asks that the linear span of the
differences of the support weights contain no root of $D_7$. The two
conditions are closely related. For the four words
$\varnothing,1234,1256,3456$ of Lemma~\ref{lem:s14-four-to-two}, which are not
reduced, the root $e_1+e_2=\frac12\bigl[(\eta_\varnothing-\eta_{1234})
+(\eta_\varnothing-\eta_{1256})-(\eta_\varnothing-\eta_{3456})\bigr]$ lies in
that span. In fact the enumeration in Appendix~\ref{app:s14-enumeration} shows
that on all fifteen Weyl classes of root-distinct supports in seven modes the
two conditions coincide.
\end{remark}

\subsection{Classification by annihilator incidence}
\label{sec:s14-incidence}

The classification rests on a double count. Let a reduced support
have $k$ constituents $\psi_1,\ldots,\psi_k$. Among the
fourteen null basis directions $a_j,a_j^\dagger$ ($1\le j\le7$), let $n_r$ be
the number that belong to exactly $r$ of the annihilators
$M(\psi_a)$. For each mode $j$, exactly one of $a_j,a_j^\dagger$ annihilates
each constituent; if $a_j$ belongs to $r$ annihilators, then $a_j^\dagger$
belongs to the complementary $k-r$. Hence
\begin{equation}
 n_r=n_{k-r},\qquad \sum_{r=0}^k n_r=14.
 \label{eq:s14-nr}
\end{equation}
Put $e_{ab}=\dim(M(\psi_a)\cap M(\psi_b))$ and $E=\sum_{a<b}e_{ab}$. Counting
pairs consisting of a null basis direction and an unordered pair of
annihilators containing it gives
\begin{equation}
 E=\sum_{r=0}^k\binom r2 n_r ,
 \label{eq:s14-incidence}
\end{equation}
and reducedness means
\begin{equation}
 n_r=0\qquad(r\ge4).
 \label{eq:s14-no-four}
\end{equation}
These identities are the ones used by Bhoja and Krasnov
\cite{BhojaKrasnov2025}. Now the condition
\eqref{eq:s14-no-four} follows from
Proposition~\ref{prop:root-distinct-normal-forms}, rather than from an
explicit reduction.

\begin{proposition}[Nine reduced support types]
\label{prop:s14-nine}
Up to permutation of the constituents and the Weyl group of $D_7$, the
reduced supports have the following edge-label patterns:
\begin{center}
\normalfont\small
\begin{tabular}{cccl}
\toprule
type & $k$ & carrier & labels $e_{ab}$\\
\midrule
I & 1 & $A_1$ & no edge\\
II, IV & 2 & $2A_1$, $A_2$ & $3$; $1$\\
III, VI & 3 & $3A_1$, $A_2{+}A_1$ & $(3,3,3)$; $(3,3,1)$\\
V, VII, IX & 4 & $4A_1$, $A_2{+}2A_1$, $2A_2$ & six $3$'s; five $3$'s and one $1$;
      four $3$'s and two opposite $1$'s\\
VIII & 5 & $A_2{+}3A_1$ & nine $3$'s and one $1$\\
-- & $k\ge6$ & -- & none\\
\bottomrule
\end{tabular}
\end{center}
Each pattern is realized by exactly one support up to the Weyl group.
\end{proposition}

\begin{proof}
For $k=1$ there is nothing to prove, and for $k=2$ the two labels are $3$ and
$1$ by \eqref{eq:s14-intersection}.

For $k=3$, \eqref{eq:s14-nr} gives $n_3=n_0$, $n_2=n_1$ and $n_3+n_2=7$, so
$E=3n_3+n_2=7+2n_3\ge7$. A sum of three labels from $\{1,3\}$ is odd and at
most $9$, hence $E\in\{7,9\}$, giving $(3,3,1)$ and $(3,3,3)$.

For $k=4$, \eqref{eq:s14-no-four} and \eqref{eq:s14-nr} give $n_4=n_0=0$,
$2n_3+n_2=14$ and $E=14+n_3$. Six labels from $\{1,3\}$ have an even sum of at
most $18$, so $E\in\{14,16,18\}$ with two, one or no labels equal to $1$. If
$E=14$, then $n_3=0$ and every null direction belongs to exactly two
annihilators, and its dual to the complementary two. Thus opposite edges
have equal labels, and the two labels $1$ are on opposite edges.

For $k=5$, the same conditions leave only $n_2=n_3=7$, so
$E=n_2+3n_3=28$: nine labels $3$ and one label $1$. For $k=6$ they force
$n_3=14$ and $E=42$, which contradicts the odd parity of a sum of fifteen odd
labels. For $k\ge7$ the vanishing ranges $r\ge4$ and $r\le k-4$ cover all $r$,
contradicting $\sum_rn_r=14$.

Existence and uniqueness of each pattern are proved in
Section~\ref{sec:s14-reconstruction}.
\end{proof}

\SpinFourteenDynkinFigure

\subsection{Reconstruction up to Weyl equivalence}
\label{sec:s14-reconstruction}

It remains to show that each pattern is realized, and by only one support.
Send the first constituent to $\ket\varnothing$; the Weyl stabilizer of
$\eta_\varnothing$ then permutes the modes. For each mode form the occupation
column $u\in\{0,1\}^k$, with $u_1=0$, and identify it with the set
$S\subseteq\{2,\ldots,k\}$ of rows in which it is occupied. With $m_S$ the
multiplicity of $S$ and $\lvert I_a\rvert=7-e_{1a}$,
\begin{align}
 \sum_Sm_S&=7,\label{eq:s14-columns-one}\\
 \sum_{S\ni a}m_S=\lvert I_a\rvert,\qquad
 \sum_{S\supseteq\{a,b\}}m_S&=\lvert I_a\cap I_b\rvert
 =\tfrac12\bigl(\lvert I_a\rvert+\lvert I_b\rvert-7+e_{ab}\bigr).
 \label{eq:s14-column-intersections}
\end{align}
By reducedness no column is constant on four rows. For $k\le4$ this only
excludes $S=\varnothing$ when $k=4$, and \eqref{eq:s14-columns-one},
\eqref{eq:s14-column-intersections} determine the $m_S$ by M\"obius inversion
on the lattice of subsets. For $k=5$ only $\lvert S\rvert\in\{2,3\}$ remain,
and the eleven equations determine the ten multiplicities; they are
consistent exactly when $E=28$. The explicit formulas are in
Appendix~\ref{app:occupation-data}. Substituting the patterns of
Proposition~\ref{prop:s14-nine} gives the nonnegative integers of
Table~\ref{tab:s14-columns}, whose rows are even words with no quadruple
intersection. This proves existence and uniqueness up to the Weyl group. For
type VIII, ordering the seven columns as
$\mathtt{01110},\mathtt{01101},\mathtt{01100},\mathtt{01001},\mathtt{01010},
\mathtt{01011},\mathtt{00111}$ and reading across the rows gives
$\varnothing,123456,1237,1567,2467$, the support in
Table~\ref{tab:s14-orbits}.

\subsection{Proof of Theorem~\ref{thm:s14}}

By Proposition~\ref{prop:root-distinct-normal-forms}, every nonzero orbit has
a representative with a reduced, affinely independent support. By
Proposition~\ref{prop:s14-nine} and Section~\ref{sec:s14-reconstruction}
there are nine such supports up to the Weyl group; by
Lemma~\ref{lem:weyl-fock} and Lemma~\ref{lem:torus-rescaling} all choices of
nonzero coefficients on one support lie in one orbit. Hence there are at most
nine orbits, represented by the vectors of Table~\ref{tab:s14-orbits}. The
graph $\Gamma$ of Lemma~\ref{lem:dynkin-coefficients} consists of the edges
labelled $1$, so the lemma shows that these vectors are critical, with
$\norm\beta^2=\frac2C-\frac14$ for $C=1,2,3,4,4,5,6,7,8$; moving
$\beta=\sum_ap_a\eta_{I_a}$ to the positive chamber gives $\beta^+$. The nine
chamber values are distinct, so Proposition~\ref{prop:ness-uniqueness} shows
that the nine orbits are distinct. \qed

\subsection{Orbit dimensions and comparison with the literature}
\label{sec:popov-comparison-integrated}

The projective orbit dimensions in the last column of
Table~\ref{tab:s14-orbits} are computed by \eqref{eq:exact-orbit-rank}
(Appendix~\ref{app:orbit-dimensions}). The orbits of vectors are one dimension
larger, except in type IX. Indeed, if $\beta\ne0$, the critical-point equation
$\widehat\beta\psi=\norm\beta^2\psi$ shows that $\psi$ itself lies in the
tangent space $\mathfrak g\psi$, so rescaling $\psi$ does not leave its orbit.
If $\beta=0$, every $X\in\mathfrak g$ satisfies $\langle\psi,X\psi\rangle=0$,
so $\psi$ is orthogonal to $\mathfrak g\psi$. The resulting vector-orbit
dimensions $22,35,43,44,50,54,59,63,63$ agree with \cite[Table~1]{Gatti_dim13};
this is an independent check of the rank computations. In the notation of
Popov \cite{Popov_dim14RUS}, as quoted in
\cite{BhojaKrasnov2025}, the orbits II to IX are the cases b, c, d, e, f, g, h,
i. The affine orbits in the class IX form a one-parameter family separated by
the invariant of degree eight; after projectivization they form one orbit
\cite[Section~2.1]{Gatti_dim13}.

\begin{remark}[Non-reduced supports]
\label{rem:s14-nonreduced}
Root-distinct supports need not be reduced. In seven modes there are fifteen
Weyl classes of root-distinct supports, of sizes one to eight; all are
affinely independent and have positive Gram solutions, hence give critical
points. The six non-reduced classes give no new chamber values: their
critical points lie in the orbits IV, VI, VII, IX, VIII and IX. For example,
the uniform superposition of the eight codewords of the $[7,3,4]$ simplex
code,
\[
 \ket\varnothing+\ket{1234}+\ket{1256}+\ket{3456}
 +\ket{1357}+\ket{2457}+\ket{2367}+\ket{1467},
\]
has $\mu=0$ and therefore lies in the generic orbit IX, and dropping the last
word gives a critical point in the orbit VIII. This agrees with
Proposition~\ref{prop:ness-uniqueness}, and the same supports give the
degenerations used in Proposition~\ref{prop:s14-closure}.
\end{remark}

\section{Odd-dimensional spinor orbits}
\label{sec:odd-orbits}

Let $N=2n+1$ and
\[
 S_N=\bigwedge\nolimits^\bullet\C^n,
 \qquad G_N=\Spin(2n+1,\C).
\]
All occupation strings occur.  The long roots of $B_n$ change two
occupations and the short roots change one.  Hence two weights are
root-neighbouring if and only if their binary strings have Hamming distance
one or two.  A root-distinct support is therefore a binary code of minimum
distance at least three.

The $B_n$ Weyl group permits arbitrary sign changes and permutations.  In
occupation language it permits translating every word by a fixed binary
word and then permuting coordinates.  
We may therefore normalize one support word to $\varnothing$. For
$n\le5$ the remaining words are then determined below from their sizes
and the sizes of their pairwise intersections.

\subsection{Restriction from even-dimensional spin groups}
\label{sec:odd-exhaustiveness}

We bound the odd-dimensional orbit counts by restricting the
even-dimensional spin action. The critical representatives below attain
these bounds.

\begin{lemma}
\label{lem:spin7-quadric}
The complex spin representation of $\Spin(7,\C)$ has two projective orbits. Moreover, $\Spin(7,\C)$ is transitive on each nonzero level of its invariant quadratic form.
\end{lemma}

\begin{proof}
Use the $F_4$ roots
$\pm\varepsilon_i\pm\varepsilon_j$, $\pm\varepsilon_i$ and
$\frac12(\pm\varepsilon_1\pm\varepsilon_2\pm\varepsilon_3
\pm\varepsilon_4)$, with all signs in the last family, graded by
$2x_4$. Then $\mathfrak l_0=\mathfrak{so}_7\oplus\C$ and
$\mathfrak l_1$ is the eight-dimensional spin module. Its degree-two
component has dimension seven. The carrier argument of
Proposition~\ref{prop:root-distinct-normal-forms}, using
Lemma~\ref{lem:shallow-carriers}, supplies root-distinct representatives.
In three binary coordinates the only such supports, up to the Weyl group,
are $\{\varnothing\}$ and $\{\varnothing,123\}$.
Cartan rescaling removes their coefficient ratios.
The invariant symmetric bilinear form pairs opposite weights
\cite{Chevalley1954}; its quadratic form vanishes on $\ket\varnothing$
and is nonzero on $\ket\varnothing+\ket{123}$. The two orbits are
therefore distinct. On a fixed nonzero quadratic level a projective
identification has scalar $\pm1$, and the central element $-1$ removes
this sign.
\end{proof}

Let $G=\Spin(2m,\C)$ act on its vector module $V$ with quadratic form
$q$, and fix $v_0$ with $q(v_0)=1$. Its stabilizer is
$H=\Spin(2m-1,\C)$. The restriction of either half-spin module to $H$
is the odd spin module. For a fixed $G$-orbit of a spinor line $[\psi]$,
there is a bijection
\begin{equation}
 H\backslash G[\psi]\ \simeq\
 G_{[\psi]}\backslash Q,
 \qquad Q=\{v\in V:q(v)=1\}.
 \label{eq:odd-double-cosets}
\end{equation}
Writing $L=G_{[\psi]}$, the map is
$HgL\mapsto Lg^{-1}v_0$. Its inverse sends $Lv$, with
$v=g^{-1}v_0$, to $HgL$; both maps are independent of the chosen
representatives because $H=G_{v_0}$.
It suffices to exhibit subgroups of $G_{[\psi]}$ with few orbits on $Q$.
Inside $G$, each general linear group below is understood as the inverse
image of its standard subgroup of $\mathrm{SO}(V,q)$.

\begin{proposition}
\label{prop:odd-orbit-bound}
There are at most three projective spinor orbits for $\Spin(9,\C)$
and at most five for $\Spin(11,\C)$.
\end{proposition}

\begin{proof}
The even-dimensional classifications already proved give two spinor
orbits for $\Spin(10)$ and four for $\Spin(12)$. We compute their
contributions to \eqref{eq:odd-double-cosets}.

For a pure spinor, its line stabilizer is the parabolic preserving a
maximal isotropic subspace. In a decomposition $V=W\oplus W^*$,
write $v=(w,f)$ and $q(v)=2f(w)$. On $Q$, both $w$ and $f$ are nonzero.
If $(w',f')$ is another such pair, the element of $\mathrm{GL}(W)$ that
sends $w$ to $w'$ and a basis of $\ker f$ to a basis of $\ker f'$ maps
$(w,f)$ to $(w',f')$. Thus $\mathrm{GL}(W)$ is transitive on $Q$, and this orbit contributes one $H$-orbit.

For $\psi=\ket\varnothing+\ket{1234}$, let
$R=M(\psi)$, of dimension $m-4$, and write
$V=R\oplus U\oplus R^*$, where $\dim U=8$.
The line stabilizer contains the unipotent radical $N$ of the parabolic
preserving $R$, together with $\mathrm{GL}(R)$ and a $\Spin(7)$
acting on $U$ in its spin representation. Indeed, $R\wedge U$
and $\bigwedge^2R$ annihilate $\psi$, $\mathrm{GL}(R)$ scales it,
and the stabilizer of the nonisotropic half-spinor
$\ket\varnothing+\ket{1234}$ in $\Spin(U)$ is $\Spin(7)$;
its action on $U$ is the spin action by Cartan triality
\cite{Cartan1966,Chevalley1954}.

To verify the action of $N$, write $q(r,u,f)=2f(r)+q_U(u)$ and let
$b_U$ be the symmetric bilinear form with $b_U(u,u)=q_U(u)$.
For $A:U\to R$, define $A^*:R^*\to U$ by
$b_U(A^*f,u)=f(Au)$. If $B:R^*\to R$ is skew, the corresponding
element of $N$ acts by
\begin{equation}
 (r,u,f)\longmapsto
 \bigl(r+Au+(B-\tfrac12AA^*)f,\ u-A^*f,\ f\bigr).
 \label{eq:odd-unipotent-action}
\end{equation}
Substitution preserves $q$, and the differential has components in
$R\wedge U$ and $\bigwedge^2R$.
If $f\ne0$, take $r_0$ with $f(r_0)=1$ and set
$Av=b_U(u,v)r_0$, so that $A^*f=u$. After this eliminates $u$,
the equation $q=1$ gives $f(r)=1/2$. For any $s\in\ker f$, the
skew map $B\xi=\xi(r_0)s-\xi(s)r_0$ satisfies $Bf=s$.
These maps remove the remaining freedom in $r$.
The group $\mathrm{GL}(R)$ then makes all such pairs $(r,f)$ equivalent.
If $f=0$, then $q_U(u)=1$; choose $A$ with $Au=-r$, and apply
Lemma~\ref{lem:spin7-quadric} to $u$.
Thus this spinor orbit contributes at most two $H$-orbits.
This proves the bound $1+2=3$ for $\Spin(9)$.

For $\Spin(12)$ there are two more spinor orbits. The three-term orbit
has the representative
$\omega=\ket{12}+\ket{34}+\ket{56}$, a nondegenerate bivector on
$W=\C^6$, obtained from the listed support by translation by $34$ and Cartan
rescaling. Its line stabilizer contains
$\mathrm{Sp}(W,\omega)$ and the contractions
\begin{equation}
 N_\omega=\{\exp(\iota_B):B\in\bigwedge^2W^*,\
                         \iota_B\omega=0\}.
\end{equation}
They act on $V$ as shears $(w,f)\mapsto(w,f+B(w,\cdot))$,
up to a fixed nonzero normalization of $B$.
For fixed $w\ne0$, choose a symplectic basis
$e_1,f_1,e_2,f_2,e_3,f_3$ with $e_1=w$.
Given $\ell(w)=0$, the two-form $B_0=e_1^*\wedge\ell$ satisfies
$B_0(w,\cdot)=\ell$. Adding a multiple of $e_2^*\wedge f_2^*$
makes $\iota_B\omega=0$ without changing $B(w,\cdot)$.
The symplectic group is transitive on nonzero $w$, and these shears are
transitive on $f(w)=1/2$. Hence this orbit contributes one $H$-orbit.

Finally, $\ket\varnothing+\ket{123456}$ is fixed by
$\mathrm{SL}(W)$. This subgroup is transitive on the pairs
$f(w)=1/2$: first map the vector and covector by $\mathrm{GL}(W)$,
then correct the determinant on $\ker f$. Thus the last orbit also
contributes one. The bound for $\Spin(11)$ is $1+2+1+1=5$.
\end{proof}

It remains to construct two, three and five distinct critical orbits in
dimensions seven, nine and eleven, respectively.

\subsection{Pure spinors in dimensions one, three and five}
\label{sec:odd-pure-cases}

The odd pure-spinor variety is also a projectively embedded maximal
isotropic Grassmannian, hence is closed. For $n=0,1,2$, it has dimension
$n(n+1)/2=2^n-1=\dim\PP(S_N)$, and is therefore the whole projective
space. 
(The dimension $n(n+1)/2$ comes from the local description of a maximal
isotropic $n$-plane in $\C^{2n+1}$ by a skew $n\times n$ matrix and $n$
further coordinates.) Thus every spinor is pure and $G_N$ is transitive.
Correspondingly, any two distinct Fock strings have distance at most two,
so a root-distinct support is a singleton.
The vacuum gives the unique critical orbit.  Its momentum value and projective orbit
dimension are
\[
\begin{array}{c|ccc}
N&1&3&5\\ \hline
\beta^+&0&(\frac12)&((\frac12)^2)\\
\norm\beta^2&0&\frac14&\frac12\\
\dim G_N[\ket\varnothing]&0&1&3.
\end{array}
\]

\subsection{Dimension seven}
\label{sec:odd-seven}

For $n=3$, the only word at distance at least three from the vacuum is
$123$.  The unique nontrivial root-distinct support is therefore
\[
 \{\varnothing,123\},
 \qquad
 \frac12(\eta_\varnothing+\eta_{123})=0.
\]
It produces the critical representative
$\ket\varnothing+\ket{123}$.  Together with the pure orbit this gives two
distinct critical orbits, which exhaust the representation by
Lemma~\ref{lem:spin7-quadric}.

\subsection{Dimension nine}
\label{sec:odd-nine}

For $n=4$, a partner of the vacuum must have size three or four.  These
sizes are inequivalent under the stabilizer of the vacuum and give
\[
\begin{aligned}
 \{\varnothing,123\}:&
 &\beta^+&=\left(\frac12,0,0,0\right),\\
 \{\varnothing,1234\}:&
 &\beta^+&=0.
\end{aligned}
\]
Any two distinct words of size at least three in a four-element set have
mutual distance at most two.  No root-distinct triple exists.  Hence the
vacuum and these two equal-coefficient pairs give three critical
candidates.

\subsection{Dimension eleven}
\label{sec:odd-eleven}

For $n=5$, the three possible partners of the vacuum have sizes three,
four and five.  They give the pair supports
\[
 \{\varnothing,123\},\qquad
 \{\varnothing,1234\},\qquad
 \{\varnothing,12345\},
\]
with positive-chamber midpoint values
\[
 \left(\left(\frac12\right)^2,0^3\right),
 \qquad
 \left(\frac12,0^4\right),
 \qquad
 0,
\]
respectively.

There is one root-distinct triple type. A five-letter word is incompatible
with every other possible partner of the vacuum, and two four-subsets
have distance two. Thus a triple containing the vacuum must have a
three-letter word; normalize it to $A=123$.  A second word $B$ of size at least three
and distance at least three from $A$ must either be a three-set meeting
$A$ once or a four-set meeting $A$ twice.  These two descriptions are
equivalent under a translation and coordinate permutation, and both have
the distance multiset $\{3,3,4\}$.  We choose
\[
 \Sigma_{11}=\{\varnothing,123,345\}.
\]
The Gram matrix of its weights is
\[
 G_{\Sigma_{11}}
 =\frac14
 \begin{pmatrix}
 5&-1&-1\\
 -1&5&-3\\
 -1&-3&5
 \end{pmatrix}.
\]
Solving the bordered system gives
\[
 (p_\varnothing,p_{123},p_{345})
 =\left(\frac14,\frac38,\frac38\right).
\]
Thus a critical representative and its
positive-chamber value are
\begin{equation}
 \psi_{11,\triangle}
 =\sqrt2\,\ket\varnothing+\sqrt3\,\ket{123}
                  +\sqrt3\,\ket{345},
 \qquad
 \beta^+=\left(\frac14,
                    \left(\frac18\right)^4\right),
 \qquad
 \norm\beta^2=\frac18 .
 \label{eq:B5-triangle-critical}
\end{equation}
The coefficients have been multiplied by a common factor; their squared
ratios are $2:3:3$.

To bound larger supports, use the normalized triple
$\{\varnothing,123,145\}$, which is equivalent to the displayed one.
A further three-subset would have to meet both $123$ and $145$ in
at most one position. If it contains $1$, its other two positions are
forbidden; if it does not, three elements cannot occupy at most one
position in each of $\{2,3\}$ and $\{4,5\}$. Thus none exists.
A further four-subset must omit $1$ in order to meet both triples in
two positions, and hence is $2345$. This gives the unique four-word
extension
\[
 \{\varnothing,123,145,2345\};
\]
no fifth word is compatible with it. Its equally weighted barycentre is
zero. Its weights are affinely independent, so
Lemma~\ref{lem:torus-rescaling} makes all four nonzero coefficients equal.

The four-word support is again of four-to-two type, and the trick of
Lemma~\ref{lem:s14-four-to-two}(2) still works. Regard mode $1$ as one qubit,
and the pairs $(2,3)$ and $(4,5)$, each empty or doubly occupied, as two more. The four-term sum becomes
$\ket{000}+\ket{110}+\ket{101}+\ket{011}$. Put
\[
 R_1(\theta)=\exp\!\bigl(\theta(a_1^\dagger P-Pa_1)\bigr),
 \qquad P=(-1)^F.
\]
Its generator is $(a_1^\dagger+a_1)P$, a short-root combination,
and is anti-Hermitian. On the indicated subspace it acts as
$\bigl(\begin{smallmatrix}0&-1\\1&0\end{smallmatrix}\bigr)$ on the
first factor. The long-root rotations $R_{23}$ and $R_{45}$ from
Lemma~\ref{lem:s14-four-to-two} act on the other two factors; the three
generators commute. The tensor identity in that lemma gives
\begin{equation}
 R_1(\pi/4)R_{23}(\pi/4)R_{45}(\pi/4)
 (\ket\varnothing+\ket{123}+\ket{145}+\ket{2345})
 =\sqrt2(\ket\varnothing+\ket{12345}).
 \label{eq:B5-four-to-two}
\end{equation}
The four-word support therefore gives the already listed zero-momentum
orbit.

\subsection{Odd-dimensional orbit representatives}
\label{sec:odd-summary}

\begin{table}[t]
\centering
\renewcommand{\arraystretch}{1.15}
\small
\begin{tabular}{c l c c c}
\toprule
$N$ & critical representative $\psi$ & $\beta^+$ &
$\norm\beta^2$ & $\dim G_N[\psi]$\\
\midrule
1 & $\ket\varnothing$ & $0$ & $0$ & 0\\
3 & $\ket\varnothing$ & $(\frac12)$ & $\frac14$ & 1\\
5 & $\ket\varnothing$ & $((\frac12)^2)$ & $\frac12$ & 3\\
\midrule
7 & $\ket\varnothing$ & $((\frac12)^3)$ & $\frac34$ & 6\\
7 & $\ket\varnothing+\ket{123}$ & $(0^3)$ & $0$ & 7\\
\midrule
9 & $\ket\varnothing$ & $((\frac12)^4)$ & $1$ & 10\\
9 & $\ket\varnothing+\ket{123}$ & $(\frac12,0^3)$ & $\frac14$ & 14\\
9 & $\ket\varnothing+\ket{1234}$ & $(0^4)$ & $0$ & 15\\
\midrule
11 & $\ket\varnothing$ & $((\frac12)^5)$ & $\frac54$ & 15\\
11 & $\ket\varnothing+\ket{123}$ & $((\frac12)^2,0^3)$ & $\frac12$ & 22\\
11 & $\ket\varnothing+\ket{1234}$ & $(\frac12,0^4)$ & $\frac14$ & 24\\
11 & $\sqrt2\ket\varnothing+\sqrt3\ket{123}+\sqrt3\ket{345}$
   & $(\frac14,(\frac18)^4)$ & $\frac18$ & 30\\
11 & $\ket\varnothing+\ket{12345}$ & $(0^5)$ & $0$ & 31\\
\bottomrule
\end{tabular}
\caption{Critical representatives, chamber values and projective orbit
dimensions for the odd-dimensional spin modules with
finitely many orbits. The list agrees with Igusa's classification
\cite{Igusa1970uptodim12}.}
\label{tab:finite-odd}
\end{table}

The support analysis produces $1,1,1,2,3,5$ candidates in dimensions
$N=1,3,5,7,9,11$.  Their positive-chamber momentum values are distinct
within each dimension, so Ness's theorem separates their complex orbits.
The pure cases, Lemma~\ref{lem:spin7-quadric} and
Proposition~\ref{prop:odd-orbit-bound} give the same numbers as upper
bounds for all orbits. Hence the table is exhaustive. In particular,
every orbit has a critical representative with root-distinct support.
The completed list agrees with Igusa \cite{Igusa1970uptodim12}.

\section{Dimensions with infinitely many projective orbits}
\label{sec:finite-boundary}

In dimension thirteen the invariant algebra is freely generated by
homogeneous polynomials $P_4,Q_8$ of degrees four and eight
\cite[Theorem~1]{Gatti_dim13}. Their algebraic independence
implies that the degree-zero rational invariant
\[
 \mathcal J([\psi])=\frac{Q_8(\psi)}{P_4(\psi)^2},
 \qquad P_4(\psi)\ne0,
\]
is nonconstant: otherwise $Q_8-cP_4^2=0$ would be a polynomial relation.
It is unchanged by both the spin action and scalar multiplication, so
there are infinitely many projective orbits.

In our Fock convention the Cartan subspace of
\cite[Proposition~10]{Gatti_dim13} is spanned by
\[
 \psi_{\alpha,b}=\alpha(\ket\varnothing+\ket{123456})
                 +b(\ket{123}+\ket{456}).
\]
In these coordinates the basic invariants may be normalized to restrict to
$P_4=\alpha^4+b^4$ and $Q_8=\alpha^4b^4$
\cite[Sections~3.2--3.3]{Gatti_dim13}; the identification of their basis
with our Fock basis may change signs, which does not affect $\mathcal J$. On the line $(\alpha,b)=(1,t)$,
\[
 \mathcal J=\frac{t^4}{(1+t^4)^2},\qquad
 \frac{d\mathcal J}{dt}=\frac{4t^3(1-t^4)}{(1+t^4)^3}>0
 \quad(0<t<1).
\]
So the points of this interval are pairwise inequivalent. The four support words have pairwise distances three or six,
and occur in antipodal pairs with equal moduli, so
$\mu([\psi_{\alpha,b}])=0$ whenever $(\alpha,b)\ne(0,0)$.
Once continuous families appear, even the full momentum value no longer
separates the orbits: all these spinors have $\mu=0$.

The same conclusion can be reached without the invariants of Kac and
Vinberg. All we need is the uniqueness theorem and a $K$-invariant function
that is not constant along the family.

\begin{proposition}
\label{prop:spin13-moment}
For $t>0$ let $\psi_t=\ket\varnothing+\ket{123456}+t(\ket{123}+\ket{456})\in
S_{13}$. Then $\mu([\psi_t])=0$, and the function
\begin{equation}
 F_3([\psi])=\frac1{\norm\psi^4}\sum_{1\le r<s<u\le13}
 \bigl\lvert\langle\psi,\gamma_r\gamma_s\gamma_u\psi\rangle\bigr\rvert^2
 \label{eq:clifford-moment}
\end{equation}
is $K$-invariant and satisfies $F_3([\psi_t])=16t^2/(1+t^2)^2$. Consequently
the points $[\psi_t]$, $0<t<1$, lie in pairwise distinct
$\Spin(13,\C)$-orbits.
\end{proposition}

\begin{proof}
The support words have pairwise distances three or six and come in two
antipodal pairs with equal moduli, so $\mu([\psi_t])=0$ as shown above. For
$k\in K$ with rotation $R_k$ we have $k\gamma(v)k^{-1}=\gamma(R_kv)$, so the
numbers $\langle\psi,\gamma_r\gamma_s\gamma_u\psi\rangle$, $r<s<u$, are the
components of an element of $\bigwedge^3\C^{13}$ on which $K$ acts through
$SO(13)$; the sum of their squared moduli is therefore $K$-invariant, and
\eqref{eq:clifford-moment} is also invariant under scaling. A product of
three distinct generators changes the occupations of at most three modes, and
of exactly three only if it does not contain $\gamma_{13}=P$ and involves
three different modes. The support words have mutual distances three and six,
and the diagonal contributions of a word $I$ and of its complement cancel,
because the two have equal moduli, equal parity and opposite occupations.
Hence only the four pairs of support words at Hamming distance three
contribute, and a direct count gives
$\sum\lvert\langle\psi_t,\gamma_r\gamma_s\gamma_u\psi_t\rangle\rvert^2=64t^2$
and $\norm{\psi_t}^4=4(1+t^2)^2$ (Appendix~\ref{app:spin13-computation}).
Each $[\psi_t]$ is critical with $\mu=0$. If $[\psi_s]$ and $[\psi_t]$ were
$G$-equivalent, Proposition~\ref{prop:ness-uniqueness} would make them
$K$-equivalent, and $F_3$ would take the same value at both. But
$16t^2/(1+t^2)^2$ is strictly increasing on $(0,1)$.
\end{proof}

For $N\ge15$ a dimension count suffices: $\PP(S)$ has larger dimension
than $G$, and an orbit cannot be larger than the group, so finitely many
orbits cannot fill $\PP(S)$. The numbers are as follows.

For odd dimension $N=2n+1$,
\[
 \dim\PP(S_N)=2^n-1,
 \qquad
 \dim\Spin(N,\C)=n(2n+1).
\]
At $n=7$, or $N=15$, these numbers are $127$ and $105$.  Moreover,
if
$d_{\mathrm{odd}}(n)=2^n-1-n(2n+1)$, then
\[
 d_{\mathrm{odd}}(n+1)-d_{\mathrm{odd}}(n)
 =2^n-(4n+3)>0
 \qquad(n\geq7).
\]
Thus the dimension obstruction persists in every odd dimension from fifteen
onward.

For even dimension $N=2n$, a half-spin module has dimension $2^{n-1}$,
and therefore
\[
 \dim\PP(S_N^+)=2^{n-1}-1,
 \qquad
 \dim\Spin(N,\C)=n(2n-1).
\]
At $n=8$, or $N=16$, these numbers are $127$ and $120$.  With
$d_{\mathrm{even}}(n)=2^{n-1}-1-n(2n-1)$, one has
\[
 d_{\mathrm{even}}(n+1)-d_{\mathrm{even}}(n)
 =2^{n-1}-(4n+1)>0
 \qquad(n\geq8).
\]
Hence all even dimensions from sixteen onward also have infinitely many
projective orbits. Consequently the finite projective cases are
\[
 N=1,2,\ldots,12,\quad\text{and}\quad N=14,
\]
with the appropriate irreducible spin module in odd dimension and either
half-spin module in even dimension.

\begin{proof}[Proof of Theorems~\ref{thm:finite-orbit-main}
and~\ref{thm:finite-range}]
Parts (1) and (2) of Theorem~\ref{thm:finite-orbit-main} are proved in
Section~\ref{sec:even-low-summary} for even $N\le12$, in
Theorem~\ref{thm:s14} for $N=14$, and in Section~\ref{sec:odd-summary} for odd
$N$; negative chirality follows from the chirality remark in the
introduction. For part (3), every orbit contains a listed critical point;
Proposition~\ref{prop:ness-uniqueness} shows that the critical points in one
orbit form a single $K$-orbit, and \eqref{eq:minimum-on-orbit-closure} shows
that $\norm\mu^2$ attains its minimum over the closure there. 
Theorem~\ref{thm:finite-range} follows from
Proposition~\ref{prop:spin13-moment} (or from the invariant $\mathcal J$) for
$N=13$, and from the dimension count above for $N\ge15$.
\end{proof}


\section{Momentum norm and orbit degenerations}
\label{sec:momentum-orbit-geometry}

In this section we ask what happens to the momentum map when an orbit
degenerates into a smaller one, and we use the answer to find the closure
order in dimension fourteen. Let $G=K^\C$ be the complex spin group acting on
$S$, and put $X=\PP(S)$ and $f=\norm\mu^2$. Orbit closures are algebraic
objects, while limits and minima of $f$ are analytic ones, so we need both
the Zariski topology and the usual topology of $X$. The following lemma says
that for orbits the two give the same closures. All dimensions in this section
are complex dimensions.

\begin{lemma}[Orbit closures]
\label{lem:orbit-closures}
Let a connected complex algebraic group $G$ act on a variety $X$.
For an orbit $\mathcal O$, its Zariski closure $Y$ is irreducible,
$\mathcal O$ is open in $Y$, and $Y\setminus\mathcal O$ is a closed
$G$-invariant subvariety of smaller dimension. The closure of
$\mathcal O$ in the usual topology is also $Y$.
\end{lemma}

\begin{proof}
The first three assertions are the closed orbit lemma
\cite{Borel1991}. For the last one, $Y$ is irreducible of
dimension $\dim\mathcal O$, so the lower-dimensional closed subset
$Y\setminus\mathcal O$ contains no nonempty open subset of $Y$ in the usual
topology. Hence $\mathcal O$ is dense in $Y$ in the usual topology, and $Y$,
being defined by polynomial equations, is closed in it.
\end{proof}

In particular, an orbit is a locally closed variety; it need not be
closed in $X$. For a projective orbit define
\begin{equation}
 m(\mathcal O)=\inf_{x\in\mathcal O}f(x).
\end{equation}
Its closure $Y$ is compact in the usual topology. By the lemma and
continuity of $f$,
\begin{equation}
 m(\mathcal O)=\min_{y\in\overline{\mathcal O}}f(y).
 \label{eq:minimum-on-orbit-closure}
\end{equation}
Indeed, each $y\in Y$ is a limit of points of $\mathcal O$, so
$f(y)\ge m(\mathcal O)$; the reverse inequality for the minimum
follows from $\mathcal O\subset Y$.

We need two results of Kirwan--Ness theory. First, if $x\in\mathcal O$
is critical, then $f(x)\le f(gx)$ for all $g\in G$, and consequently
$f(x)=m(\mathcal O)$ \cite[Theorem~6.2]{Ness1984}; see also
\cite[Corollary~6.2]{GeorgoulasRobbinSalamon2021}.
Second, the points at which $f$ attains its minimum on
$\overline{\mathcal O}$ form a single $K$-orbit
\cite[Theorem~6.5]{GeorgoulasRobbinSalamon2021}.
The classification above supplies a critical point in every orbit in
the finite spinor range, so in these cases the minimum in
\eqref{eq:minimum-on-orbit-closure} is attained in $\mathcal O$ itself.
(Lemma~\ref{lem:orbit-closures} requires $G$ to be connected, which
holds for $N\ge2$; for $N=1$ the projective space is a point.)

\begin{proposition}[Monotonicity under degeneration]
\label{prop:momentum-degeneration}
If $\mathcal O'\subset\overline{\mathcal O}$ and
$\mathcal O'\ne\mathcal O$, then
\begin{equation}
 \dim\mathcal O'<\dim\mathcal O,\qquad
 m(\mathcal O')\ge m(\mathcal O).
 \label{eq:degeneration-dimension-momentum}
\end{equation}
If both orbits contain critical points, the second inequality is strict.
\end{proposition}

\begin{proof}
Distinct orbits are disjoint, so $\mathcal O'$ lies in
$\overline{\mathcal O}\setminus\mathcal O$. The dimension inequality
follows from Lemma~\ref{lem:orbit-closures}. Moreover,
$\overline{\mathcal O'}\subset\overline{\mathcal O}$, and
\eqref{eq:minimum-on-orbit-closure} gives
\begin{equation}
 m(\mathcal O')=\min_{\overline{\mathcal O'}}f
       \ge\min_{\overline{\mathcal O}}f=m(\mathcal O).
\end{equation}
Suppose equality holds and choose critical points $x\in\mathcal O$
and $y\in\mathcal O'$. Both minimize $f$ on
$\overline{\mathcal O}$. The second uniqueness theorem gives
$y\in Kx\subset\mathcal O$, a contradiction.
\end{proof}

To find closure relations we need a way to produce degenerations. The
simplest one is to switch off some of the constituents with the torus.

\begin{lemma}[Torus degenerations]
\label{lem:torus-degeneration}
Let $\psi=\sum_{\eta\in\Sigma}c_\eta\ket\eta$ have an affinely independent
support $\Sigma$, and let $\varnothing\ne\Sigma'\subset\Sigma$. Then
$\bigl[\sum_{\eta\in\Sigma'}c_\eta\ket\eta\bigr]$ lies in the closure of
$G[\psi]$.
\end{lemma}

\begin{proof}
Fix $\eta_0\in\Sigma'$. Since the differences $\eta-\eta_0$,
$\eta\in\Sigma\setminus\{\eta_0\}$, are linearly independent, there is
$x\in\R^n$ with $(\eta-\eta_0,x)=0$ for $\eta\in\Sigma'$ and $(\eta-\eta_0,x)=-1$
for $\eta\in\Sigma\setminus\Sigma'$. The operator
$H=\sum_jx_j\widehat H_j$ lies in $\rho_*(\mathfrak g)$ and acts on
$\ket\eta$ by $(\eta,x)$, so
\[
 e^{sH}\psi=e^{s(\eta_0,x)}\Bigl(\sum_{\eta\in\Sigma'}c_\eta\ket\eta
 +e^{-s}\sum_{\eta\in\Sigma\setminus\Sigma'}c_\eta\ket\eta\Bigr),
\]
whose class tends to $\bigl[\sum_{\eta\in\Sigma'}c_\eta\ket\eta\bigr]$ as
$s\to\infty$. Lemma~\ref{lem:orbit-closures} identifies the closure in the
usual topology with the Zariski closure.
\end{proof}

\begin{proposition}
\label{prop:s14-closure}
For the nine nonzero orbits in dimension fourteen, numbered as in
Table~\ref{tab:s14-orbits},
\[
 \mathrm I\subset\overline{\mathrm{II}},\quad
 \mathrm{II}\subset\overline{\mathrm{III}},\quad
 \mathrm{III}\subset\overline{\mathrm{IV}}\cap\overline{\mathrm V},\quad
 \mathrm{IV}\cup\mathrm V\subset\overline{\mathrm{VI}},\quad
 \mathrm{VI}\subset\overline{\mathrm{VII}},\quad
 \mathrm{VII}\subset\overline{\mathrm{VIII}},\quad
 \mathrm{VIII}\subset\overline{\mathrm{IX}},
\]
and IV and V are incomparable. These are all closure relations
(Figure~\ref{fig:s14-closure}).
\end{proposition}

\begin{proof}
The idea is to find a single nested family of supports that passes through
almost all the types; the $[7,3,4]$ simplex code provides one. Its eight words
\[
 \varnothing,\ 1234,\ 1256,\ 3456,\ 1357,\ 2457,\ 2367,\ 1467
\]
differ pairwise in exactly four positions. Let
$\Sigma_k$ consist of the first $k$ of them. By
Lemma~\ref{lem:dynkin-coefficients} (with $\Gamma$ empty) the uniform
superposition over $\Sigma_k$ is critical, with $\norm\beta^2=\frac2k-\frac14$,
that is $\frac74,\frac34,\frac5{12},\frac14,\frac3{20},\frac1{12},\frac1{28},0$
for $k=1,\ldots,8$. By Theorem~\ref{thm:s14} and
Proposition~\ref{prop:ness-uniqueness}, a critical point lies in the orbit
whose listed chamber value equals its own. The values of $\norm\beta^2$ identify the orbits
I, II, III, VI, VII, VIII, IX for $k\ne4$, and for $k=4$ the chamber value
$(\frac12,0^6)$ identifies IV. Lemma~\ref{lem:torus-degeneration} applied to
$\Sigma_{k-1}\subset\Sigma_k$ gives the chain
$\mathrm I\subset\overline{\mathrm{II}}$, \dots,
$\mathrm{VIII}\subset\overline{\mathrm{IX}}$ with IV in place of V. The support
$\{\varnothing,1234,1256,1357\}$ of type V contains $\Sigma_3$ and is
contained in $\Sigma_5$, which gives
$\mathrm{III}\subset\overline{\mathrm V}$ and
$\mathrm V\subset\overline{\mathrm{VI}}$. The orbits IV and V have the same
momentum minimum $\frac14$ and both contain critical points, so by
Proposition~\ref{prop:momentum-degeneration} neither lies in the closure of
the other. Every other pair is comparable by transitivity, and by
Proposition~\ref{prop:momentum-degeneration} a closure relation can only go
from the lower-dimensional orbit to the higher-dimensional one.
\end{proof}

In Popov's notation our orbits $\mathrm{IV}$ and $\mathrm{V}$ are
$d$ and $e$, respectively. Figure~1 of his 1977 announcement
\cite[Theorem~1, p.~199]{Popov_dim14RUS} places $d$ in the closure of $e$. Such an affine degeneration to a nonzero vector would also give $\mathrm{IV}\subset\overline{\mathrm{V}}$ after projectivization. Both orbits contain critical points with momentum minimum $1/4$, so Proposition~\ref{prop:momentum-degeneration} excludes this inclusion. Instead, $\mathrm{III}$ lies in the closure of each of the two orbits, and both lie in $\overline{\mathrm{VI}}$.

\SpinFourteenClosureFigure

Along the arrows, from I to IX, the momentum minimum decreases strictly,
as Proposition~\ref{prop:momentum-degeneration} requires; in other words,
each degeneration leads to an orbit whose critical states are closer to
Gaussian states. The diagram is a fermionic analogue of the degeneration
diagrams of SLOCC classes of qubits: IX, with $\mu=0$, plays the role of
the GHZ class, and VIII, the open orbit in the null cone of the invariant
of degree eight, that of the W class.


\subsection{The open orbit and zero momentum}

A finite orbit decomposition of the irreducible variety $X$ has a
unique open orbit $\mathcal O_{\mathrm{op}}$. To see this, write $X$
as the finite union of the orbit closures. Irreducibility forces one
of these closures to equal $X$, and Lemma~\ref{lem:orbit-closures}
makes that orbit open. Two nonempty open subsets of an irreducible
variety intersect, which proves uniqueness. Equation~\eqref{eq:minimum-on-orbit-closure}
now gives
\begin{equation}
 m(\mathcal O_{\mathrm{op}})=\min_X f.
\end{equation}
Since $\mathcal O_{\mathrm{op}}$ contains a critical point, it attains
this minimum. It therefore meets $\mu^{-1}(0)$ if and only if
$\mu^{-1}(0)$ is nonempty. In the finite spinor range this occurs for
$N=7,8,9,11,12,14$, and trivially for $N=1$. The open orbit for
$N=10$ instead has $\beta^+=(\tfrac12,0^4)$ and $m=1/4$;
the single orbits for $N=2,3,4,5,6$ also have positive minima.

Closedness of a vector orbit in $S$ is a different condition from
closedness of its projective orbit in $\PP(S)$. For a critical spinor
with $\mu([\psi])=\beta\ne0$, the critical-point equation gives
\begin{equation}
 \exp(-t\widehat\beta)\psi
       =e^{-t\norm\beta^2}\psi\longrightarrow0
       \qquad(t\longrightarrow+\infty).
\end{equation}
Here $\widehat\beta\in i\rho_*(\mathfrak k)\subset\rho_*(\mathfrak g)$,
so $\exp(-t\widehat\beta)\in\rho(G)$.
Thus $0\in\overline{G\psi}\subset S$. The projective class along
this curve is constant, so this limit produces no projective boundary
orbit. If instead $\mu([\psi])=0$, the Kempf--Ness theorem implies
that $G\psi$ is closed in $S$ and disjoint from $0$
\cite{KempfNess1979}.

For example, in $S_8^+$ let $q$ be the invariant quadratic form.
The projective orbit $\{q\ne0\}$ is open and has the pure-spinor
quadric $\{q=0\}$ as its boundary. If $q(\psi)\ne0$, the affine orbit
of $\psi$ is the closed level set $\{v:q(v)=q(\psi)\}$, whereas the affine
orbit of a pure spinor has $0$ in its closure. The projective degeneration
\begin{equation}
 [\vac+t\ket{1234}]\longrightarrow[\vac]\qquad(t\longrightarrow0)
\end{equation}
runs from the open orbit, for $t\ne0$, to the closed pure orbit.
Their dimensions are $7$ and $6$, and their momentum minima are $0$
and $1$, respectively.

\section{Conclusion}
\label{sec:conclusion}

We have revisited the classification of spinors with finitely many
projective orbits from the point of view of the momentum map. In every
orbit we found the critical points of $\norm\mu^2$, which form a single
$K$-orbit and, in even dimension, maximize the associated generalized
entanglement, and we computed their momentum value and the orbit
dimension.
The representatives are built from a few Fock states, chosen
according to Vinberg's carrier algebras, and in dimensions ten to fourteen
their coefficients can be read off from the Dynkin diagram of the carrier.
The momentum values separate the orbits, and the minimum of $\norm\mu^2$
grows strictly at every degeneration. In dimension fourteen this observation,
together with torus degenerations inside the simplex code, gives the complete
closure order.

Figure~\ref{fig:all-dimensions} shows the resulting critical values for all
dimensions at once.
\AllDimensionsFigure

Several questions remain open. In dimension thirteen Kac and Vinberg describe
twenty types of orbits, including continuous families \cite{Gatti_dim13}. The closed orbits of vectors are exactly those that
meet $\mu^{-1}(0)$, and one could follow the critical sets of
$\norm\mu^2$ along these families.
In dimension sixteen,
$\Spin(16,\C)$ acting on $S^+_{16}$ is the $\theta$-representation of the
$\mathbb Z_2$-grading $\mathfrak e_8=\mathfrak{so}_{16}\oplus S^+_{16}$,
whose orbits were classified by Antonyan and \`Elashvili
\cite{AntonyanElashvili1982}; the next step would be a momentum-map
description of its null cone, with the closure order computed by the
methods of \cite{deGraaf2011}. Finally,
real forms $\Spin(p,q)$, mixed states and the corresponding entanglement
polytopes \cite{WalterEtAl2013}, and the role of $\norm\mu^2$ as a measure of
non-Gaussianity, deserve further study.

\section*{Acknowledgements}
This work was initiated through discussions with Paweł Nurowski on spinors and entanglement. The authors thank him for these inspiring discussions.

The research was partially funded from the Norwegian Financial Mechanism 2014–2021 with project registration number 2019/34/H/ST1/00636.

\appendix

\section{Carrier algebras and coordinate supports}
\label{app:finite-calculations}

We prove Proposition~\ref{prop:root-distinct-normal-forms} using Vinberg's
carrier-algebra theorem. The proof reduces to the finite root-system
calculation in Lemma~\ref{lem:finite-grading-calculation}.
The final subsection gives the explicit occupation-column formulas used
in Section~\ref{sec:s14-reconstruction}.

\subsection{Carrier algebras and root-distinct normal forms}
\label{app:carriers}

Here we give the argument behind
Proposition~\ref{prop:root-distinct-normal-forms}, which
Section~\ref{sec:root-distinct-exhaustiveness} only sketches.

We use Vinberg's theory of carrier algebras \cite{Vinberg1979}; see also
\cite[Section~1]{Gatti_dim13}, \cite{deGraaf2011}, \cite[Theorem~25]{Manivel2013}
and \cite{DietrichGlobkeOriglia2022}. Let
$\mathfrak l=\bigoplus_{j\in\mathbb Z}\mathfrak l_j$ be a $\mathbb Z$-graded
complex semisimple Lie algebra and $L_0$ the connected subgroup of its
adjoint group with Lie algebra $\mathfrak l_0$. Every $e\in\mathfrak l_1$ is
nilpotent, because $\operatorname{ad}e$ raises the degree.

\begin{definition}
\label{def:carrier}
A graded semisimple subalgebra
$\mathfrak c=\bigoplus_j\mathfrak c_j$, $\mathfrak c_j=\mathfrak c\cap\mathfrak l_j$, of
$\mathfrak l$ is
\emph{regular} if it is normalized by a Cartan subalgebra of $\mathfrak l_0$,
\emph{locally flat} if $\dim\mathfrak c_0=\dim\mathfrak c_1$, and
\emph{complete} if it is not a proper graded subalgebra of a regular graded
semisimple subalgebra of $\mathfrak l$ of the same rank. An element
$e\in\mathfrak c_1$ is in \emph{general position} if
$[\mathfrak c_0,e]=\mathfrak c_1$.
\end{definition}

\begin{theorem}[Vinberg]
\label{thm:vinberg-carrier}
Every nonzero $e\in\mathfrak l_1$ is $L_0$-conjugate to an element in general
position in $\mathfrak c_1$, for some regular, locally flat and complete
graded semisimple subalgebra $\mathfrak c\subset\mathfrak l$ (a \emph{carrier
algebra} of $e$).
\end{theorem}
\begin{lemma}[Gradings of depth at most two]
\label{lem:shallow-carriers}
Suppose $\mathfrak l$ has type $E_6$, $E_7$, $E_8$ or $F_4$ and
$\mathfrak l_j=0$ for $|j|>2$. Every carrier algebra in
$\mathfrak l$ is a direct sum of principally graded factors of type
$A_1$ and $A_2$. Consequently its degree-one component is the sum of its
simple-root spaces.
\end{lemma}

\begin{proof}
The grading derivation preserves every simple ideal
$\mathfrak c^{(s)}$ of $\mathfrak c$. Write $e=\sum_s e_s$.
Genericity implies
$[\mathfrak c^{(s)}_0,e_s]=\mathfrak c^{(s)}_1$, hence
$\dim\mathfrak c^{(s)}_0\ge\dim\mathfrak c^{(s)}_1$.
Since the sum of these differences is zero, each factor is locally flat.
It also inherits depth at most two. Choose its simple roots with
nonnegative degrees. If $r$ is its rank,
$\Phi^+$ its positive roots, and $d=(d_1,\ldots,d_r)$ its simple-root
degrees, local flatness is the finite condition
\begin{equation}
 r+2\#\{a\in\Phi^+:a\cdot d=0\}
   =\#\{a\in\Phi^+:a\cdot d=1\},
 \qquad 0<\max_{a\in\Phi^+}a\cdot d\le2.
 \label{eq:shallow-flatness}
\end{equation}
Here $a$ denotes the integral simple-root coordinates. Necessarily
$d_i\in\{0,1,2\}$. For a regular subalgebra of an $E$-type algebra,
the factors have type $A_r$ ($r\le8$), $D_r$ ($4\le r\le8$),
or $E_6,E_7,E_8$. In $F_4$ one must also allow $B_r,C_r$
($2\le r\le4$) and $F_4$; $G_2$ is excluded by the ratio of root
lengths. Lemma~\ref{lem:finite-grading-calculation} in
Appendix~\ref{app:shallow-enumeration} solves
\eqref{eq:shallow-flatness} for these types: the only possibilities
are $A_1$ with $d=(1)$ and $A_2$ with $d=(1,1)$.
\end{proof}

The gradings we need for the half-spin modules are
\begin{equation}
\begin{array}{c|ccc}
\mathfrak l&\mathfrak l_0&\mathfrak l_1&\dim\mathfrak l_2\\\hline
\mathfrak e_6&\mathfrak{so}_{10}\oplus\C&S_{10}^+&0\\
\mathfrak e_7&\mathfrak{so}_{12}\oplus\C&S_{12}^+&1\\
\mathfrak e_8&\mathfrak{so}_{14}\oplus\C&S_{14}^+&14.
\end{array}
\label{eq:exceptional-spin-gradings}
\end{equation}
The $E_8$ realization also appears in \cite{AbuafManivel2021}.
We check all three identifications directly. In orthonormal coordinates
$\varepsilon_1,\ldots,\varepsilon_8$, the $E_8$ roots are
\begin{equation}
 \{\pm\varepsilon_i\pm\varepsilon_j:i<j\}
 \ \cup\
 \left\{\tfrac12\sum_{i=1}^8\sigma_i\varepsilon_i:
 \sigma_i=\pm1,\ \prod_i\sigma_i=1\right\}.
 \label{eq:ambient-E8-roots}
\end{equation}
Take the roots satisfying $x_{n+1}=\cdots=x_8$, for $n=5,6,7$,
and give a root degree $2x_8$. The degree-zero roots are $D_n$;
the degree-one roots restrict to the positive half-spin weights on the
first $n$ coordinates. This proves \eqref{eq:exceptional-spin-gradings}.
The central $\C^*$ acts on $\mathfrak l_1$ by scalars, so the nonzero
$L_0$-orbits in $\mathfrak l_1$ correspond exactly to the projective spinor
orbits.

\begin{proof}[Proof of Proposition~\ref{prop:root-distinct-normal-forms}]
Choose a carrier algebra and apply Lemma~\ref{lem:shallow-carriers}.
Its degree-one space is spanned by simple-root vectors
$E_{\rho_1},\ldots,E_{\rho_k}$, and genericity means that all their
coefficients in $e$ are nonzero. Distinct simple roots cannot differ by
an ambient root: otherwise
$[E_{\rho_a},E_{-\rho_b}]\ne0$ would put that root space in the carrier,
contradicting the choice of simple roots. Since $\rho_a-\rho_b$ has degree zero, the
restricted spin weights are root-distinct. The $\rho_a$ are linearly
independent. Write $\rho_a=(\eta_a;u)$ in the coordinates of
\eqref{eq:ambient-E8-roots}, where the final coordinates $u$ are common
to all degree-one roots. An affine relation
$\sum_a t_a\eta_a=0$, $\sum_a t_a=0$ would give
$\sum_a t_a\rho_a=0$, hence $t_a=0$ for every $a$.

For $N=14$, suppose four of these words have a common annihilator.
The elementary six-mode argument in
Lemma~\ref{lem:s14-four-to-two} puts them in the form
$\varnothing,1234,1256,3456$, with mode seven empty. Their lifted
$E_8$ roots satisfy
\begin{equation}
 \tfrac12(\rho_1+\rho_2+\rho_3+\rho_4)
   =\varepsilon_7+\varepsilon_8.
\end{equation}
Put $\mathcal R=\operatorname{span}_{\mathbb R}\{\rho_1,\ldots,\rho_k\}$.
The set $\Phi(\mathfrak e_8)\cap\mathcal R$ is a closed root subsystem:
it is stable under root reflections and under addition whenever the
sum is an ambient root. Its root spaces and coroots generate a graded
semisimple algebra containing $\mathfrak c$ and of the same rank.
The displayed root is not a root of $\mathfrak c$, since its
coordinates in the simple-root basis of $\mathfrak c$ include $1/2$.
The containment is therefore proper, contradicting completeness.
\end{proof}

\begin{remark}
\label{rem:carrier-dynkin}
The lifts $\rho_a=(\eta_{I_a};\frac12,\ldots,\frac12)$ of the support weights
used in Lemma~\ref{lem:dynkin-coefficients} are roots of degree one in
\eqref{eq:exceptional-spin-gradings}, and by \eqref{eq:shifted-gram} their Gram
matrix is $2I-A_\Gamma$, the Cartan matrix of $aA_1+bA_2$, where $a$ is the
number of isolated vertices of $\Gamma$ and $b$ the number of edges. For a
support that comes from a carrier algebra, the $\rho_a$ are the simple roots of
the carrier, so $\Gamma$ is its Dynkin diagram.
\end{remark}

\subsection{The finite grading calculation}
\label{app:shallow-enumeration}

Table~\ref{tab:shallow-gradings} lists the Dynkin types of the possible
simple factors of a regular carrier algebra, in the usual notation
$A_r=\mathfrak{sl}_{r+1}$, $B_r=\mathfrak{so}_{2r+1}$,
$C_r=\mathfrak{sp}_{2r}$, $D_r=\mathfrak{so}_{2r}$.

\begin{lemma}[Finite grading calculation]
\label{lem:finite-grading-calculation}
For each root system in Table~\ref{tab:shallow-gradings}, every
nontrivial integral grading of depth at most two satisfies
$\dim\mathfrak c_0\ge\dim\mathfrak c_1$.
Equality holds only for $A_1$ and $A_2$, with all simple roots of
degree one, after a Weyl transformation.
\end{lemma}

\begin{proof}
Choose simple roots with nonnegative degrees
$d=(d_1,\ldots,d_r)$. The depth bound forces
$d_i\in\{0,1,2\}$, and a positive root with simple-root coordinates
$a$ has degree $a\cdot d$. Thus it suffices to test
\begin{equation}
 d\in\{0,1,2\}^r,\qquad
 0<\max_{a\in\Phi^+}a\cdot d\le2.
\end{equation}
For each such tuple,
\begin{equation}
 D(d):=\dim\mathfrak c_0-\dim\mathfrak c_1
 =r+2\#\{a\in\Phi^+:a\cdot d=0\}
     -\#\{a\in\Phi^+:a\cdot d=1\}.
\end{equation}
The first term is the Cartan dimension; degree-zero roots occur in
opposite pairs, while every degree-one root is positive.

To generate $\Phi^+$, take the Cartan matrix
$C_{ij}=2(\alpha_i,\alpha_j)/(\alpha_j,\alpha_j)$.
Start with $e_1,\ldots,e_r\in\mathbb Z^r$ and repeatedly adjoin
\begin{equation}
 s_i(a)=a-\left(\sum_j a_jC_{ji}\right)e_i
 \qquad(1\le i\le r).
\end{equation}
The process terminates because every new vector is a root. Its output
is the full root system, since every root is Weyl-conjugate to a
simple root. The vectors with nonnegative coordinates give $\Phi^+$.
We evaluated $D(d)$ by computer on this finite set of tuples, in integer
arithmetic; the numbers of tuples and the minima are listed in
Table~\ref{tab:shallow-gradings}. All $257$ admissible gradings have
$D(d)\ge0$, and the only zeros are the two stated in the lemma.

For type $A_r$ the same conclusion follows without enumeration.
Write the grading element on the standard module in decreasing order,
with multiplicities $b_0,\ldots,b_h$ at consecutive levels. Missing
levels have multiplicity zero, $b_0,b_h>0$, and $1\le h\le2$.
Then
\begin{equation}
 D=\sum_{j=0}^h b_j^2-1-\sum_{j=0}^{h-1}b_jb_{j+1},
 \qquad
 2D=(b_0^2-1)+(b_h^2-1)
       +\sum_{j=0}^{h-1}(b_j-b_{j+1})^2.
\end{equation}
Hence $D=0$ forces every $b_j=1$, giving the principal $A_1$ and
$A_2$ gradings.
\end{proof}

\begin{remark}
Table~\ref{tab:shallow-gradings} includes every possible simple
type in the proof of Lemma~\ref{lem:shallow-carriers}. Local flatness
excludes all but $A_1$ and $A_2$, so no further restrictions on their
embeddings are needed for that lemma.
The same root data give the graded pieces of
the ambient gradings. Counting roots by degree, with the Cartan subalgebra in
degree zero, gives
$(\dim\mathfrak l_0,\dim\mathfrak l_1,\dim\mathfrak l_2)=(46,16,0)$, $(67,32,1)$
and $(92,64,14)$ for the gradings \eqref{eq:exceptional-spin-gradings} of
$E_6$, $E_7$ and $E_8$, and $(22,8,7)$ for the grading of $F_4$ in
Lemma~\ref{lem:spin7-quadric}, in agreement with $\dim\mathfrak l_0+
2\dim\mathfrak l_1+2\dim\mathfrak l_2=78,133,248,52$.
\end{remark}

\begin{table}[ht]
\centering
\small
\renewcommand{\arraystretch}{1.12}
\begin{tabular}{lrrl}
\toprule
candidate type & number of $d$ & least $D(d)>0$ & degrees with $D(d)=0$\\\midrule
$A_1$ & 2 & 1 & $(1)$\\
$A_2$ & 5 & 2 & $(1,1)$\\
$A_3,A_4,A_5$ & $9,14,20$ & $1,2,3$ & none\\
$A_6,A_7,A_8$ & $27,35,44$ & $4,6,8$ & none\\
$D_4,D_5,D_6$ & $10,11,12$ & $2,3,6$ & none\\
$D_7,D_8$ & $13,14$ & $7,10$ & none\\
$E_6,E_7,E_8$ & $8,5,2$ & $8,14,28$ & none\\
$B_2,B_3,B_4$ & $3,4,5$ & $1,1,3$ & none\\
$C_2,C_3,C_4$ & $3,4,5$ & $1,3,6$ & none\\
$F_4$ & 2 & 8 & none\\\bottomrule
\end{tabular}
\caption{Candidate carrier factors and their nontrivial integral
gradings of depth at most two. Here $D(d)=\dim\mathfrak c_0-\dim\mathfrak c_1$;
local flatness requires $D(d)=0$. All computed differences are nonnegative.
Grouped counts and minima follow the order of the listed types.}
\label{tab:shallow-gradings}
\end{table}

\subsection{Occupation columns of the reduced supports}
\label{app:occupation-data}

We record the solutions of \eqref{eq:s14-columns-one} and
\eqref{eq:s14-column-intersections}. With $I_1=\varnothing$, put
$s_a=\lvert I_a\rvert$ and $t_{ab}=\lvert I_a\cap I_b\rvert=t_{ba}$ for
distinct $a,b\in\{2,\ldots,k\}$. Identify a column with the set
$S\subseteq\{2,\ldots,k\}$ of its occupied rows, and let $m_S$ be its
multiplicity.

For $k=2$ there are $s_2$ columns $01$ and $7-s_2$ columns $00$. For $k=3$
the multiplicities of $011,010,001,000$ are
\[
 t_{23},\quad s_2-t_{23},\quad s_3-t_{23},\quad 7-s_2-s_3+t_{23}.
\]
For $k=4$ let $U=\{2,3,4\}$. The column $0000$ is excluded, and
inclusion--exclusion gives
\[
 z:=m_U=7-\sum_{a\in U}s_a+\sum_{a<b}t_{ab},\qquad
 m_{\{a,b\}}=t_{ab}-z,\qquad
 m_{\{a\}}=s_a-\sum_{b\ne a}t_{ab}+z .
\]
For $k=5$ let $U=\{2,3,4,5\}$. Only subsets with two or three elements
occur. With $x_{ab}=m_{\{a,b\}}$, $y_a=m_{U\setminus\{a\}}$ and
$Y=\sum_ay_a$,
\[
 Y=\sum_as_a-14,\qquad
 y_a=s_a-\sum_{b\ne a}t_{ab}+Y,\qquad
 x_{ab}=t_{ab}-Y+y_a+y_b ,
\]
and the eleven equations are consistent exactly when
$2\sum_as_a-\sum_{a<b}t_{ab}=21$, which is equivalent to $E=28$.
Substituting the patterns of Proposition~\ref{prop:s14-nine} gives
Table~\ref{tab:s14-columns}.

\begin{table}[ht]
\centering
\renewcommand{\arraystretch}{1.15}
\small
\begin{tabular}{cc*{7}{c}}
\toprule
type & $k$ & $u^{(1)}$ & $u^{(2)}$ & $u^{(3)}$ & $u^{(4)}$
     & $u^{(5)}$ & $u^{(6)}$ & $u^{(7)}$\\
\midrule
I    & 1 & $0$ & $0$ & $0$ & $0$ & $0$ & $0$ & $0$\\
II   & 2 & $00$ & $00$ & $00$ & $01$ & $01$ & $01$ & $01$\\
III  & 3 & $000$ & $001$ & $001$ & $010$ & $010$ & $011$ & $011$\\
IV   & 2 & $00$ & $01$ & $01$ & $01$ & $01$ & $01$ & $01$\\
V    & 4 & $0001$ & $0010$ & $0011$ & $0100$ & $0101$ & $0110$ & $0111$\\
VI   & 3 & $001$ & $010$ & $010$ & $010$ & $011$ & $011$ & $011$\\
VII  & 4 & $0011$ & $0100$ & $0101$ & $0101$ & $0110$ & $0110$ & $0111$\\
VIII & 5 & $00111$ & $01001$ & $01010$ & $01011$ & $01100$ & $01101$ & $01110$\\
IX   & 4 & $0011$ & $0101$ & $0101$ & $0101$ & $0110$ & $0110$ & $0110$\\
\bottomrule
\end{tabular}
\caption{Occupation columns of the nine reduced supports. The $a$-th entry of
a column is the occupation of that mode in constituent $a$; the first
constituent is the vacuum. The columns may be permuted, and reading their
$a$-th entries gives the word $I_a$.}
\label{tab:s14-columns}
\end{table}

\section{Exact computations}
\label{app:exact-verification}

This appendix gives the orbit-dimension calculations, the support and
weight enumerations used in the remarks, and the Clifford-moment
calculation in dimension thirteen. The finite grading calculation used
in the completeness proof is given separately in
Appendix~\ref{app:shallow-enumeration}.

All calculations use exact arithmetic. By
\eqref{eq:creation-on-fock-basis}--\eqref{eq:annihilation-on-fock-basis}, each
$\gamma_r$ permutes the Fock basis up to factors in $\{\pm1,\pm i\}$.
Thus a spinor with coefficients in $\mathbb Z[i]$ gives orbit matrices
with entries in $\mathbb Z[i]$. Their ranks can be computed over
$\mathbb Q$ using
\begin{equation}
 \operatorname{rank}_{\mathbb Q(i)}(R+iS)
 =\frac12\operatorname{rank}_{\mathbb Q}
 \begin{pmatrix}R&-S\\S&R\end{pmatrix},
 \qquad R,S\in\operatorname{Mat}_{m\times d}(\mathbb Q).
\end{equation}
We used SymPy~\cite{SymPy2017} to evaluate Clifford expectation values symbolically and to verify the momentum-map values and the critical-point equation for the tabulated representatives. We used
\texttt{python-flint}~\cite{PythonFlint}, the Python interface to FLINT, to compute the ranks of the integer block matrices above and to solve the bordered Gram systems over $\mathbb{Q}$

\subsection{Orbit dimensions}
\label{app:orbit-dimensions}

Let $\psi$ be one of the representatives in
Tables~\ref{tab:finite-even-low}, \ref{tab:s14-orbits}
and~\ref{tab:finite-odd}. Its support is affinely independent, so by
Lemma~\ref{lem:torus-rescaling} replacing all nonzero coefficients by $1$
changes neither $\dim G\psi$ nor $\dim G[\psi]$. We take for
$X_1,\ldots,X_d$ in \eqref{eq:orbit-matrices} the operators
$\gamma_r\gamma_s$, $1\le r<s\le N$, restricted to $S_N$ (or to $S_N^+$), and
compute the ranks of $A_\psi$ and $B_\psi$. Because these are the operators
that actually act on the module, \eqref{eq:exact-orbit-rank} is valid also
when the action has a kernel.

For $N\le12$ and for odd $N$ the resulting dimensions are the last columns of
Tables~\ref{tab:finite-even-low} and~\ref{tab:finite-odd}. For $N=14$ the
ranks are
\[
\begin{array}{c|ccccccccc}
 &\mathrm I&\mathrm{II}&\mathrm{III}&\mathrm{IV}&\mathrm V&\mathrm{VI}
 &\mathrm{VII}&\mathrm{VIII}&\mathrm{IX}\\\hline
 \operatorname{rank}A_\psi&22&35&43&44&50&54&59&63&63\\
 \operatorname{rank}B_\psi&22&35&43&44&50&54&59&63&64
\end{array}
\]
and give the orbit dimensions in Table~\ref{tab:s14-orbits}. The first row
agrees with the vector-orbit dimensions of \cite[Table~1]{Gatti_dim13}.

\subsection{Supports in dimension fourteen}
\label{app:s14-enumeration}

Proposition~\ref{prop:s14-nine} was proved by counting incidences. The
following enumeration confirms it. By
Definition~\ref{def:reduced-coordinate}, a reduced support is a set of even
subsets of $\{1,\ldots,7\}$, any two of which differ in at least four
elements, such that no element lies in four of them and no element lies
outside four of them.

\begin{proposition}
\label{prop:s14-enumeration}
The numbers of reduced supports that contain $\varnothing$ and have
$1,2,3,4,5$ elements are $1$, $42$, $525$, $2170$ and $2100$, and no reduced
support has six elements. Under $W(D_7)$ these supports form $1,2,2,3,1$
classes, namely the supports of the types I--IX.
\end{proposition}

\begin{proof}
We adjoin even subsets to $\{\varnothing\}$ in increasing binary order and
keep a subset only if both conditions still hold. Both conditions pass to
subsets, so every reduced support containing $\varnothing$ is produced exactly
once. The group $W(D_7)$ acts on even subsets by permutations of
$\{1,\ldots,7\}$ and by symmetric difference with a fixed even subset, and
the subgroup fixing $\varnothing$ is the permutation group. Hence two supports
are equivalent if and only if they have the same canonical form, defined as
follows: move one of the elements to $\varnothing$ by a symmetric difference,
order the others, write the result as a $0/1$ matrix with one row for each
element and one column for each mode, sort its columns, and take the smallest
matrix obtained over all choices of the element and all orders of the others.
Computing the canonical forms gives the stated numbers of classes.
\end{proof}

If the condition on four elements is dropped, the same procedure gives
fifteen classes of root-distinct supports, with one to eight elements. For
each of them the bordered Gram system has a unique positive solution, and
reducedness holds exactly when property (A) holds. These are the data used in
Remarks~\ref{rem:property-A} and~\ref{rem:s14-nonreduced}. The chamber values
of the six non-reduced classes are those of the types IV, VI, VII, IX, VIII
and IX.

\subsection{Minimal combinations of weights}
\label{app:minimal-combinations}

A minimal combination of weights is the point closest to $0$ of the convex hull
of a nonempty set $\Sigma$ of weights
\cite{Kirwan_cohomology_84,MaciazekSawicki2015}. By Carath\'eodory's theorem
we may shrink $\Sigma$ until it is affinely independent and the point lies in
the relative interior of its convex hull; the point is then the point of the
affine hull of $\Sigma$ closest to $0$. Since the Weyl group acts transitively
on the weights, we may also assume that $\Sigma$ contains the highest weight. For every such $\Sigma$ we solved the bordered Gram system,
kept the solutions with positive weights, and moved the resulting points into
the positive chamber. This gives $6$, $12$ and $26$ minimal combinations for
$D_4$, $D_5$ and $D_6$, and $5$, $10$ and $26$ for $B_3$, $B_4$ and $B_5$, the
numbers quoted in Remark~\ref{rem:qubit-comparison}.

\subsection{Dimension thirteen}
\label{app:spin13-computation}

For $\psi_t$ as in Proposition~\ref{prop:spin13-moment} all
$\binom{13}{2}=78$ components of the momentum map vanish identically in $t$.
Of the $\binom{13}{3}=286$ products $\gamma_r\gamma_s\gamma_u$, exactly four
have a nonzero expectation value in $\psi_t$, and each of these values has
modulus $4t$. Hence
$\sum_{r<s<u}\lvert\langle\psi_t,\gamma_r\gamma_s\gamma_u\psi_t\rangle\rvert^2
=64t^2$, and with $\norm{\psi_t}^2=2(1+t^2)$ this gives
$F_3([\psi_t])=16t^2/(1+t^2)^2$.

\subsection{Checks of hand computations}

We also verified by direct substitution the identities used in the proofs of
Lemmas~\ref{lem:weyl-fock} and~\ref{lem:s14-four-to-two}, of the identity
\eqref{eq:B5-four-to-two} and of Proposition~\ref{prop:odd-orbit-bound}, and, for every representative in the
tables, all components of the momentum map and the identity
$\widehat\mu\psi=\norm\mu^2\psi$. 

\begingroup
\small
\bibliographystyle{alphaurl}
\bibliography{bibl}
\endgroup
\end{document}